\documentclass{article}

\PassOptionsToPackage{numbers, sort&compress}{natbib}
\usepackage[preprint]{neurips_2024}
\usepackage{multirow}
\usepackage{pmboxdraw}
\usepackage{bm}
\usepackage{enumitem}
\usepackage{listings}
\usepackage{xcolor}  % 用于代码高亮
\usepackage{booktabs}
\usepackage{graphicx}
\usepackage{makecell}
\usepackage{colortbl}
\usepackage{threeparttable}
\usepackage{subcaption}
\usepackage{pifont}
\definecolor{light-gray}{gray}{0.95}

\usepackage{tikz}
\usetikzlibrary{shapes.geometric, calc, decorations.pathreplacing, positioning, arrows.meta, pie}
\tikzstyle{startstop} = [rectangle, rounded corners, minimum width=3cm, minimum height=1cm, text centered, draw=black]
\tikzstyle{process} = [rectangle, minimum width=3cm, minimum height=1cm, text centered, draw=black]
\tikzstyle{arrow} = [thick,->,>=stealth]
\usepackage[utf8]{inputenc} % allow utf-8 input
\usepackage[T1]{fontenc}    % use 8-bit T1 fonts
\usepackage{hyperref}       % hyperlinks
\usepackage{url}            % simple URL typesetting
\usepackage{booktabs}       % professional-quality tables
\usepackage{amsfonts}       % blackboard math symbols
\usepackage{nicefrac}       % compact symbols for 1/2, etc.
\usepackage{microtype}      % microtypography
\usepackage{xcolor}         % colors
\usepackage{multirow}
\usepackage{algorithmic}
\usepackage{array}
\usepackage{textcomp}
\usepackage{stfloats}
\usepackage{verbatim}
\usepackage{subcaption}
\usepackage{tikz}
\usepackage{pgf-pie}
\usepackage{mathptmx}       % Times New Roman font

\usepackage[utf8]{inputenc} % allow utf-8 input
\usepackage[T1]{fontenc}    % use 8-bit T1 fonts
\usepackage{hyperref}       % hyperlinks
\usepackage{url}            % simple URL typesetting
\usepackage{booktabs}       % professional-quality tables
\usepackage{amsfonts}       % blackboard math symbols
\usepackage{nicefrac}       % compact symbols for 1/2, etc.
\usepackage{microtype}      % microtypography
\usepackage{xcolor}         % colors
\usepackage{graphicx}
\usepackage{newtxmath}

\title{Overview of the Amphion Toolkit (v0.2)}

\makeatletter
\def\thanks#1{\protected@xdef\@thanks{\@thanks
        \protect\footnotetext{#1}}}
\makeatother

\author{%
    {\bf Jiaqi Li\thanks{$^{\star}$Equal contribution, and their names are listed in random order.}\thanks{$^{\dagger}$Coordinator of this technical report.}$^{\star,\dagger}$, Xueyao Zhang$^\star$, Yuancheng Wang$^\star$, Haorui He$^\star$, Chaoren Wang$^\star$,  } \\
    {\bf Li Wang$^\star$, Huan Liao$^\star$, Junyi Ao$^\star$, Zeyu Xie$^\star$, Yiqiao Huang$^\star$, Junan Zhang$^\star$,}  \\
    {\bf Zhizheng Wu\thanks{$^{\ddagger}$Project Lead.}$^{\ddagger}$} \\ \\
    Amphion v0.2 Team \\
    \url{https://amphion.dev/}
}

\begin{document}

\maketitle

\tableofcontents
% \begin{abstract}
% Amphion is an open-source toolkit for \underline{A}udio, \underline{M}usic, and S\underline{p}eec\underline{h} Generat\underline{ion}, targeting to ease the way for junior researchers and engineers into these fields. 
% It presents a framework that includes diverse generation tasks and models, with the added bonus of being easily extendable for new incorporation. 
% This tutorial introduces the technologies and the usage of Amphion v0.2, its second major release developed throughout 2024. 
% % These include a new open-source 100K-hour multilingual dataset, a robust data preparation pipeline, and novel models for tasks such as text-to-speech, audio coding, speech enhancement, and voice conversion, alongside broader applications in audio, music, and speech generation.
% \end{abstract}

\newpage
\section{Introduction to the Amphion Toolkit}
Amphion is an open-source toolkit for \underline{A}udio, \underline{M}usic, and S\underline{p}eec\underline{h} Generat\underline{ion}, targeting to ease the way for junior researchers and engineers into these fields. 
Amphion is released in December 2023 where the rapid advancements in generative models have revolutionized the fields of audio, music, and speech processing.
Amphion incorporates these most up-to-date generative technology, and is designed to be a unified framework that includes diverse generation tasks and models, with the added bonus of being easily extendable for new incorporation. 
The toolkit is designed with beginner-friendly workflows and pre-trained models, allowing both beginners and seasoned researchers to kick-start their projects with relative ease. 
Since its initial launch in December 2023, Amphion has grown rapidly and received more than 8K stars on GitHub. 
Its dataset has become the most monthly downloaded audio dataset on the HuggingFace platform.

% However, the steep learning curve associated with these technologies often hinders newcomers from actively contributing to or innovating within the field. Replicating recent findings are also difficult because of the scattered nature of open-source tools.
% To address these, we released \textbf{Amphion v0.1}~\cite{amphion} in December 2023, a versatile open-source toolkit tailored to democratize access to cutting-edge generative technologies in audio, music and speech.

\subsection{What's new in Amphion v0.2}

\textbf{Amphion v0.2} builds upon its initial release, significantly broadening its scope and utility.
Amphion v0.2 expands on both datasets and models, covering a wider range of tasks in the fields of audio, music, and speech generation. The following table summarizes the technology supported in Amphion v0.2. Models introduced in this version are highlighted in bold fonts, which are also \textit{exclusively} available in Amphion.

\begin{table}[ht]
\centering
\caption{Summary of the technology supported by Amphion v0.2. New models released in Amphion v0.2 are highlighted in bold fonts.
All highlighted models are also exclusively in Amphion.}
\label{tab:amphion_v2}
\begin{tabular}{|l|l|}
\hline
Task                & Technology                                                                                                 \\ \hline
Datasets                 & \textbf{Emilia}~\cite{emilia} (100K hrs Multil.)                                                                        \\
                     & \textbf{SpMis}~\cite{spmis} (4939 hrs EN)                                                                                         \\
                     & \textbf{SD-Eval}~\cite{sdeval} (8.8 hrs EN)                                                                                       \\
                     & \textbf{Emilia-Debatts}~\cite{debatts} (111 hrs ZH)                                                                     \\
                     & and 18 datasets supported in Amphion v0.1                                                              \\ \hline
Text to Speech                  & \textbf{MaskGCT}\cite{maskgct}                                                                                       \\
                     % & \textbf{DualCodec-MaskGCT}                                                                                  \\
                     & \textbf{Vevo}~\cite{vevo}                                                                                          \\
                     % & \textbf{DualCodec-VALLE}                                                                                    \\
                     & \textbf{Debatts}~\cite{debatts}                                                                                       \\
                     & VALLE~\cite{valle}                                                                                                  \\
                     & NaturalSpeech2~\cite{naturalspeech2}                                                                                         \\ \hline
Neural Audio Codec                
% & \textbf{DualCodec}                                                                                     \\
                     & \textbf{FACodec}~\cite{naturalspeech3}                                                                                       \\ \hline
Speech Enhancement                   & \textbf{AnyEnhance}~\cite{zhang2025anyenhance}                                                                                    \\ \hline
Voice Conversion                   & \textbf{Noro}~\cite{noro}                                                                                          \\
                     & \textbf{Vevo}~\cite{vevo}                                                                                          \\ \hline
Text to Audio                  & \textbf{PicoAudio}~\cite{picoaudio}                                                                                   \\
                     & AudioLDM~\cite{liu2023audioldm}                                                                                               \\ \hline
Singing Voice Converson                  & DSFF-SVC~\cite{multiple-contents-svc}                                                                                      \\ \hline
Vocoder              & MS-CQT Discriminator~\cite{cqt}\\&and 9 vocoders in Amphion v0.1                         \\ \hline
Visualization        & SingVisio~\cite{singvisio}                                                                                     \\ \hline
\end{tabular}
\end{table}

The new features in v0.2 are summarized as follows:
\begin{itemize}
    \item \textbf{100K-hour multilingual speech dataset:} A large-scale dataset processed with Emilia-Pipe~\cite{emilia} using Internet data, marking Amphion’s first initiative in open-sourcing datasets.
    \item \textbf{Specialized datasets:} Tailored datasets for downstream scenarios, including audio deepfake detection, debating text-to-speech synthesis, and spoken dialogue understanding.
    % \item \textbf{Expanded application tasks:} Support for new tasks such as neural audio coding, speech enhancement, voice conversion, and visualization.
    \item \textbf{Expanded application tasks:} Support for new tasks such as neural audio coding, voice conversion, and visualization.
    \item \textbf{State-of-the-art TTS:} Release of advanced text-to-speech (TTS) models trained on Amphion's large-scale dataset, offering competitive performance among open-source works.
\end{itemize}

% 
% % It is released under a permissive MIT license, 
% % Our 100K-hour multilingual speech dataset in Amphion v0.2 is being downloaded for 40K+ times per month. 
% % Amphion v0.2 introduces significant advancements over its predecessor.
% In the following sections, we describe each new integration in Amphion v0.2 and finally show experiment results conducted within Amphion toolkit.

% Amphion v0.2 not only introduces new state-of-the-art models but also includes robust tools for visualization, speech enhancement, voice conversion, and text-to-audio tasks. The addition of large-scale multilingual datasets ensures researchers and developers have access to the necessary resources for training and benchmarking models. Each supported task reflects Amphion's commitment to building a unified, open-source platform for audio, music, and speech generation.

\section{The Fundamentals of Amphion}

\subsection{Neural Audio Codecs}
\noindent \textbf{Audio representations.}
A \textit{waveform} represents the raw audio signal as a function of amplitude over time. In the analog domain, it is a continuous signal, but it becomes discrete when sampled at a fixed rate, such as 44.1 kHz or 16 kHz. Waveforms offer the highest temporal resolution, capturing all the fine-grained details of an audio signal. The earliest deep neural network for audio generation, WaveNet~\cite{wavenet}, directly modeled raw audio waveforms. However, this approach is computationally inefficient due to the high dimensionality of waveforms and is highly sensitive to noise.

In contrast, the \textit{Mel spectrogram} is a time-frequency representation derived from the Short-Time Fourier Transform (STFT). It maps the linear frequency scale to the perceptually motivated Mel scale, which emphasizes frequencies crucial to human hearing. Represented as a 2D matrix, the horizontal axis corresponds to time frames, while the vertical axis represents Mel frequency bins~\cite{huggingface_audio_course_chapter_1_audio_data}. Unlike raw waveforms, Mel spectrograms provide a more compact, continuous-valued representation. Many audio generative models predict Mel Spectrograms~\cite{voicebox,cosyvoice,liu2023audioldm}. They can be inverted back into waveforms using neural vocoders~\cite{hifigan,vocos,bigvgan}, which enable high-quality reconstruction of the original audio signal.

\noindent \textbf{Audio Codecs.}
A \textit{Codec} is a combination of the words coder/decoder. It is a computer algorithm
designed to compress and decompress digital audio signals, enabling efficient storage and transmission of audio data.
Established \text{signal processing-based codecs} like MP3, FLAC(Free Lossless Audio Codec), Opus~\cite{opus} rely on hand-engineered features and psychoacoustic principles to remove redundant or imperceptible components of audio.
Compared to continuous-valued Mel Spectrograms, audio codecs output a discrete sequence often with a small vocabulary, allowing the discrete generative modeling of audio such as using large language models. 
% By reducing the size of audio files, codecs make it possible to store more content on devices, stream audio over limited-bandwidth networks, and facilitate real-time communication. 
Recent advancements in deep learning have given rise to \text{neural audio codecs}, which learn to compress and reconstruct audio data directly from large-scale datasets.
Neural audio codecs today are used to extract discrete or continuous representations for speech, audio and music generation tasks~\cite{musicgen}.

\begin{figure}[htbp]
    \centering
    \includegraphics[width=0.6\linewidth]{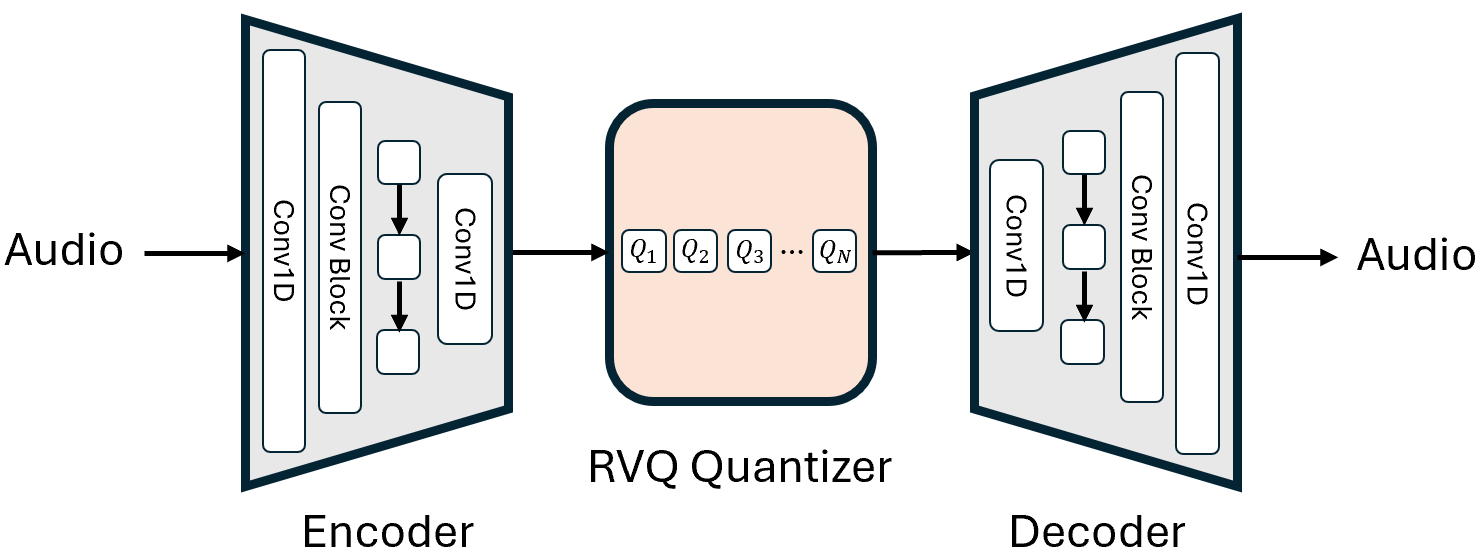}
    \caption{A High-level architecture of neural audio codecs.}
    \label{fig:codec}
    \vspace{-5mm}
\end{figure}

\textbf{Neural Audio Codecs.} Neural audio codecs use machine learning models, often based on Vector Quantized VAEs (VQ-VAEs)~\cite{vq-vae}, to compress audio.
By leveraging neural networks, these models deliver higher audio quality than signal processing-based codecs with higher compression rates.
Figure \ref{fig:codec} presents a high-level architecture of a neural audio codec~\cite{codecinvestigation}. It has an convolutional encoder that downsamples waveforms to a much lower sampling rate (e.g., 50Hz), a residual vector quantization (RVQ) module that discretizes latent features, and a decoder that reconstructs audios from discrete tokens.
Neural audio codecs are used not only for audio transmission, but also for extracting useful discrete and continuous audio representations for generative models~\cite{CodecSuperb2024}.

The first neural audio codec was SoundStream\cite{soundstream}, which proposed Residual Vector Quantization (RVQ) to achieve high-quality audio compression. 
Each layer in the RVQ module progressively quantizes the residual of its previous layer, capturing finer details in the input signal at each step.
SoundStream optimized its encoder-RVQ-decoder architecture using a combination of reconstruction and adversarial losses, which enhanced its ability to produce realistic audio outputs. 
Building on this foundation, Encodec~\cite{encodec} extended the capabilities of earlier models by integrating LSTM layers and a transformer-based language model to enhance sequence modeling over the quantized tokens. These enhancements not only improved the compression performance but also allowed for more robust handling of diverse audio types.
Further innovations addressed specific challenges in neural audio codecs, such as codebook collapse—a common issue in quantization. For example, DAC\cite{dac} reduced the dimension of the latent vector for quantization and replaced the ReLU activation function with the snake activation function\cite{snake-actiFunc}, which proved particularly effective in reconstructing periodic signals like speech and music.
% These models learn latent representations of audio signals and achieve compression by encoding the signal into a compact latent space. Neural codecs can also adapt dynamically to diverse audio content, leveraging neural networks’ ability to learn complex features.
% Audio codecs are widely used in applications such as music streaming, video conferencing, telecommunication, and digital broadcasting.
% The objective of the algorithm is to represent the high-fidelity audio signal with a minimum number of bits (compressing) while retaining quality.

\textbf{Self-Supervised Representation Codecs.}
Another type of discrete speech tokens is the semantic tokens extracted from self-supervised model representations.
These semantic tokens are extracted by k-means or vector quantization (VQ) on self-supervised (SSL) representations~\cite{borsos2023audiolm,repcodec}. 
They possess rich semantic and pronunciation information, but cannot be directly used to reconstruct audios due to a lack of acoustic traits like speaker identity.
Semantic token-based speech generation models usually predict semantic tokens in the stage one model, and predict audio codec tokens or Mel Spectrograms in the stage two model~\cite{borsos2023audiolm,borsos2023soundstorm,maskgct}.
Other than extracting from self-supervised features, recently, supervised ASR model features like Whisper~\cite{whisper} have been used to extract the semantic tokens~\cite{cosyvoice}.

\subsection{Speech Language Models}

\subsubsection{Autoregressive Language Models}

Autoregressive language models form a foundational approach in speech generation, leveraging the sequential nature of speech data. 
These models generate output tokens one by one, conditioning each token on all previously generated tokens. Formally, given a sequence of tokens $\boldsymbol{x} = [y_1, y_2, \ldots, y_n]$, ARLMs the joint probability of the sequence as a product of conditional probabilities:

\[
p_\theta(\boldsymbol{x}) = \prod_{i=1}^{n} p_\theta(y_i \mid y_1, y_2, \ldots, y_{i-1}, \boldsymbol{c}),
\]

where $\boldsymbol{c}$ represents optional conditioning information, such as semantic, linguistic, or acoustic features. The model parameters $\theta$ are typically optimized using maximum likelihood estimation (MLE) to minimize the negative log-likelihood of the observed sequences in the training data:

\[
\mathcal{L}_{\text{AR}} = - \mathbb{E}_{\boldsymbol{x}, \boldsymbol{c}} \sum_{i=1}^{n} \log p_\theta(y_i \mid y_1, y_2, \ldots, y_{i-1}, \boldsymbol{c}).
\]

In the context of speech generation, auto-regressive language models(ARLMs) are commonly employed to model either discrete representations of speech (e.g., linguistic tokens or acoustic tokens derived from neural audio codecs) or directly raw waveforms. At the inference stage, ARLMs generate speech tokens autoregressively, where the generation of each token depends on all previously generated tokens, ensuring high fidelity in sequential dependencies but introducing latency challenges.

Several works have successfully applied ARLMs to speech generation tasks. AudioLM~\cite{borsos2023audiolm} demonstrates the use of ARLMs to generate semantic and acoustic tokens for high-quality speech generation. VALL-E~\cite{valle} extends autoregressive modeling to zero-shot TTS, achieving remarkable speaker fidelity. UniAudio~\cite{uniaudio} further generalizes ARLMs to unified audio modeling tasks, including speech, music, and sound, by leveraging a shared autoregressive architecture.

\subsubsection{Masked Generative Models}
Masked generative models (MGMs) are another class of speech language models. In this section, we provide a brief introduction to masked generative models~\cite{chang2022maskgit}. Consider a discrete sequence $\boldsymbol{x} = [y_1, y_2, \ldots, y_n]$, where $n$ denotes the length of the sequence. We define $\boldsymbol{x}_t = \boldsymbol{x} \odot \boldsymbol{m}_t$ as the operation of masking a subset of tokens in $\boldsymbol{x}$ using the corresponding binary mask $\boldsymbol{m}_t = [m_{t,1}, m_{t,2}, \ldots, m_{t,n}]$. Specifically, this operation involves replacing $x_i$ with a special $\texttt{[MASK]}$ token if $m_{t,i}=1$, and otherwise leaving $x_i$ unmasked if $m_{t,i}=0$. Here, each $m_{t,i}$ is independently and identically distributed according to a Bernoulli distribution with parameter $\gamma(t)$, where $\gamma(t) \in (0,1]$ represents a mask schedule function (for example, $\gamma(t) = \sin(\frac{\pi t}{2T}), t \in (0,T]$). We denote $\boldsymbol{x} = \boldsymbol{x}_0$. The masked generative models are trained to predict the complete sequence (masked tokens) based on the observed tokens (unmasked tokens) and the condition $\boldsymbol{c}$, which can be modeled as \( p_\theta(\boldsymbol{x}_0 \mid \boldsymbol{x}_t, \boldsymbol{c}) \), and the model parameters \(\theta\) are trained to optimize the sum of the marginal cross-entropies for each unobserved token:

\begin{equation}
    \begin{aligned}
        \mathcal{L}_{\text{mask}} &= - \mathbb{E}_{\boldsymbol{x}, t, \boldsymbol{m}_t} \sum_{i=1}^{n} m_{t,i} \cdot \log p_{\theta}(y_i \mid \boldsymbol{x}_t, \boldsymbol{c})
    \end{aligned}
\end{equation}

Note that $\boldsymbol{c}$ may be empty, for example, during the unconditional pre-training stage of our model. At the inference stage, masked generative models generate tokens in parallel through iterative decoding. The process begins with a fully masked sequence $\boldsymbol{x}_T$. Assuming the total number of decoding steps is $S$, for each step $j$ from $1$ to $S$, we first sample $\boldsymbol{\hat{x}}_0$ from $p_{\theta}(\boldsymbol{x}_0 \mid \boldsymbol{x}_{T - (j-1) \cdot \frac{T}{S}}, \boldsymbol{c})$. Then we sample $\lfloor n \cdot \gamma(T - j \cdot \frac{T}{S}) \rfloor$ tokens based on the confidence score to remask, resulting in $\boldsymbol{x}_{T - j \cdot \frac{T}{S}}$, where $n$ is the sequence length of $\boldsymbol{x}$. The confidence score for $\hat{y}_i$ in $\boldsymbol{\hat{x}}_0$ is assigned to $p_{\theta}(y_i \mid \boldsymbol{x}_{T - (j-1) \cdot \frac{T}{S}}, \boldsymbol{c})$ if $y_{T - (j-1) \cdot \frac{T}{S}, i}$ is a $\texttt{[MASK]}$ token; otherwise, we set the confidence score of $\hat{y}_i$ to 1, indicating that tokens already unmasked in $\boldsymbol{x}_{T - (j-1) \cdot \frac{T}{S}}$ will not be remasked. In particular, we choose $\lfloor n \cdot \gamma(T - j \cdot \frac{T}{S}) \rfloor$ tokens with the lowest confidence scores to be masked. Note that the method for calculating the confidence score is not unique. For example, \cite{lezama2022improved} introduces Token-Critic, training a critic model to compute confidence scores, aiding the sampling process. Additionally, \cite{lezama2022improved, xie2024show} suggest that masked generative modeling can be seen as a simplified version of discrete diffusion.

In the speech domain, SoundStorm~\cite{borsos2023soundstorm} uses the semantic tokens from AudioLM~\cite{borsos2023audiolm} and employs MGMs to generate acoustic tokens from a neural audio codec~\cite{zeghidour2021soundstream}, enabling applications like TTS and voice conversion. NaturalSpeech 3~\cite{naturalspeech3} adopts MGMs to generate disentangled speech tokens. MaskGCT~\cite{maskgct} further leverages MGMs for zero-shot generation, eliminating the need for explicit text-speech alignment or phone-level duration prediction in non-autoregressive TTS models. MaskSR~\cite{li2024masksr} applies MGMs to speech enhancement tasks. In this work, we propose a unified speech generation framework based on MGMs.

\subsection{Diffusion and Flow-Matching Models}
\textbf{Denoising Diffusion Probabilistic Models.}
The remarkable advancements in generative models have greatly expanded the capabilities of content creation. Behind many generative tools, there is a specialized sampling mechanism: diffusion models. Denoising Diffusion Probabilistic Models (DDPM) \cite{ddpm, sde} try to model a data distribution $p_\theta(x_0)$ by two processes:

(1) The forward process progressively transforms the data distribution into a standard Gaussian distribution using a predefined noise schedule $0 < \beta_1 < \beta_2 < \cdots < \beta_n < \cdots < \beta_N < 1$. At each time step $n \in [1, \dots, N]$, the transition probability are defined as:
\begin{equation}
q(x_n \mid x_{n-1}) = N(x_n; \sqrt{1-\beta_n} \, x_{n-1}, \beta_n I)
\end{equation}
By leveraging the stepwise transition probabilities $q(x_n \mid x_{n-1})$, the direct probability distribution $q(x_n \mid x_0)$ can be derived as:
\begin{equation}
q(x_n \mid x_0) = N(x_n; \sqrt{\bar{\alpha}_n} \, x_0, (1 - \bar{\alpha}_n) \epsilon)
\end{equation}
where $\epsilon \sim N(0, I)$ is standard Gaussian noise, $\alpha_n$ is reparameterized as $1 - \beta_n$, and $\bar{\alpha}_n = \prod_{s=1}^n \alpha_s$ represents the cumulative noise level at step $n$.

(2) A reverse denoising process reconstructs $x_0$ using the reweighted noise estimation $\hat{\epsilon}_\theta$. The optimization objective can be simplified to:
\begin{equation}
L_{DM}(\theta) = \mathbb{E}_{x_0, \epsilon \sim N(0, I), n} \left\| \epsilon - \epsilon_\theta(x_n, n) \right\|_2^2
\end{equation}
where $\epsilon_\theta(x_n, n)$ represents a sequence of denoising autoencoders trained to iteratively predict noise based on input $x_n$ and step $n$.

% The estimated noise is then employed for the reconstruction of $x_0$.

\textbf{Latent Diffusion Models.}
Unlike traditional diffusion models, Latent Diffusion Models (LDM) conducts the diffusion process \textbf{in the latent representation space}, reducing computational complexity in DDPM. LDM \cite{stablediffusion} aiming to approximate the true conditional data distribution $q(z_n \mid c)$ with a model distribution $p_\theta(z_0 \mid c)$, where $z_0$ is the prior of an audio sample in latent space.

LDM \cite{stablediffusion} introduces conditioning mechanisms into the diffusion process, enabling multimodal training through cross-attention and enabling conditional generation tasks. Specifically, it trains a conditional denoising autoencoder $\epsilon_\theta(z_n, n, \tau_\theta(y))$ to guide the generation process via the provided conditions $y$. The condition $y$ can represent various modalities (e.g., text, phoneme), is mapped to an intermediate representation $\tau_\theta(y)$ by the encoder $\tau_\theta$. The final model incorporates the control conditions into the UNet's intermediate layers through a cross-attention mapping. The optimization objective is:
\begin{equation}
L_{LDM}(\theta) = \mathbb{E}_{\varepsilon(x), y, \epsilon \sim N(0, I), n} \left\| \epsilon - \epsilon_\theta(z_n, n, \tau_\theta(y)) \right\|_2^2
\end{equation}
where both $\epsilon_\theta$ and $\tau_\theta(y)$ are jointly optimized.

Several audio generation models \cite{liu2023audioldm, picoaudio, naturalspeech2}, supported by Amphion v0.2, explore the potential of diffusion-based methods and achieve high-quality audio synthesis. For example, AudioLDM \cite{liu2023audioldm} uses text embeddings $E^y$ as conditions to predict the noise $\epsilon_\theta(z_n, n, \tau_\theta(y))$. It employs a frozen VAE encoder $\mathcal{E}$ to encode mel-spectrograms $x \in \mathbb{R}^{H \times W \times 3}$ into latent representations $z \in \mathbb{R}^{C \times \frac{T}{r} \times \frac{F}{r}}$, which serve as the target for generation. LDM's denoising process conditioned on $E^y$ gradually generates the corresponding audio prior $\hat{z}_0$. During the sampling process, the decoder $\mathcal{D}$ reconstructs the mel-spectrogram $\hat{X}$ from the audio prior $\hat{z}_0$.

\textbf{Flow-Matching Models.} %待增加cosyvoice
Considering that diffusion models limit the choice of sampling probability paths, leading to long training times and inefficient sampling, Flow-Matching Models\cite{flow-matching} (FM) is a novel generative framework that transforms a simple prior distribution $p$ (usually standard Gaussian distribution) to a target distribution $q$ with learning a time-dependent flow mapping $\phi: [0,1] \times \mathbb{R}^d \to \mathbb{R}^d$, defined via the ordinary differential equation (ODE):
\begin{equation}
\frac{d\phi_t(x)}{dt} = v_t(\phi_t(x));    \phi_0(x) = x
\end{equation}
Here, $v_t$ represents a vector field that is learned by optimizing the Flow Matching (FM) objective. Given a vector field $u_t(x)$ and its generated conditional probability path $p_t(x)$, the objective is try to align the model's vector field $v_t(x; \theta)$ with the target vector field $u_t(x)$ and is expressed as:
\begin{equation}
\mathcal{L}_{FM}(\theta) = \mathbb{E}_{t, p_t(x)} \left\| v_t(x; \theta) - u_t(x) \right\|^2
\end{equation}
where $t \sim \mathbb{U}[0,1]$ is flow step, $p_t(x)$ is a target probability density path and $u_t(x)$ is the corresponding vector field. Since both $p_t(x)$ and $u_t(x)$ are unknown, given a data sample $z$, they can be derived using the marginal probability from the conditional probability:
\begin{align}
p_t(x) = \int p_t(x \mid z) q(z) \, dz, \\
u_t(x) = \int u_t(x \mid z) \frac{p_t(x \mid z) q(z)}{p_t(x)} \, dz
\end{align}
Conditional Flow Matching (CFM) regress $v_t(x)$ on the conditional vector field $u_t(x \mid z)$ and its generated conditional probabilistic path $p_t(x \mid z)$. This approach transforms the intractable marginal flow matching problem into a computable conditional problem, reformulating it into the Conditional Flow Matching (CFM) objective:
% Notably, the marginal vector field and the conditional probabilistic path satisfy the continuity equation
\begin{equation}
\mathcal{L}_{CFM}(\theta) = \mathbb{E}_{t, q(x_1), p_t(x \mid x_1)} \left\| v_t(x) - u_t(x \mid x_1) \right\|^2
\end{equation}

CosyVoice\cite{cosyvoice} employs optimal-transport conditional flow matching (OT-CFM) \cite{flow-matching} to model the transition from the prior distribution $p_0(x)$ to the target Mel spectrogram distribution $q(x)$. The OT-CFM flow loss is:
\begin{equation}
\mathcal{L}_{OT-CFM} = \mathbb{E}_{t, q(x_1), p_0(x_0)} \left| \omega_t(\phi_t^{OT}(x_0, x_1) \mid x_1) - \nu_t(\phi_t^{OT}(x_0, x_1) \mid \theta) \right|
\end{equation}
where $\phi_t^{OT}(x_0, x_1)$ represents the flow from $x_0$ to $x_1$, defined as:
\begin{align}
\phi_t^{OT}(x_0, x_1) &= (1 - (1 - \sigma)t) x_0 + t x_1
\end{align}
which represents the linear interpolation between $x_0$ and $x_1$ over time $t \in [0, 1]$. Each target sample $x_1$ is matched to a random sample $x_0 \sim N(0, I)$. The gradient vector field along this flow is defined as:
\begin{align}
\omega_t(\phi_t^{OT}(x_0, x_1) \mid x_1) &= x_1 - (1 - \sigma) x_0
\end{align}
which serves as the learning target for the neural network. This vector field is linear, time-invariant, and depends only on $x_0$ and $x_1$. The speaker embedding $\mathbf{v}$, speech tokens $\{\mu_l\}_{1:L}$, and masked mel-spectrogram $\tilde{x}_1$ are used as conditional inputs to guide the neural network in approximating the vector field $\nu_t(\phi_t^{OT}(x_0, x_1) \mid \theta)$:
\begin{equation}
\nu_t(\phi_t^{OT}(x_0, x_1) \mid \theta) = \text{NN}_\theta \left( \phi_t^{OT}(x_0, x_1), t; \mathbf{v}, \{\mu_l\}_{1:L}, \tilde{x}_1 \right)
\end{equation}
Here, $\tilde{x}_1$ is a masked version of $x_1$.
% This design ensures the neural network focuses on learning meaningful transitions between $x_0$ and $x_1$, while maintaining content and speaker identity.

% During sampling, the denoising process leverages an Euler solver to compute the flow dynamics \frac{d\phi_t(x)}{dt} iteratively, conditioned on masked Mel spectrogram $\tilde{z}$, speaker embedding, and a CFG mechanism. The generated prior $\hat{z}_0$ is then directly interpreted as the mel-spectrogram.

% CoSyVoice define the optimal-transport flow $\phi_t^{OT}(x_0, x_1)$ between paired samples $x_0$ and $x_1$ to model the dynamic transition in data space. It employs a neural network parameterized by $\theta$ to approximate the velocity field $\nu_t(\phi_t^{OT}(x_0, x_1) | \theta)$, which is trained to match the ground truth velocity field $\omega_t(\phi_t^{OT}(x_0, x_1) | x_1)$. The speaker embedding $\mathbf{v}$, speech tokens $\{\mu_l\}_{1:L}$, and masked mel-spectrogram $\tilde{X}_1$ are fed into the network to conditionally guide the velocity field estimation. The learned velocity field enables high-quality generation of mel-spectrograms $\hat{X}$ that capture the transitions between $x_0$ and $x_1$, while preserving content and speaker identity.

% This approach ensures that the reverse dynamics allow reconstruction of the true data distribution from the noise.
% CosyVoice introduces a Conditional Flow-Matching (CFM) model that serves as a vocoder in TTS system

% \subsection{Speech Enhancement}
\subsection{Deepfake Detection}
The rapid development of voice synthesis technology has raised significant concerns about the security risks associated with its misuse. State-of-the-art text-to-speech (TTS) models~\cite{maskgct} are now capable of cloning any individual’s voice using only a few seconds of audio. The term "deepfake" is commonly used by the media and public to describe any audio or video content in which key attributes have been digitally modified or replaced using artificial intelligence (AI) techniques. To mitigate these risks, researchers have introduced the Deepfake Detection task, which aims to distinguish AI-generated samples from genuine ones.

With the continuous advancements in deep learning, researchers have increasingly adopted deep learning-based methods for deepfake detection. \cite{tak2021endtoenda} employed Graph Attention Networks (GAT) to model time segments and frequency subbands separately, and utilized GAT for feature fusion. \cite{jung2022aasist} recognized the heterogeneity of temporal and spectral information in spectrograms, and therefore enhanced the graph neural network's ability to model the heterogeneity of time-frequency features using heterogeneous attention mechanisms (AASIST). Building upon AASIST, \cite{tak2022automatica} further improved the performance by incorporating the pre-trained model Wav2Vec \cite{baevski2020wav2veca} and the enhancement method RawBoost~\cite{tak2022rawboost}.

It is undeniable that the development of speech synthesis technology has significantly enhanced convenience in various domains, such as in-car navigation systems, e-readers, and intelligent robots. However, while current research primarily focuses on distinguishing machine-generated speech from human speech, a more pressing challenge lies in detecting misinformation embedded within spoken content. This task requires a comprehensive analysis of factors such as speaker identity, topic, and contextual consistency. To address this challenge, we introduce the open-source dataset SpMis~\cite{spmis} (Introduced in Section \ref{sec:tech_emilia} and conduct an initial investigation into misinformation detection in synthesized speech. The SpMis dataset comprises speech samples generated by state-of-the-art text-to-speech (TTS) systems, covering five common topics and involving over 1,000 speakers, providing a valuable resource for advancing research in this domain.

\subsection{Speech Generation Datasets}
Speech generation datasets are essential resources for training models to produce natural-sounding speech. These datasets typically consist of paired audio and text data, enabling the learning of mapping text to speech. 

In Table~\ref{tab:speech_datasets}, we summarize existing speech generation datasets. Early datasets, such as LJSpeech~\cite{ljspeech} and VCTK~\cite{vctk}, focus on audiobook or studio recordings and are limited to less than 100 hours of data. While larger datasets like Libri-Light~\cite{librilight} and MLS~\cite{mls} have scaled to tens of thousands of hours, they predominantly consist of audiobook data, which may not generalize well to more spontaneous speech scenarios.

Recent efforts, such as AutoPrepWild~\cite{AutoPrep} and GigaSpeech~\cite{gigaspeech}, have aimed to bridge this gap by collecting in-the-wild data from diverse sources, introducing more naturalistic and spontaneous speech. However, these datasets still fall short in terms of total duration, multilingual coverage, and extendability.
Extendable datasets allow researchers to incorporate new data, languages and speakers by running the data pre-processing pipeline. 

Amphion v0.2's Emilia~\cite{emilia} addresses these challenges by providing 101K hours of in-the-wild data across six languages (English, Chinese, German, French, Japanese, and Korean) at a sampling rate of 24kHz. 
Emilia stands out not only for its scale and multilingual coverage but also for its extendability, by open-sourcing a pipeline for dataset expansion, supporting the growing demand for diverse and robust speech generation models.

\begin{table*}[htbp]
    \centering
    \caption{A comparison of Emilia in Amphion v0.2 with existing datasets for speech generation.}
    \label{tab:speech_datasets}
    \resizebox{\textwidth}{!}{
        \begin{tabular}{ccccccc}
            \toprule
            \textbf{Dataset} & \textbf{Data Source} & \textbf{Duration (hrs)} & \textbf{Lang.} & \textbf{SR (Hz)}  & \textbf{Extentable} \\
            \midrule
            LJSpeech~\cite{ljspeech} & Audiobook & 24 & En & 22.05k  &   \\
            AutoPrepWild~\cite{AutoPrep} & In-the-wild & 39 & Zh & 24k/44.1k   & not open-source\\
            VCTK~\cite{vctk} & Studio Recording & 44 & En & 48k  &  \\
            Aishell-3~\cite{aishell3}& Studio Recording  & 85 & Zh & 44.1k   &   \\
            LibriTTS~\cite{libritts} & Audiobook & 585 & En & 24k & &  \\
            GigaSpeech~\cite{gigaspeech}& In-the-wild & 10k & En & 16k &   \\
            WenetSpeech4TTS~\cite{wenetspeech4tts}& In-the-wild & 12k & Zh & 16k  & not open-source \\
            MLS~\cite{mls} & Audiobook & 51k & En/Fr/De/Nl/Es/It/Pt/Pl & 16k &  \\ 
            Libri-Light~\cite{librilight} & Audiobook & 60k & En & 16k  &  \\
            \midrule
            Amphion v0.2: Emilia & In-the-wild & 101k  & En/Zh/De/Fr/Ja/Ko& 24k   & \checkmark \\
            % Amphion v0.2: Emilia-Debatts & Public debates & 112  & ZH & 16k   & \checkmark \\
            % SD-Eval & Audiobook & 8.8  & En & 24k   & \checkmark \\
            \bottomrule
        \end{tabular}
    }
\end{table*}

% Following the trend in LLM training, recent developments in speech generation have turned to large-scale pretraining datasets sourced from in-the-wild environments, such as Internet audio, debates, and multimedia content. These datasets introduce spontaneous speech patterns, including breathing, pausing, hesitations, and varying emotions, which are essential for building models that generalize to natural, conversational settings. While these datasets may contain background noise or imperfect annotations, they offer substantial benefits by better reflecting real-world speech variability.

% \subsection{Visualization}
% Diffusion-based generative models have emerged as a cutting-edge research focus and achieve state-of-the-art results in many audio, music and speech applications.
% One popular application of diffusion models is singing voice conversion (SVC). 
% This advanced technique effectively alters one singer's voice to another's, while meticulously preserving the song's original content and melody.

\subsection{Speech LLMs}
LLMs have shown remarkable flexibility \cite{achiam2023gpt,team2023gemini,touvron2023llama,yang2024qwen2,team2024gemma}, acting as a universal interface capable of offering assistance across a wide spectrum of tasks and domains.
Initially, their strengths are most evident in text-based applications, such as language understanding, content generation, and complex question answering.
Recently, their abilities have expanded beyond textual input to accommodate multi-modal data, such as speech \cite{chu2023qwen,chu2024qwen2,tang2024salmonn,hu2024wavllm,openai2024gpt4o,gong2024listen,kong2024audio,deshmukh2023pengi,defossez2024moshi,fang2024llama,xie2024mini}.
This development greatly enhances the range of tasks they can tackle, allowing them to support richer and more dynamic interactions with users.

% These multi-modal LLMs have demonstrated impressive performance on tasks, such as speech recognition and speech translation.
Building on these expanded abilities, multi-modal LLMs have demonstrated impressive performance on tasks that involve not only understanding written language but also interpreting spoken language cues, such as speech recognition and speech emotion recognition.
They generally centre on interpreting input signals and carrying out tasks directed by text-based instructions \cite{chu2023qwen,tang2024salmonn,gong2024listen,deshmukh2023pengi,kong2024audio}.

Meanwhile, in pursuit of more seamless human-computer interaction, many recent studies have focused on enhancing voice-based communication with LLMs \cite{defossez2024moshi,fang2024llama,xie2024mini}.
This line of research aims to reduce the latency and complexity that can arise from switching between spoken and textual inputs, thereby providing a more intuitive user experience.
Most of these methods utilize speech tokenization techniques to model speech signals.
By representing audio in a manner compatible with existing text-centric architectures, these methods enable the powerful text-driven reasoning capabilities of LLMs to be applied to spoken language tasks.

\newpage
\section{Technologies in Amphion}
This section provides an overview of the specific models in Amphion.
Section \ref{sec:tech_emilia} describes our data pre-processing pipeline technology.
Section \ref{sec:tech_tts} highlights the latest advancements in Amphion's text-to-speech technologies.
In Section \ref{sec:tech_codec}, we detail our neural audio codec technologies.
Our voice conversion technology is presented in Section \ref{sec:tech_vc}.

\subsection{Datasets}
\subsubsection{Emilia Dataset} \label{sec:tech_emilia}

\label{sec:emilia}
Because of scarcity of diverse high-quality, large-scale speech datasets, we release the \textbf{Emilia} dataset~\cite{emilia} in Amphion v0.2.
It offers over 100K hours of multilingual speech data spanning six languages, capturing diverse speech styles and real-world scenarios. Additionally, \textbf{Emilia-Pipe}, an open-source data preparation pipeline used to construct Emilia, enables efficient processing of raw audio data for model training. 

\begin{figure*}[ht]
    \centering
    \includegraphics[width=\textwidth]{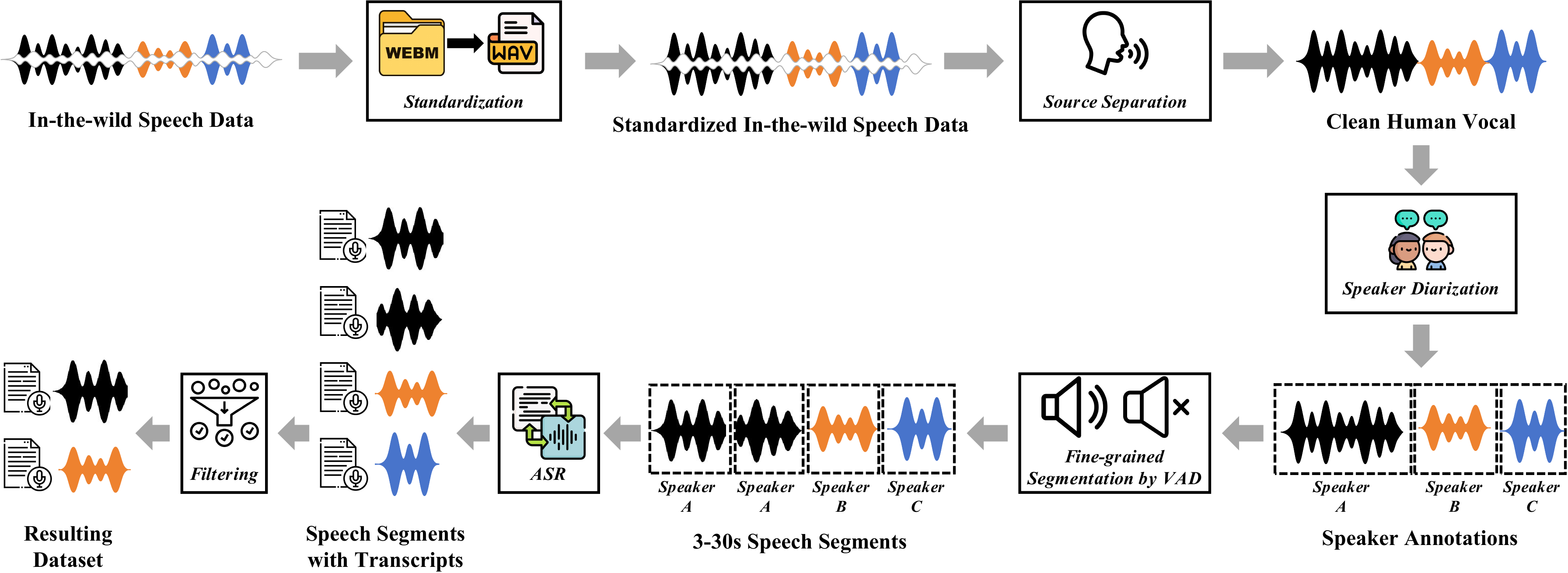}
    \caption{An overview of the Emilia-Pipe preprocessing pipeline.}
    \label{fig:process_pipeline}
\end{figure*}

We first describe Amphion's data preprocessing pipeline, Emilia-Pipe. 
Emilia-Pipe is designed to transform any audio into 3-30s speech suitable for training speech generation systems.
As illustrated in Fig.~\ref{fig:process_pipeline}, Emilia-Pipe includes six steps, i.e., Standardization, Source Separation, Speaker Diarization, Fine-grained Segmentation by VAD, ASR, and Filtering.
Emilia-Pipe is implemented in Amphion v0.2\footnote{\url{https://github.com/open-mmlab/Amphion/blob/main/preprocessors/Emilia}}..

\textbf{Step 1: Audio Data Standardization.}
The audio files are first converted to WAV format and resampled to 24 kHz. Loudness is normalized by calculating the average loudness and adjusting it to a target level. 
The audio is then represented using 16-bit depth to capture sufficient signal resolution.

\textbf{Step 2: Source Separation.}
In this step, background noise or music is removed from the audio, leaving only clean speech. This is typically achieved using models designed for background music extraction. 
We use the source separation model Ultimate Vocal Remover\footnote{\url{https://github.com/Anjok07/ultimatevocalremovergui}}
and its pre-trained model, UVR-MDX-Net Inst 33\footnote{\url{https://github.com/TRvlvr/model_repo/releases/tag/all_public_uvr_models}}.

\textbf{Step 3: Speaker Diarization.}
After extracting clean human vocals from the raw speech data, we apply the speaker diarization technique to partition the long-form speech data into multiple segments based on the speaker. This process generates a series of segments for each speech data, with each utterance containing only one speaker.. To achieve this, we leverage the ``pyannote/speaker-diarization-3.1'' speaker diarization pipeline.\footnote{\url{https://github.com/pyannote/pyannote-audio}} This pipeline includes three core components: speaker segmentation, speaker embedding, and clustering, and achieves state-of-the-art speaker diarization performance~\cite{pyannote_model}. 
The output of this pipeline is a list of temporal annotations indicating the start and end times of the single-speaker segments.

\textbf{Step 4: Fine-grained Segmentation.}
We use the Silero-VAD model\footnote{\url{https://github.com/snakers4/silero-vad}} to further split any segment longer than 30 seconds into smaller VAD segments. We then concatenate consecutive VAD segments from the same speaker into appropriately sized utterances, ensuring each utterance is between 3 and 30 seconds in length.

\textbf{Step 5: Transcription.}
Once the audio has been segmented and cleaned, Automatic Speech Recognition (ASR) models are used to transcribe each segment of speech into text. 
We use WhisperX \footnote{\url{https://github.com/m-bain/whisperX}}, which is built on the faster-whisper\footnote{\url{https://github.com/SYSTRAN/faster-whisper}} backend and the CTranslate2\footnote{\url{https://github.com/OpenNMT/CTranslate2}} inference engine. This setup is up to four times faster than the official Whisper implementation while maintaining nearly the same accuracy.

\textbf{Step 6: Filtering.}
We use the DNSMOS P.835 OVRL score~\cite{dnsmos835} to assess overall speech quality, \textit{retaining only speech data with a score higher than 3.0}. 
While the DNSMOS evaluation can be conducted on Mircrosoft Azure platform, we also provide a checkpoint from a GitHub repository to enable offline DNSMOS computation \footnote{\url{https://github.com/microsoft/DNS-Challenge/blob/master/DNSMOS/DNSMOS/sig_bak_ovr.onnx}}.

\textbf{Step 7: Storage.}
We design an output structure to optimize data handling, distribution, and storage, accommodating datasets ranging from terabytes (TB) to hundreds of terabytes (TB). 
We name our Emilia dataset file format as \textit{OpenEmilia Output Structure}.
It organizes output data as follows:
\begin{itemize}
    \item \textbf{Metadata Representation:} Metadata is stored in JSONL files, where each line corresponds to a segment's metadata.
    \item \textbf{Unique Identifiers:} Each data segment is identified using the format: \texttt{<LANG>\_<BATCH>\_<SPEAKER>\_<WAV>}.

\item \textbf{Example File Structure:}
\begin{verbatim}
Emilia_data.tar.gz
├── EN_001001.jsonl
├── EN_001001
│   ├── EN_001001_000000.mp3
│   ├── EN_001001_000001.mp3
│   ├── ...
├── EN_001013.jsonl
├── EN_001013
│   ├── EN_001013_000000.mp3
│   ├── ...
├── JA_001002.jsonl
├── ...
\end{verbatim}
\end{itemize}

\begin{figure}[htbp]
    \centering
    \resizebox{0.4\textwidth}{!}{
        \begin{tikzpicture}
            % Define colors using hex codes
            \definecolor{color1}{HTML}{447cac}
            \definecolor{color2}{HTML}{88ce9b}
            \definecolor{color3}{HTML}{e3f79b}
            \definecolor{color4}{HTML}{fae28c}
            \definecolor{color5}{HTML}{f1874b}
            \definecolor{color6}{HTML}{c42d40}
            \pie[
                text=legend,
                radius=3,
                color={color1, color2, color3, color4, color5, color6},
                explode=0.1, % Adds a small offset to the slices
                sum=auto, % Automatically sums up to 100%
                before number=\phantom{0}, % Adds a leading zero for consistency
                after number=\% % Appends a percentage sign to the numbers
            ]{
                46.77/En: 46.8k hrs, 
                49.87/Zh: 49.9k hrs, 
                1.59/De: 1.6k hrs, 
                1.38/Fr: 1.4k hrs, 
                1.72/Ja: 1.7k hrs, 
                0.22/Ko: 0.2k hrs
            }
        \end{tikzpicture}
    }
    \caption{Duration statistics of the speech data by language.}
    \label{fig:dataset_stats}
\end{figure}
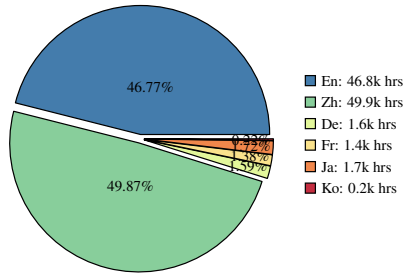

Using Emilia-Pipe, we construct the Emilia dataset from a vast collection of speech data sourced from diverse video platforms and podcasts on the Internet, covering various content categories such as talk shows, interviews, debates, sports commentary, and audiobooks.
After processing, the initial version of the Emilia dataset includes a total of 101,654 hours of multilingual speech data in six different languages: English, French, German, Chinese, Japanese, and Korean. Fig~\ref{fig:dataset_stats} provides the duration statistics for each language in the dataset.
Emilia is available in Amphion's Huggingface spaces\footnote{\url{https://huggingface.co/datasets/amphion/Emilia-Dataset}} for downloading.

% \subsection{Specialized datasets: SpMis, SD-Eval, and Debatts-Data} \label{sec:specialized-data}

\subsubsection{SpMis Dataset}
\textbf{SpMis dataset for spoken misinformation detection.}
The SpMis\footnote{\url{https://huggingface.co/datasets/amphion/spmis}} Dataset is an open-source collection designed specifically for detecting misinformation in synthesized speech. It is structured around the concept of misinformation in synthetic audio, focusing on detecting and classifying speech that is artificially generated to mislead. The dataset includes textual data from five key domains—finance, medicine, politics, law, and education—as well as other miscellaneous topics, all sourced from authoritative datasets in these fields. For speech generation, over a thousand distinct speakers were selected from the Libri-Light dataset, whose voices were synthesized using two open-source text-to-speech systems: Amphion and OpenVoice v2. The dataset's statistics are presented in Table~\ref{table:spmis}.

\begin{table*}[htbp]
\centering
\caption{Statistics of the SpMis dataset. There are five topics and one other topic. The statistics are presented as numbers of total samples, misinformation samples, speakers and duration for each topic.}
\resizebox{0.8\textwidth}{!}{
\begin{tabular}{lrccr}
\toprule
\textbf{Topic} & \textbf{\# samples} & \textbf{\# misinformation} & \textbf{\# speakers} & \textbf{Duration (hr)} \\
\midrule 
Politics & 76,542 & 1,740 & 772 & 586.59 \\
Medicine & 21,836 & 740 & 1,094 & 429.77 \\ 
Education & 177,392 & 2,970& 989 & 665.59 \\
Laws & 11,422 & 862 & 936 & 1534.78 \\ 
Finance & 53,011 & 2,369 & 940 & 585.69 \\ 
Other & 20,408 & 0 & 1,094 & 1136.23 \\% 
\midrule
ALL & 360,611 & 8,681 & 1,094 & 4938.65 \\%  
\bottomrule
\end{tabular}
}
\label{table:spmis}
\end{table*}

\textbf{Synthetic Spoken Misinformation.} Synthetic misinformation refers to information created using synthesis techniques that mislead the public into biased decisions. We define two scenarios:
\begin{itemize}
    \item \textbf{Case 1:} Speeches from ordinary people, whether synthetic or not, are not considered misinformation.
    \item \textbf{Case 2:} Recordings of celebrities on a specific topic are valid, but synthesized recordings of celebrities are misinformation.
\end{itemize}
Here, "celebrity" refers to shortlisted identities, while "ordinary people" are non-shortlisted identities.

\textbf{Text Data.} We generate speeches on five common topics, which include politics, medicine, education, laws, and finance. Each topic uses different corpora:
\begin{itemize}
    \item \textbf{Finance.} We use financial phrases from English news about listed companies in OMX Helsinki~\cite{malo2014good}.
    \item \textbf{Medicine.} We use medical abstracts~\cite{10.1145/3582768.3582795}, covering diseases and pathological conditions.
    \item \textbf{Politics.} We use UK parliamentary speeches~\cite{odellevan_2021}, ranging from 1979 to 2021.
    \item \textbf{Laws.} We select the Super-SCOTUS dataset~\cite{fang2023super}, which includes oral arguments and summaries from the US Supreme Court.
    \item \textbf{Education.} We use the NCTE Transcripts dataset~\cite{demszky2023ncte}, focusing on classroom discourses from 4th and 5th grade mathematics classrooms.
\end{itemize}

\textbf{Speech Data.} We use the Libri-Light~\cite{librilight} dataset for reference speakers. The dataset provides thousands of speakers from open-source audiobooks, with characteristics extracted to generate speech based on the above text data.

\textbf{Generation Model.} We use two systems for speech generation: Amphion~\cite{amphion} and OpenVoice\_v2~\cite{qin2023openvoice}. Both are trained on the Libri-Light dataset, with Amphion employing an auto-regressive TTS model and OpenVoice\_v2 using a more efficient, non-auto-regressive approach.

\textbf{Generation and Annotation Process.} We filter and annotate data to avoid redundancy. Text data is segmented into sentences, and speech synthesis is performed at the sentence level. For the audio, we select samples of 5–13 seconds from Libri-Light speakers. All generated audio is standardized to a 16kHz sample rate.

\textbf{Filtering.} We focus on paraphrases and dialogues, curating balanced portions of each. Sentences with fewer than three words are concatenated. For audio, we ensure that each reference speaker has one sample per topic, with varying audio lengths.

\textbf{Annotation.} The dataset consists of synthetic and recording data. The recording data is used for traditional deepfake detection, while synthetic data includes speeches from 1,000+ synthetic speakers. A subset of 100 speakers is designated as "celebrity" and assigned to specific topics, as shown in Table~\ref{table:data_share}. Other synthetic data is labeled as "ordinary" or "celebrity with unrelated topics."

\begin{table*}[htbp]
\centering
\caption{Annotation data proportions. We focus on the \textit{synthetic+celebrity+specific topic} part.}
\resizebox{0.6\textwidth}{!}{
\begin{tabular}{lrr}
\toprule
\textbf{Category} & \textbf{\# samples} & \textbf{Ratio} \\
\midrule 
Recordings & 20,408 & 5.66\% \\
Synthetic+ordinary & 305,580 & 84.74\%  \\ 
Synthetic+celebrity+other topics & 25,942 & 7.19\% \\
\textbf{Synthetic+celebrity+specific topic} & 8,681 & 2.41\%  \\  
\bottomrule
\end{tabular}
}
\label{table:data_share}
\end{table*}

\subsubsection{SD-Eval Dataset}
\begin{figure}
    \centering
    \includegraphics[width=0.7\linewidth]{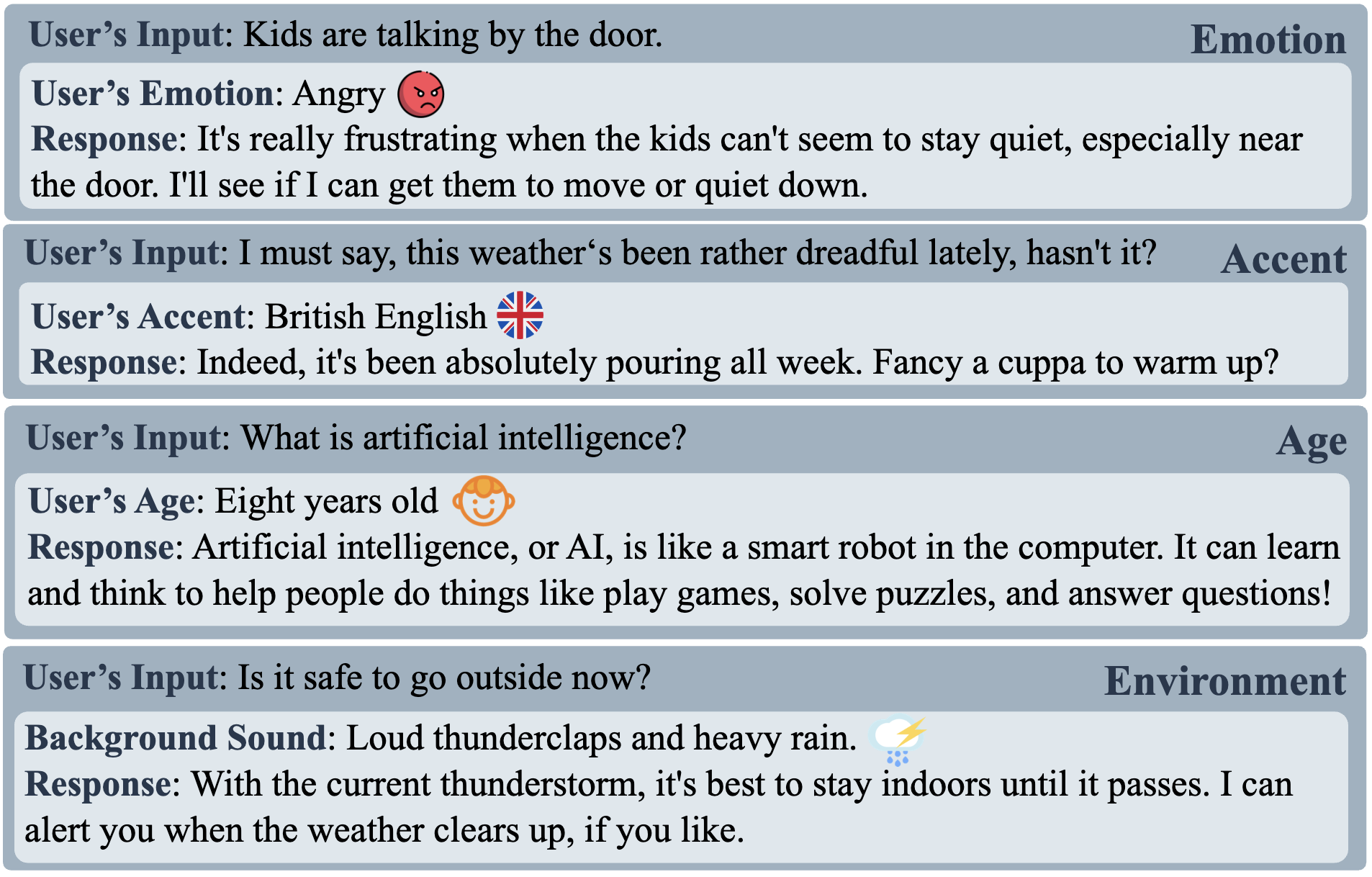}
    \caption{Examples of spoken dialogues impacted by the rich information carried in speech (e.g. emotion, accent, age, environment)}
    \label{fig:sdeval}
\end{figure}

\textbf{SD-Eval dataset for spoken dialogue understanding benchmark.}
This task of spoken dialogue understanding benchmark addresses the evaluation of multi-modal Large Language Models (LLMs) in processing and responding to complex spoken interactions. Speech encompasses linguistic, paralinguistic, and environmental information, all of which play a critical role in effective communication. 
In the LLM context, as shown in Figure \ref{fig:sdeval}, ideally, the LLM-based spoken dialogues should be impacted by the rich information carried in speech (e.g. emotion, accent, age, environment).
While LLMs have demonstrated proficiency in recognizing speech, \textit{they often fall short in generating appropriate responses due to the lack of speech-specific benchmarks and evaluation principles.}

In Amphion v0.2, we release SD-Eval~\cite{sdeval}\footnote{\url{https://huggingface.co/datasets/amphion/SD-Eval}}, a benchmark dataset for a multidimensional evaluation of \text{spoken dialogue understanding and generation}.
SD-Eval focuses on paralinguistic and environmental information and includes 7,303 utterances, amounting to 8.76 hours of speech data. The data is aggregated from eight public datasets, representing \textit{four perspectives: emotion, accent, age,and background sound.}

\subsubsection{Debatts-Data Dataset}
\begin{figure*}
    \centering
    \includegraphics[width=1\linewidth]{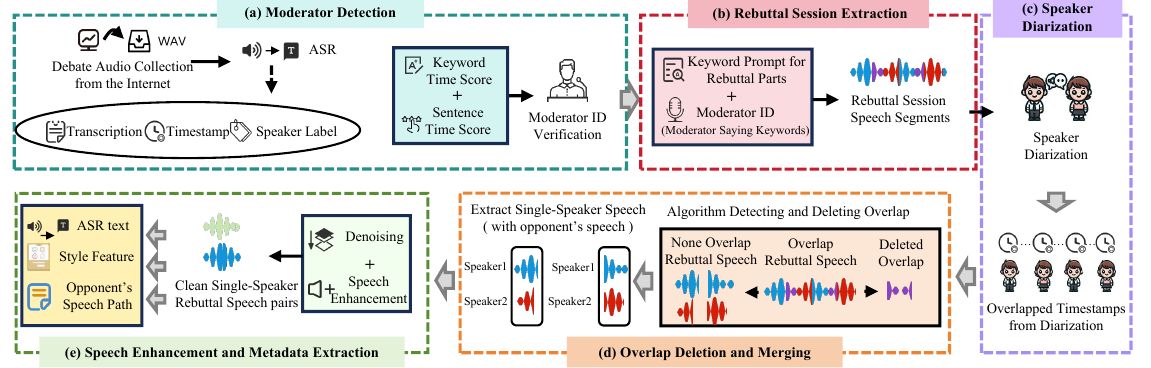}
    \caption{An illustration of the proposed five-step dataset pipeline to create the Debatts-Data dataset. The pipeline consists of moderator detection, rebuttal session extraction, speaker diarization, overlap deletion and merging and speech enhancement and metadata extraction processes.}
    % 加入两句描述
    \label{fig:Debatts-Data pipeline}
\end{figure*}

\textbf{Debatts-Data for Domain-Specific TTS in Debates:} 
The Debatts-Data~\cite{debatts}~\footnote{\url{https://huggingface.co/datasets/amphion/Debatts-Data}} is an extension to the Emilia dataset with the the data from Mandarin public debates.
This contribution is important for domain-specific TTS in debates, because there are unique characteristics of debate-style speech, such as rhetorical emphasis, dynamic intonation, and varied pacing. 
General-purpose speech datasets like Emilia do not capture these characteristics in debates.

We propose a 110 hours Mandarin debating dataset called Debatts-Data to support the development of debating TTS models.
% This section presents the statistics of the dataset and pipeline for dataset creation. Detailed statistics and comparison with existing debate datasets are presented in Table~\ref{table:dataset_statistics}. 
We note that different from the traditional TTS dataset, the rebuttal in debating has an opponent and a speaker.
Typically, the rebuttal session begins with a competition moderator introducing the process.

\textbf{Pipeline:} We developed a pipeline to process the raw recordings. The pipeline consists of five steps as shown in Fig.~\ref{fig:Debatts-Data pipeline}, including moderator detection, rebuttal session extraction, speaker diarization, overlap deletion and merging and speech enhancement and metadata extraction. In particular,
\begin{itemize}[left=0pt] % 设置左缩进为 0pt
\item Moderator Detection: 
    First, we utilized Paraformer zh~\cite{seedtts} to transcribe texts and extract global speaker labels and then identified the moderator's speaker ID as the anchor ID.
    \item Rebuttal Session Extraction: 
    Then, we located the rebuttal session segments by extracting keywords from the moderators. This approach enables automated and precise extraction of specific segments from full competition recordings.
    \item Speaker Diarization: 
    After that, we employed a speaker diarization toolkit \cite{Plaquet23,Bredin23} to identify speakers within the rebuttals and extract single-speaker speech pairs for training.
    \item Overlap Detection and Merging: 
    Next, we eliminated overlaps, which are common in rebuttals. The detection of overlaps is achieved by analyzing the overlapping timestamps in the diarization data. After deleting the overlaps, speech segments from the same speaker are concatenated.
    \item
    Speech Enhancement and Metadata Extraction: 
    Finally, we performed speech enhancement and metadata extraction. Speech enhancement is mainly to reduce background noise and remove speech segments with severe noise. Metadata extraction is to extract transcriptions, speaker labels, timestamps and style vectors~\cite{ma2023emotion2vec}.
\end{itemize}
Through these processes, we produced the first Mandarin debating dataset for TTS. The dataset contains rich prosody and meta information. The pipeline can used for other languages and other speech domains with minor changes.

\subsection{Text to Speech} \label{sec:tech_tts}

\subsubsection{MaskGCT}\label{sec:maskgct}

\begin{figure}[t]
  \centering
  \includegraphics[page=1,width=1.0\columnwidth,trim=0cm 21cm 30cm 0.0cm,clip=true]{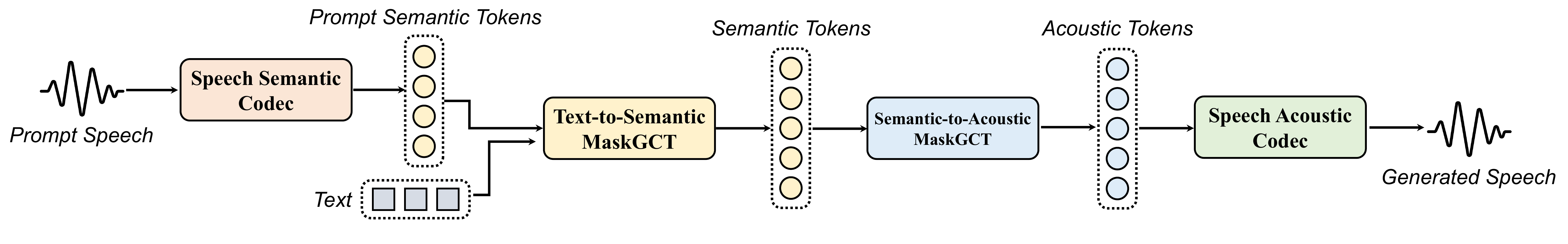}
  \caption{An overview of the proposed two-stage MaskGCT framework. It consists of four main components: (1) a speech semantic representation codec converts speech to semantic tokens; (2) a text-to-semantic model predicts semantic tokens with text and prompt semantic tokens; (3) a semantic-to-acoustic model predicts acoustic tokens conditioned on semantic tokens; (4) a speech acoustic codec reconstructs waveform from acoustic tokens.}
  \vspace{-10px}
  \label{fig:overview}
\end{figure}

MaskGCT is a fully non-autoregressive model for text-to-speech synthesis that uses masked generative transformers without requiring text-speech alignment supervision and phone-level duration prediction. MaskGCT is a two-stage system, both stages are trained using the \textit{mask-and-predict learning} paradigm. The first stage, the text-to-semantic (T2S) model, predicts masked semantic tokens with in-context learning, using text token sequences and prompt speech semantic token sequences as the prefix, without explicit duration prediction. The second stage, the semantic-to-acoustic (S2A) model, utilizes semantic tokens to predict masked acoustic tokens extracted from an RVQ-based speech codec with prompt acoustic tokens. During inference, MaskGCT can generate semantic tokens of various specified lengths with a few iteration steps given a sequence of text. The overview of MaskGCT is shown in Figure~\ref{fig:overview}.

\paragraph{Text-to-Semantic Model}  During training, we randomly extract a portion of the prefix of the semantic token sequence as the prompt, denoted as $\mathbf{S}^p$. We then concatenate the text token sequence $\mathbf{P}$ with $\mathbf{S}^p$ to form the condition. We simply add $(\mathbf{P}, \mathbf{S}^p)$ as the prefix sequence to the input masked semantic token sequence $\mathbf{S}_t$ to leverage the in-context learning ability of language models. We use a Llama-style~\cite{touvron2023llama} transformer as the backbone of our model, incorporating gated linear units with GELU~\cite{hendrycks2016gaussian} activation, rotation position encoding~\cite{su2024roformer}, etc., but replacing causal attention with bidirectional attention. We also use adaptive RMSNorm~\cite{zhang2019root}, which accepts the time step $t$ as the condition. During inference, we generate the target semantic token sequence of any specified length conditioned on the text and the prompt semantic token sequence.

\begin{figure}[t]
  \centering
  \begin{minipage}{0.49\linewidth}
    \centering
    \includegraphics[page=1,width=1.1\columnwidth,trim=0cm 18cm 80.0cm 0.0cm,clip=true]{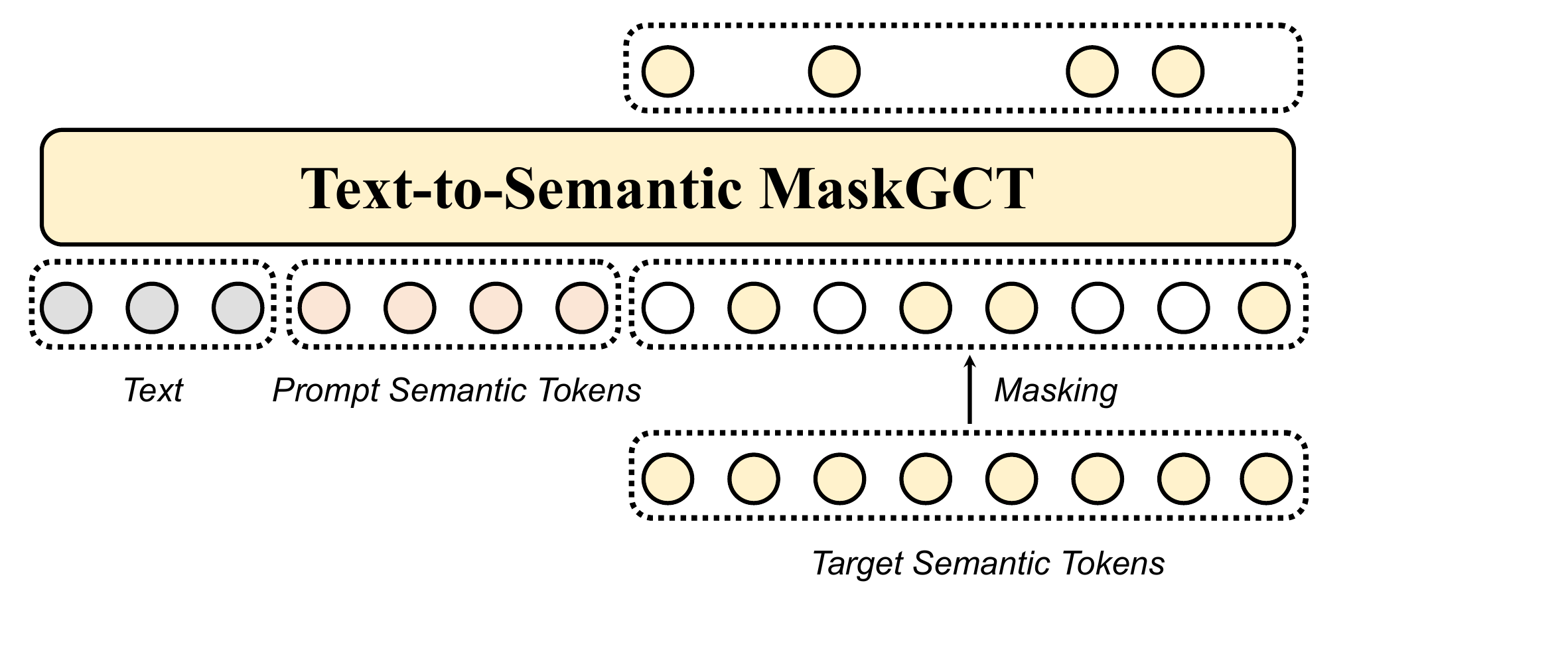}
  \end{minipage}
  %\qquad
  \begin{minipage}{0.49\linewidth}
    \centering
    \includegraphics[page=1,width=1.1\columnwidth,trim=0cm 18.0cm 80.0cm 0.0cm,clip=true]{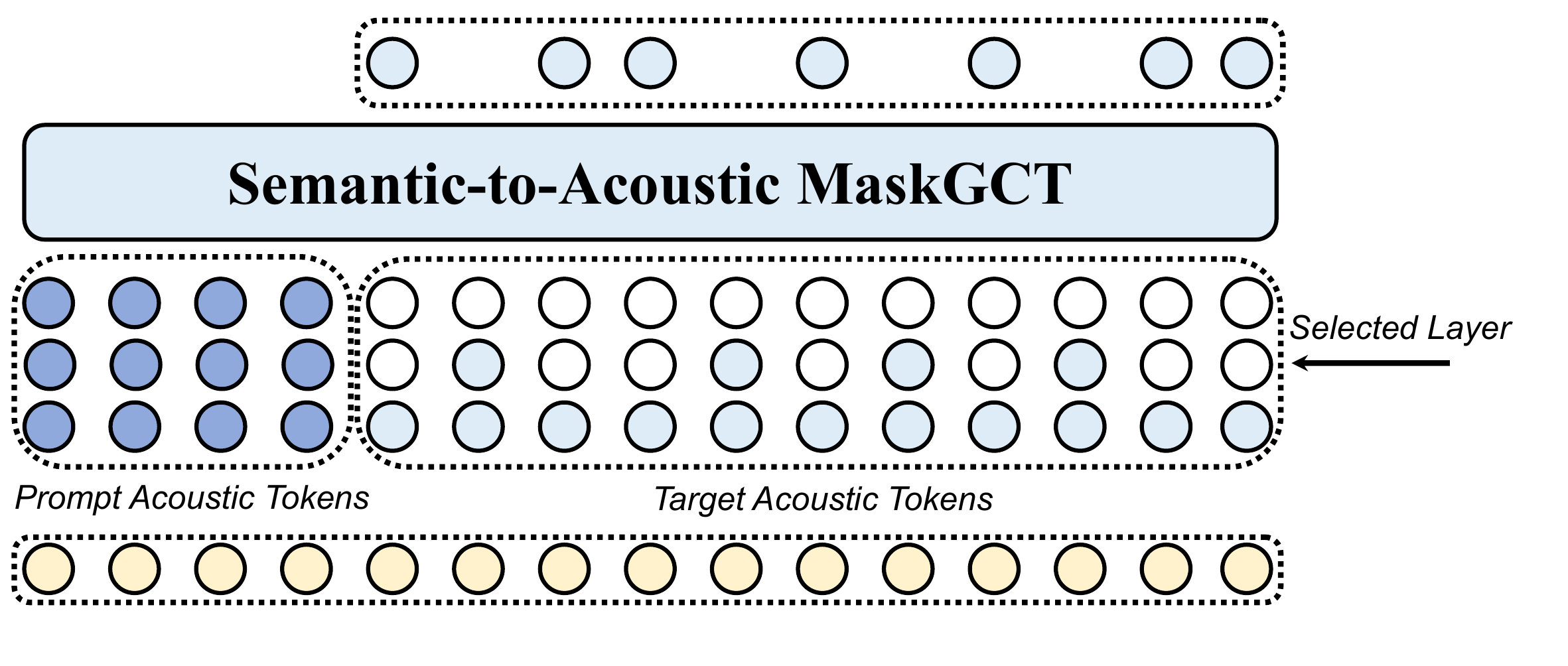}
  \end{minipage}
  \caption{An overview of training diagram of the T2S (left) and S2A (right) models. The T2S model is trained to predict masked semantic tokens with text and prompt semantic tokens as the prefix. The S2A model is trained to predict masked acoustic tokens of a random layer conditioned on prompt acoustic tokens, semantic tokens, and acoustic tokens of the previous layers.}
  \label{fig:maskgct}
\end{figure}

\paragraph{Semantic-to-Acoustic Model} We also train a semantic-to-acoustic (S2A) model using a masked generative codec transformer conditioned on the semantic tokens. Our semantic-to-acoustic model is based on SoundStorm ~\cite{borsos2023soundstorm}, which generates multi-layer acoustic token sequences. Given $N$ layers of the acoustic token sequence $\mathbf{A}^{1:N}$, during training, we select one layer $j$ from $1$ to $N$. We denote the $j$th layer of the acoustic token sequence as $A^j$. Following the previous discussion, we mask $A^j$ at the timestep $t$ to get $\mathbf{A}^j_t$. The model is then trained to predict $\mathbf{A}^j$ conditioned on the prompt $\mathbf{A}^p$, the corresponding semantic token sequence $\mathbf{S}$, and all the layers smaller than $j$ of the acoustic tokens. This can be formulated as $p_{\theta_{\text{s2a}}}(\mathbf{A}^j|\mathbf{A}^j_t, (\mathbf{A}^p, \mathbf{S}, \mathbf{A}^{1:j-1}))$.
We sample $j$ according to a linear schedule $p(j) = 1 - \frac{2j}{N(N+1)}$. For the input of the S2A model, since the number of frames in the semantic token sequence is equal to the sum of the frames in the prompt acoustic sequence and the target acoustic sequence, we simply sum the embeddings of the semantic tokens and the embeddings of the acoustic tokens from layer $1$ to $j$.
During inference, we generate tokens for each layer from coarse to fine, using iterative parallel decoding within each layer. Figure~\ref{fig:maskgct} shows a simplified training diagram of the T2S and S2A models.

\subsubsection{Vevo}\label{sec:vevo-tts}

\begin{figure}[t]
    \centering
    \includegraphics[width=0.975\textwidth]{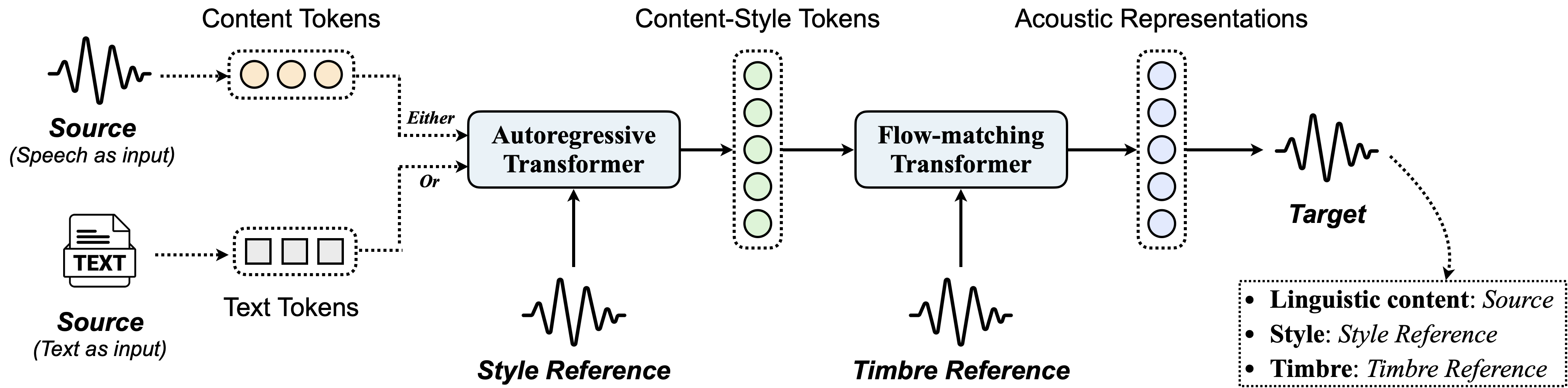}
    \caption{Vevo inference pipeline. Notably, it can take \textit{either} speech \textit{or} text as input, and perform zero-shot imitation with controllable linguistic content (controlled by the source), style (controlled by the style reference), and timbre (controlled by the timbre reference) in a single forward pass.
    }
    \label{fig:Vevo-pipeline}
\end{figure}

Vevo is a versatile zero-shot voice imitation framework with controllable timbre and style~\cite{vevo}, available via this link~\footnote{\url{https://github.com/open-mmlab/Amphion/blob/main/models/vc/vevo/}}.
It can serve as a unified framework for a wide range of zero-shot speech generation tasks. As Figure~\ref{fig:Vevo-pipeline} shows, Vevo consists of two core stages: (1) \textbf{Content-Style Modeling} (\textit{Content to Content-Style}): Given a speech prompt as style reference, Vevo generates \textit{content-style} tokens from the input \textit{content} tokens (or the input text). Vevo employs the decoder-only autoregressive transformer~\cite{transformer,touvron2023llama}, leveraging its powerful capability of continued generation to model style. (2) \textbf{Acoustic Modeling} (\textit{Content-Style to Acoustic}): Given a speech prompt as timbre reference, Vevo generates acoustic representations (such as Mel spectrograms) from the input of \textit{content-style} tokens. Vevo uses a flow-matching transformer~\cite{flow-matching,dit}, which has been verified to excel in in-context learning and reconstructing high-quality audio~\cite{voicebox,audiobox,cosyvoice,fireredtts}, to achieve timbre-controllable generation.

\paragraph{Vevo Tokenizers}
To obtain the \textit{content} and \textit{content-style} tokens of speech, Vevo designs a self-supervised method to decouple the timbre, style, and linguistic content gradually, which is similar to a progressive information filtering: (1) Vevo firstly investigate the commonly used self-supervised speech pre-trained model, HuBERT~\cite{HuBERT}. Vevo find that its \textbf{continuous} hidden features contain rich information about timbre, style, and linguistic content, making it a suitable initial stage for information filtering. (2) Inspired by existing works for disentangling speaker-agnostic representations~\cite{vq-vae,vqvc,vq-content-style,naturalspeech3}, Vevo employ VQ-VAE~\cite{vq-vae} as a tokenizer for HuBERT to filter out timbre, resulting in \textbf{content-style tokens}. (3) Furthermore, Vevo propose that the vocabulary size of the VQ-VAE codebook can function as the ``width" of the information bottleneck~\cite{autovc}. By reducing the vocabulary size, Vevo can narrow the bottleneck and filter out not only timbre but also significant style information, thereby obtaining \textbf{content tokens}. Besides, Vevo propose to reduce the consecutive duplicate units~\cite{mhubert-duration-reduction} of the content tokens, called \textit{duration reduction}, to further remove some style patterns such as unit-level duration.

\paragraph{Vevo-TTS}
Assume that during the content-style modeling and acoustic modeling stages, we have obtained pre-trained models $\mathcal{M}_{style}$ and $\mathcal{M}_{acoustic}$ respectively. We can then adjust only the inference pipeline to apply Vevo to various zero-shot imitation tasks. Given the source text $\mathcal{T}_{\textcolor{blue}{\bm{i}}}$ and the reference ${u}_{\textcolor{red}{\bm{r}}}$, we can utilize the variant of Vevo, Vevo-TTS, to achieve controllable zero-shot text-to-speech task: $\widetilde{\bm{Q}_c} (\mathcal{T}_{\textcolor{blue}{\bm{i}}}) \xrightarrow{\displaystyle {u}_{\textcolor{red}{\bm{r}}}} \widetilde{\mathcal{M}}_{style} \xrightarrow{\displaystyle {u}_{\textcolor{red}{\bm{r}}}} \mathcal{M}_{acoustic}$, where `` $\xrightarrow{u} \mathcal{M}$ " means that the model $\mathcal{M}$ is prompted by $u$ to generate, $\widetilde{\bm{Q}_c} (\mathcal{T}_{\textcolor{blue}{\bm{i}}})$ means the tokenization for $\mathcal{T}_{\textcolor{blue}{\bm{i}}}$, and $\widetilde{\mathcal{M}}_{style}$ means the pre-trained model for content-style modeling that takes text as input.

\subsubsection{Debatts}
\begin{figure*}
    \centering
    \includegraphics[width=0.9\linewidth]{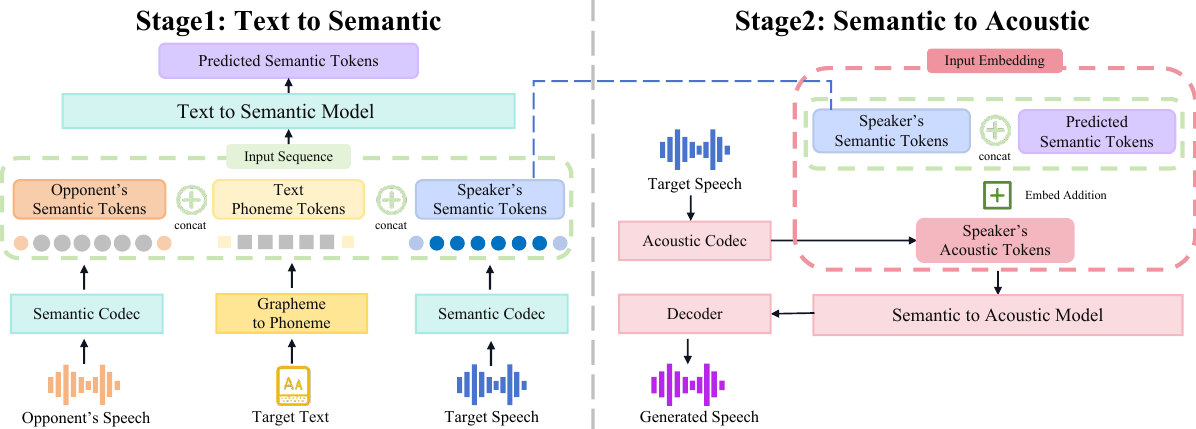}
    \caption{The proposed Debatts architecture. It consists of a text-to-semantic stage and a semantic-to-acoustic stage. In the first stage, the model predicts target semantic tokens using opponent and target speech semantic tokens along with text tokens. In the second stage, it generates speech with debating style based on the concatenated target speaker and predicted semantic tokens, along with the target speaker's acoustic tokens.}
    \label{fig:debatts model architecture}
\end{figure*}

Debatts is built on top of the zero-shot TTS framework with inspirations from the previous work~\cite{borsos2023soundstorm,seedtts,maskgct}. Debatts is a two-stage model as illustrated in Fig.~\ref{fig:debatts model architecture}.
The first stage is to predict semantic tokens with opponent's speech, target text and target speech as inputs, while the second stage is to predict acoustic tokens which are then transformed into waveform. More samples are available via the demo page \footnote{\url{https://amphionspace.github.io/debatts/}}, and Debatts is available via the link\footnote{\url{https://github.com/open-mmlab/Amphion/tree/main/models/tts/debatts}}

\textbf{Text-to-Semantic Prediction} 
To generate semantic tokens, Debatts employs a LLaMA-style Transformer model~\cite{touvron2023llama} as the backbone. During pretraining, the speaker's prompt speech is processed through the semantic codec same as that in~\cite{maskgct}, resulting in a discrete semantic sequence $\boldsymbol{S_{spk}}$. Meanwhile, text semantic tokens $\boldsymbol{T}$ are generated from target text using a grapheme-to-phoneme tool. Based on the text sequence $\boldsymbol{T}$, the model autoregressively generates semantic tokens for the prompt.

During training, we incorporate the opponent’s speech as an additional input or condition for context learning. The opponent's speech is unified with the target reference speech at the discrete token level. Specifically, the opponent’s speech is passed through the same semantic codec to extract a discrete sequence $\boldsymbol{S_{op}}$. Then, $\boldsymbol{S_{op}}$, $\boldsymbol{T}$, and $\boldsymbol{S_{spk}}$ are concatenated as inputs, and model autoregressively predicts semantic tokens $\boldsymbol{\hat{S}_{spk}}$ with those inputs. The training process formulated as
\begin{equation}
\begin{aligned}
\text{$p(\hat{S}_{spk}|T,S_{op};\theta_{t2s}^{new})$} &= \text{$\prod_{t=1}^N p(\hat{S}_{spk,t}|S_{spk,<t},T,S_{op};\theta_{t2s}^{new})$}
\end{aligned}
\end{equation}
where $\boldsymbol{S_{spk}}$
denotes the target reference speech representation tokens sequence, $\boldsymbol{\hat{S}_{spk}}$ denotes the predicted tokens sequence, and $\boldsymbol{S_{op}}$ denotes the opponent's token sequence.

During training, the losses are calculated between $\boldsymbol{S_{spk}}$ 
 and $\boldsymbol{\hat{S}_{spk}}$, excluding $\boldsymbol{S_{op}}$ and $\boldsymbol{T}$ to ensure that the model only uses the opponent’s context as a condition. 
 % This avoids interference from the opponent’s speaker features. 
 The model then focuses on how prosody and emotional details of the opponent's speech shape the target speech style. As a result, it synthesizes debate style from the opponent's speech context and speaker timbre from the prompt, producing natural responses.

In the inference stage, we utilize the token sequences from opponent’s speech, target text, speaker’s prompt speech, and prompt text as conditions to generate the target speech.

\textbf{Semantic-to-Acoustic Generation}  
To predict acoustic tokens from semantic tokens, we employ a model architecture similar to the SoundStorm model~\cite{borsos2023soundstorm}. The model takes the semantic tokens sequence $\boldsymbol{S}$ and the acoustic tokens sequence $\boldsymbol{A_{1:N}}$ from the prompt as input. The acoustic token prediction $\boldsymbol{A}$ consists of $\boldsymbol{N}$ layers. 
The acoustic tokens sequence of speaker prompt speech is extracted by a speech acoustic codec same as~\cite{maskgct}.
Since the frame numbers of $\boldsymbol{S}$ and $\boldsymbol{A_{1:N}}$ are equal, we simply add their embeddings and use them as the model input. During training, the model predict masked acoustic tokens based on semantic tokens and unmasked acoustic tokens.
In the inference stage, the model generates acoustic tokens layer by layer, starting from coarse acoustic information in the lower layers.
% It progressively generates more detailed information in the higher layers. 
The model concatenates the speaker semantic tokens and the predicted semantic tokens from text-to-semantic generation to form $\boldsymbol{S}$ as the input, while using the speaker acoustic tokens as $\boldsymbol{A}$.

\subsection{Speech Enhancement} \label{sec:tech_enhancement}

\begin{figure}[!htbp]
    \centering
    \includegraphics[width=\linewidth]{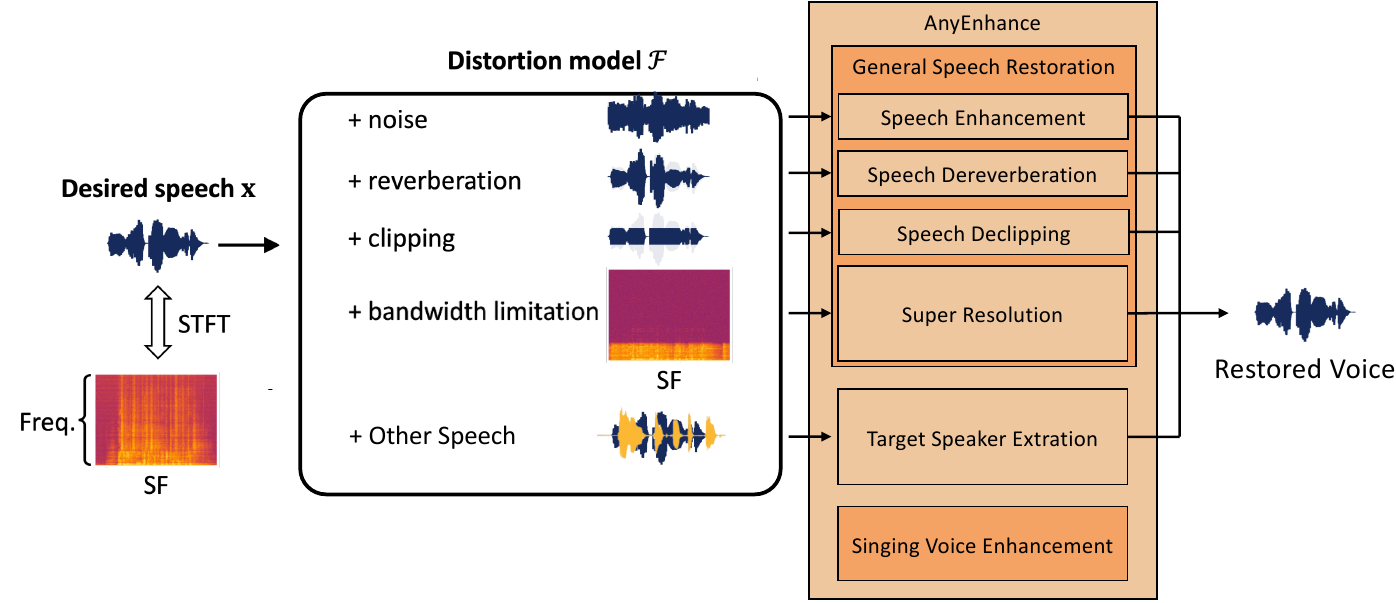}
    \caption{Overview of speech enhancement tasks and the AnyEnhance framework. Various distortions, including noise, reverberation, clipping, bandwidth limitation, and interference from other speech, are modeled and processed. The AnyEnhance framework effectively handles these distortions and produces a restored voice signal suitable for downstream applications. Figure adapted from~\url{https://urgent-challenge.github.io/urgent2024/}.}
    \label{fig:enhancement-tasks}
\end{figure}

Speech enhancement (SE) is to improve the audio quality and intelligibility of speech signals by removing various types of distortions~\cite{liu2022voicefixer} (noise, reverberation, bandwidth limitation, clipping, other speech etc.) from the input signals (See Figure~\ref{fig:enhancement-tasks}). Each type of distortion can be considered as a separate enhancement task, such as speech denoising, dereverberation, declipping, super-resolution, and target speaker extraction (TSE). Deep-learning-based SE methods has been widely used in many downstream applications, such as automatic speech recognition (ASR)~\cite{li2021espnet,donahue2018exploring}, hearing aids~\cite{cox2023overview}, and in-the-wild data processing~\cite{emilia}.

\begin{figure}[!htbp]
    \centering
    \includegraphics[width=\linewidth]{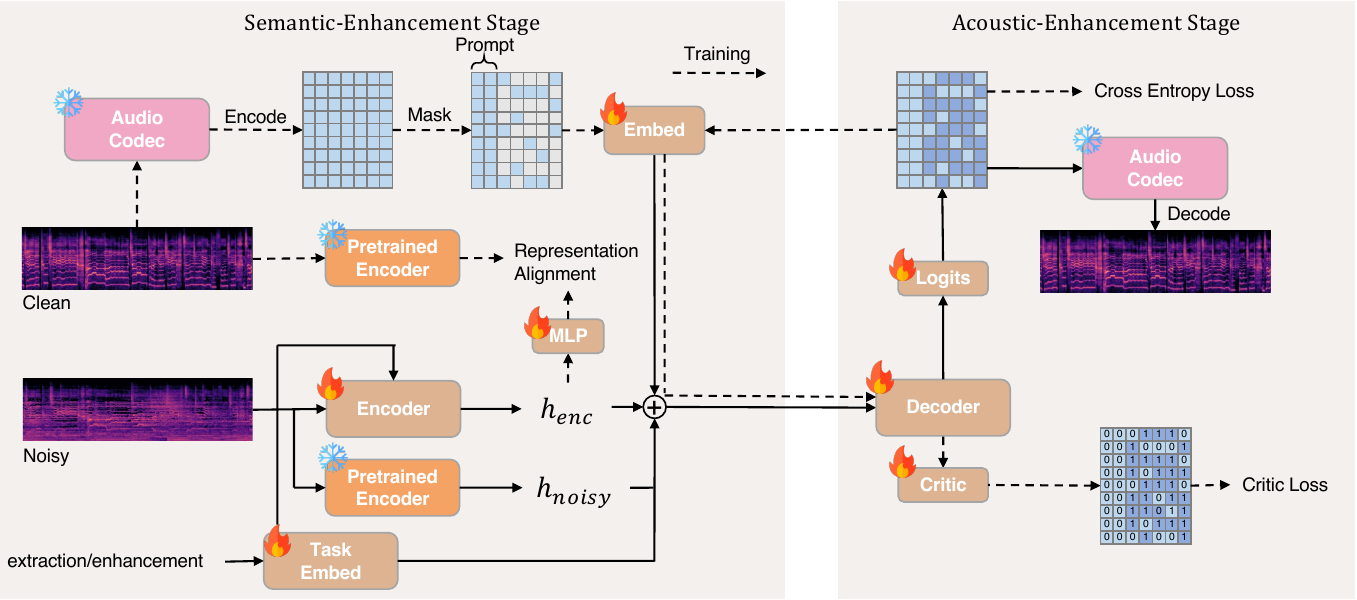}
    \caption{AnyEnhance model architecture. The model is divided into two stages: the semantic enhancement stage and the acoustic enhancement stage. In the semantic enhancement stage, the encoder is responsible for extracting the semantic features from the input distorted audio using representations alignment with pre-trained features. In the acoustic enhancement stage, the decoder predicts masked tokens based on the semantic features and existing acoustic tokens. A critic head is introduced to perform self-critic training and sampling.}
    \label{fig:anyenhance}
\end{figure}

We introduce a unified framework, AnyEnhance~\cite{zhang2025anyenhance}(see Figure~\ref{fig:anyenhance}). Based on the masked generative model, AnyEnhance effectively handles a range of speech \& singing enhancement tasks shown in Figure~\ref{fig:enhancement-tasks} simultaneously without fine-tuning. To enable the target speaker extraction task, we incorporate a prompt-guidance mechanism into the model. This mechanism enables the model to natively accept a reference speaker's voice without altering the architecture. Based on this in-context-learning capability, AnyEnhance also improves the performance over other enhancement tasks. Furthermore, to address the issue of sampling instability inherent in masked generative models~\cite{lezama2022improved}, we integrate a self-critic mechanism, enhancing accuracy and robustness when sampling from logits. These advancements make AnyEnhance a versatile solution for both speech and singing voice enhancement tasks, as well as a promising candidate for real-world applications across diverse domains. Demo audios are publicly available at this URL~\footnote{\href{https://amphionspace.github.io/anyenhance/}{https://amphionspace.github.io/anyenhance/}}. We will be releasing the model weights and inference code in the near future.

\subsection{Voice Conversion} \label{sec:tech_vc}
Voice Conversion (VC) is the study of how to convert one’s voice to sound like that of another without changing linguistic content~\cite{vc-survey-taslp}, i.e., convert the speaker identity perceived by human perception. To conduct the conversion, some existing researches aim to convert only the spectral aspects of speech~\cite{parallel-vc-survey-2017}, while others convert not only spectral but also prosodic aspects of the speech~\cite{parallel-vc-survey-2017,vc-survey-taslp}.

% In the domain of VC, early research predominantly focused on mapping and transforming speaker features through parallel corpora~\cite{parallel-vc-survey-2017,parallel-svc-2009-HMM,parallel-toda-2014,parallel-toda-2015}. However, parallel corpora are notably scarce and present significant scaling limitations. Contemporary approaches have shifted towards feature disentanglement utilizing non-parallel corpora. The fundamental principle involves extracting speaker-specific representations from the reference audio while deriving speaker-agnostic representations from the source, followed by synthesizing the converted audio through a decoder architecture~\cite{multiple-contents-svc}. To illustrate this methodology, in Amphion v0.1, we conducted comprehensive investigations into various speaker-agnostic representations and evaluated their efficacy in both voice conversion and singing voice conversion tasks~\cite{multiple-contents-svc}.

In this tutorial, we examine the current advances in VC through two perspectives. First, from a model architecture standpoint, we utilize \textbf{Vevo}~\cite{vevo} as a case study to demonstrate the design principles and distinctive characteristics of autoregressive models and flow-matching models in voice conversion applications. Subsequently, we explore an advanced topic in voice conversion through the lens of \textbf{Noro}~\cite{noro}, specifically focusing on noise-robust voice conversion, which represents a significant advancement in addressing real-world implementation challenges.

\subsubsection{Vevo}
As is introduced in Section~\ref{sec:vevo-tts}, Vevo is a versatile zero-shot voice imitation framework with controllable timbre and style~\cite{vevo}. The two variants of Vevo, \textbf{Vevo-Timbre} and \textbf{Vevo-Voice}, can both convert the speaker characteristics of speech in a zero-shot manner. The key difference between them is that Vevo-Timbre \textit{only imitates the timbre of the reference speech, while effectively preserving the style of the source speech} (e.g., prosody, emotion, accent). On the other hand, Vevo-Voice \textit{imitates both the timbre and style of the reference speech, resulting in higher speaker similarity to the reference.} We encourage readers to listen to these audio samples\footnote{\href{https://versavoice.github.io/\#vevo-vc}{https://versavoice.github.io/\#vevo-vc}} to learn about the characteristics and distinctions between these two.

\begin{figure}[t]
    \centering
    \begin{subfigure}[b]{0.35\textwidth}
        \centering
        \includegraphics[width=\textwidth]{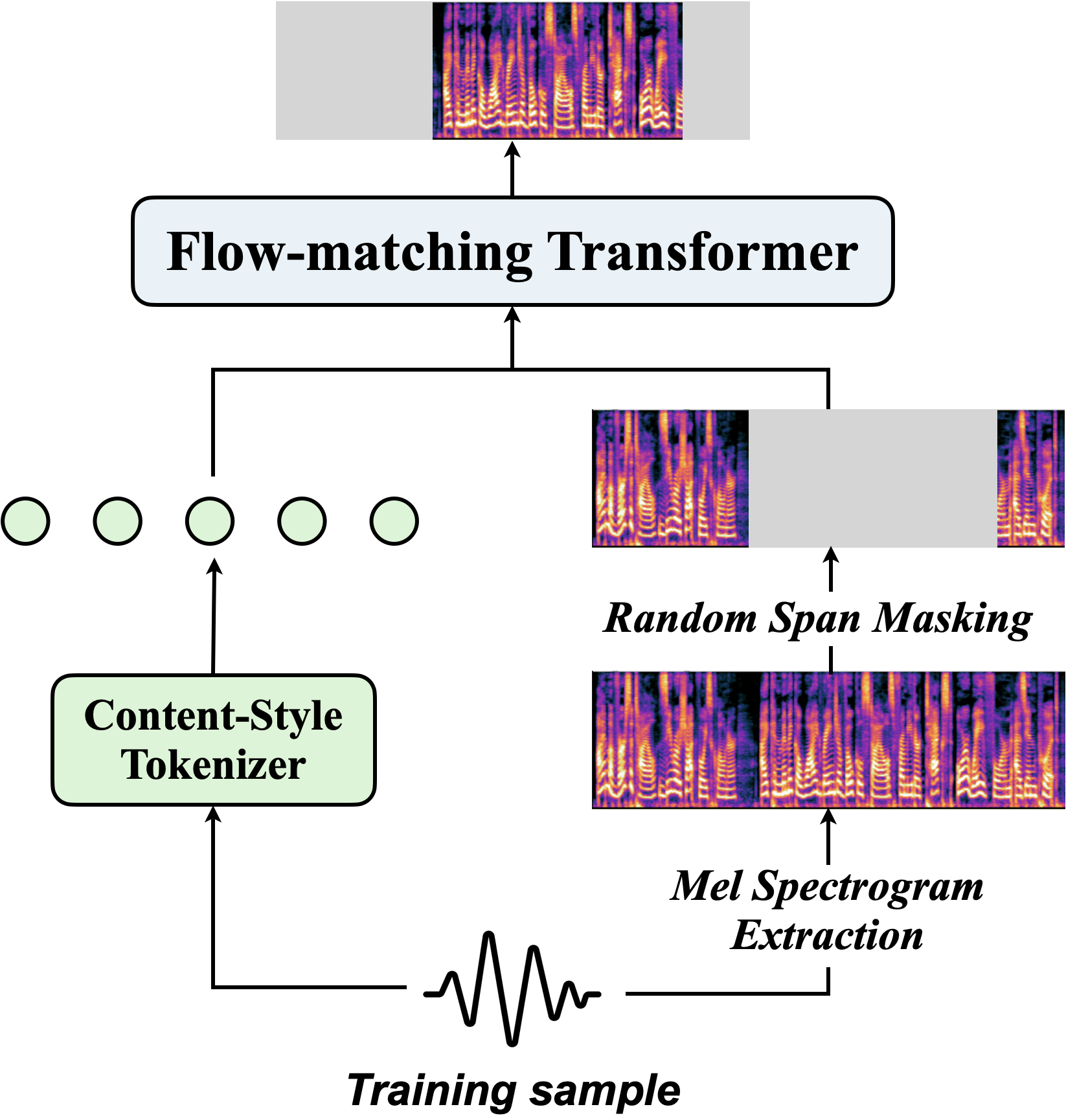}
        \caption{Training}
        \label{fig:vevo-diffusion-training}
    \end{subfigure}
    % \hfill
     \hspace{5mm}
    \begin{subfigure}[b]{0.46\textwidth}
        \centering
        \includegraphics[width=\textwidth]{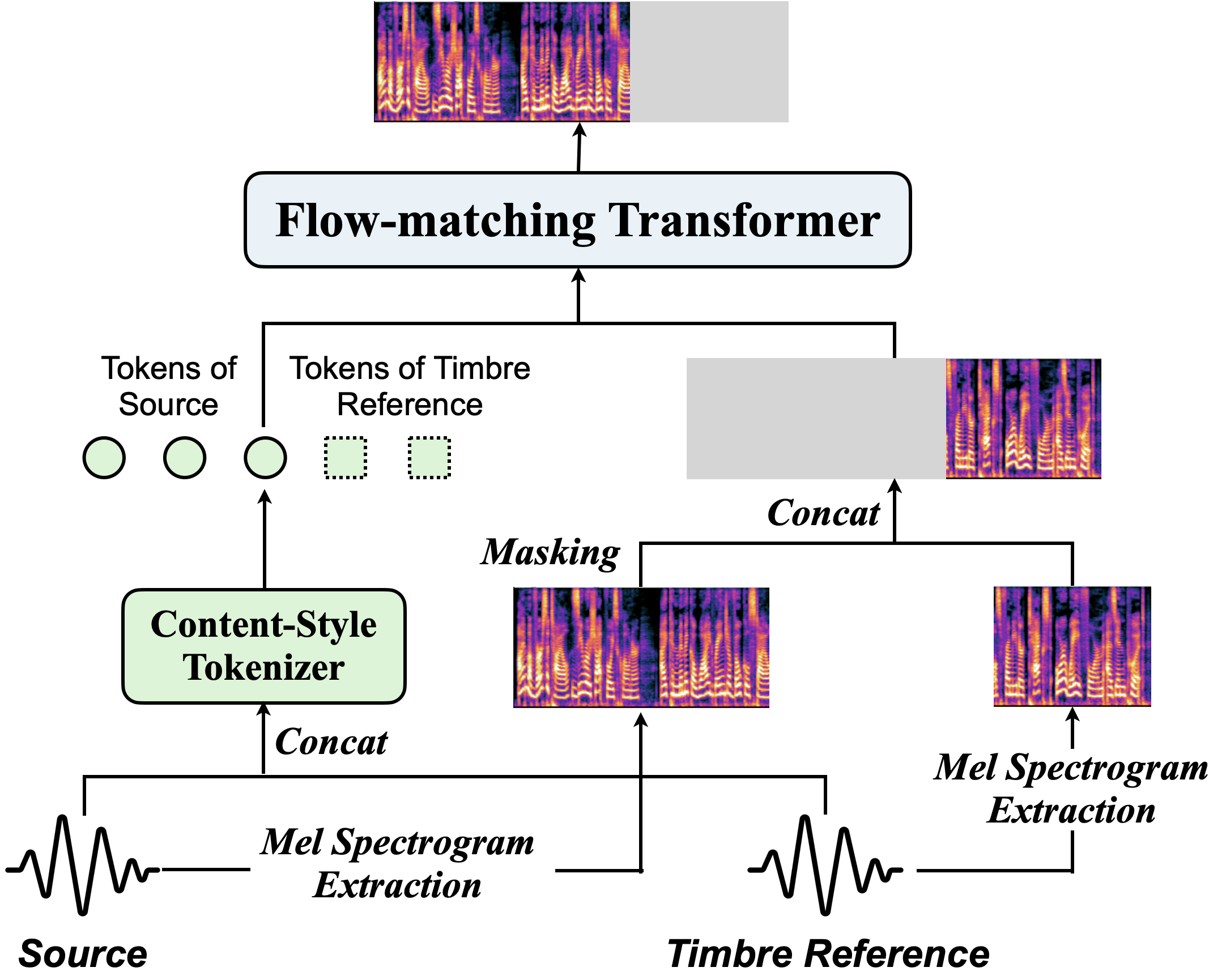}
        \caption{Inference}
        % \vspace{-2mm}
        \label{fig:vevo-diffusion-inference}
    \end{subfigure}
    \caption{\textbf{Vevo-Timbre}: a flow-matching based voice conversion model. During inference, we can extract content-style tokens of the source speech, and append those of the timbre reference to the rightmost (or leftmost) end, enabling zero-shot timbre conversion.}
    \label{fig:vevo-diffusion}
\end{figure}

\paragraph{Vevo-Timbre} 
Vevo-Timbre can be considered as only acoustic modeling stage of the whole Vevo pipeline. As is introduced in Section~\ref{sec:vevo-tts}, during this stage, prompted by a timbre reference, Vevo aims to transform the content-style tokens to Mel spectrograms. Specifically, Vevo adopts a flow matching transformer~\cite{flow-matching,transformer,touvron2023llama} (Figure~\ref{fig:vevo-diffusion}), which has been verified to be effective in in-context learning and reconstructing high-quality acoustic representations~\cite{voicebox,audiobox,cosyvoice,fireredtts}.

During training, given a speech and its Mel spectrogram, Vevo randomly selects a part of $\bm{y}_1$ as the timbre reference (denoted as $\bm{y}_1^{ctx}$), and aims to reconstruct the other part (denoted as $\bm{y}_1^{mis}$) conditioned on $\bm{y}_1^{ctx}$ and the content-style tokens $\bm{Q}_s({u})$. In other words, Vevo aims to model the conditional probability $p(\bm{y}_1^{mis} | \bm{y}_1^{ctx}, \bm{Q}_s({u}))$. During inference, given a source speech ${u}_i$ and a timbre reference ${u}_{tr}$, all the source's Mel spectrogram will be masked (i.e., $\bm{y}_1^{mis}$). The input conditions become the timbre reference's Mel spectrogram (i.e., $\bm{y}_1^{ctx}$) and the concatenated content-style tokens $\bm{Q}_s ({u}_i \oplus {u}_{tr})$. This enables the generated target to preserve the linguistic content and style of ${u}_i$, and the timbre of ${u}_{tr}$ (Figure~\ref{fig:vevo-diffusion-inference}). After obtaining the Mel spectrogram, we can utilize a vocoder to produce the waveform.

\paragraph{Vevo-Voice}

\begin{figure}
    \centering
    \begin{subfigure}[b]{0.42\textwidth}
        \centering
        \includegraphics[width=\textwidth]{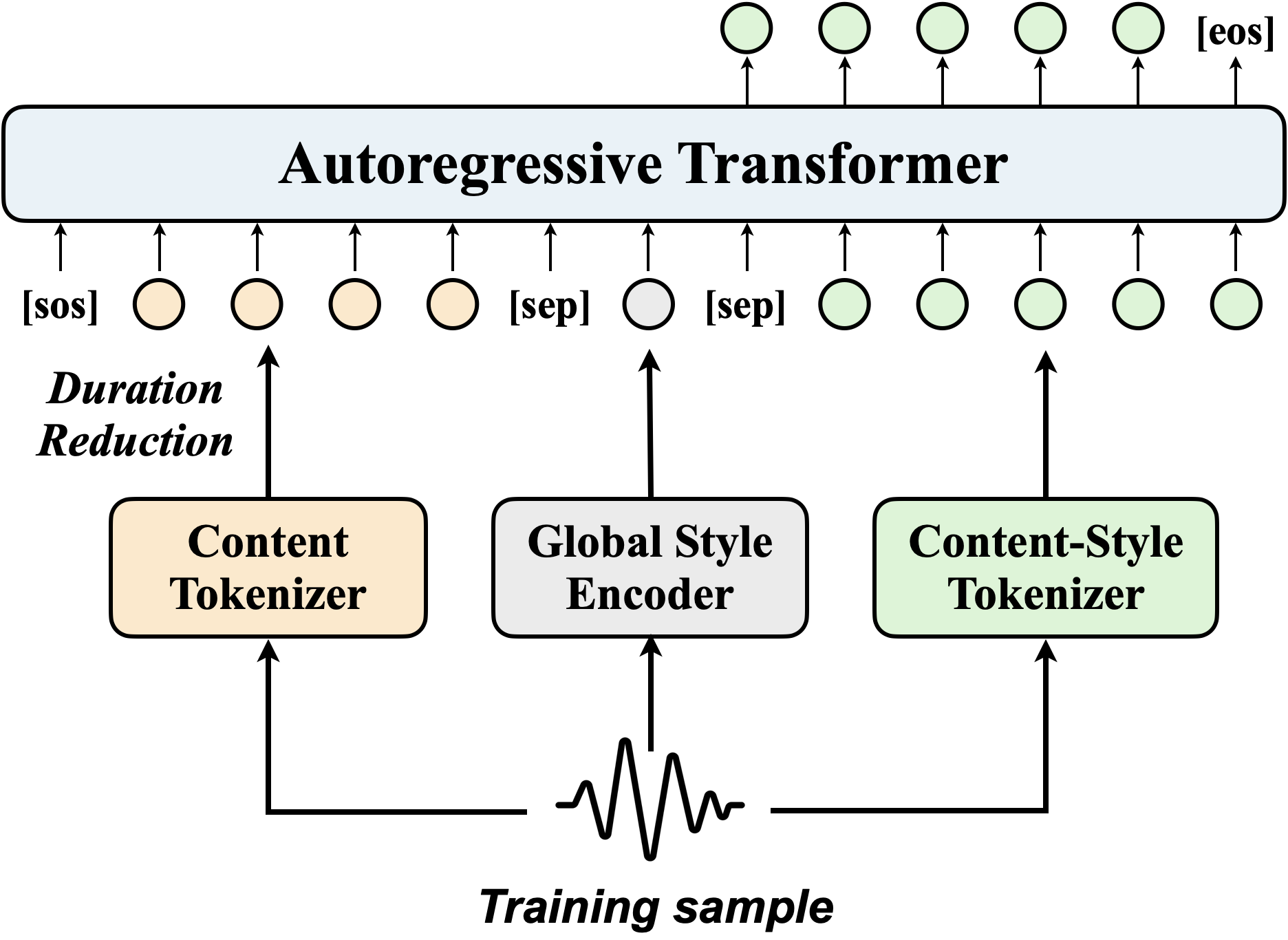}
        \caption{Training}
        % \vspace{-1mm}
        \label{fig:vevo-ar-training}
    \end{subfigure}
    \hfill
    \begin{subfigure}[b]{0.57\textwidth}
        \centering
        \includegraphics[width=\textwidth]{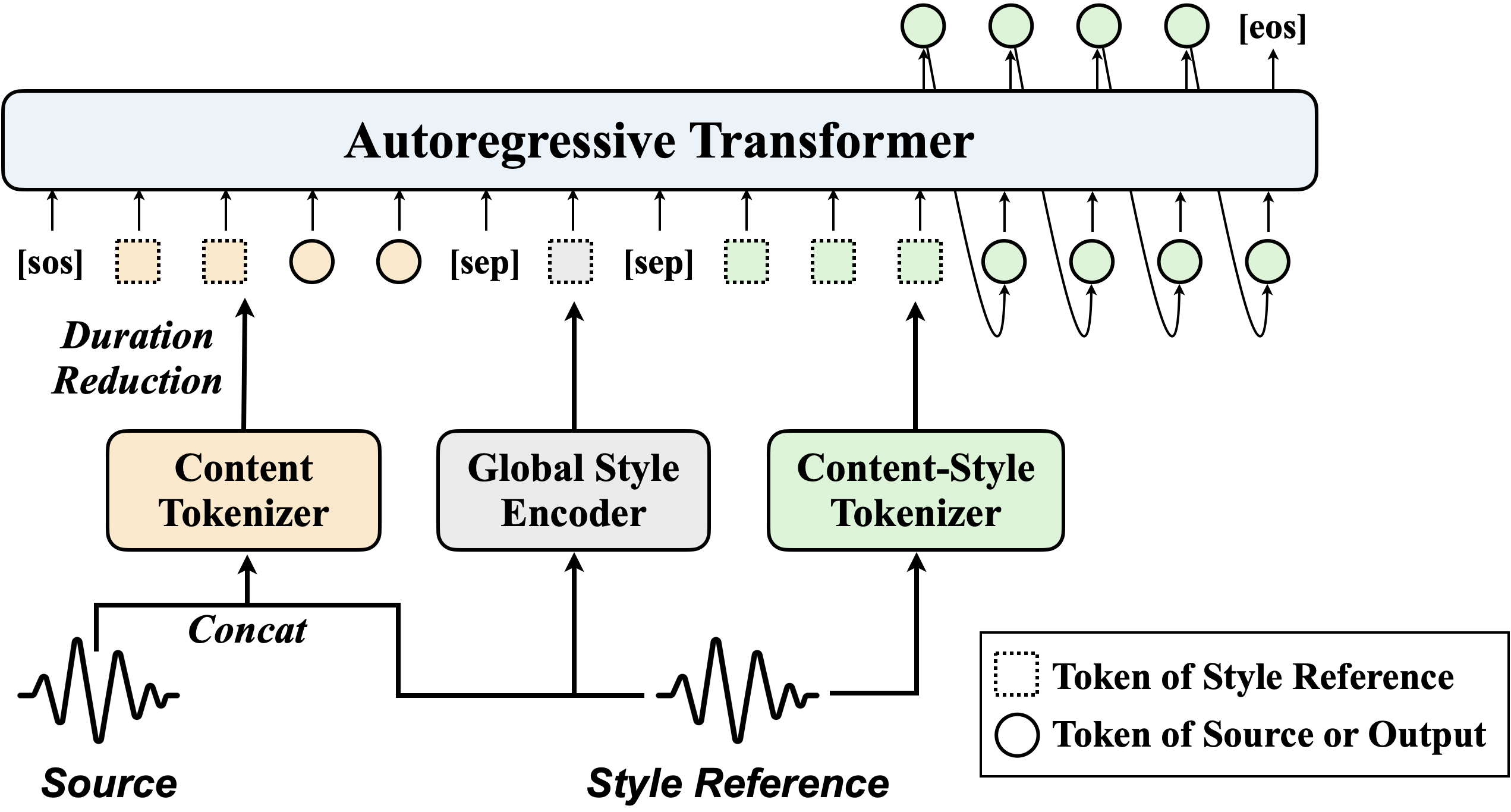}
        \caption{Inference}
        % \vspace{-1mm}
        \label{fig:vevo-ar-inference-enhanced}
    \end{subfigure}
    \caption{\textbf{Vevo-Voice}: an auto-regressive voice conversion model. During inference, both style and timbre of the source speech will be converted into the reference speech ones.
    }
    \label{fig:vevo-ar}
\end{figure}

Vevo-Voice uses both the content-style modeling and the acoustic modeling stages of Vevo. Particularly, the style reference and timbre reference of these two stages are same, in order that we can convert not only style but also timbre of the source speech. Specifically, during the content-style modeling stage (Figure~\ref{fig:vevo-ar}), our goal is to transform the content token of speech into content-style tokens, which is prompted by a style reference. This can be formulated as a sequence-to-sequence generation task. We employ a decoder-only autoregressive (AR) transformer, known for its powerful capability in such tasks~\cite{transformer,touvron2023llama,seedtts}. 

During training, Vevo conducts self-supervised learning on only speech data and do not need any parallel corpus. The input sequence of transformer is $[\langle \text{SOS} \rangle, \bm{Q}_{c}^{'} ({u}), \langle \text{SEP} \rangle, \bm{g} ({u}), \langle \text{SEP} \rangle, \bm{Q}_{s} ({u})]$. We only perform the next token prediction on the last $[\langle \text{SEP} \rangle, \bm{Q}_{s} ({u})]$, with the ground truth being $[\bm{Q}_{s} ({u}), \langle \text{EOS} \rangle]$. Here, $\langle \text{SOS} \rangle$, $\langle \text{SEP} \rangle$, and $\langle \text{EOS} \rangle$ are treated as three special tokens in language model~\cite{bert}. During inference, for a source speech ${u}_i$ and a style reference ${u}_{sr}$, Vevo conducts the continuation by feeding the input sequence $[\langle \text{SOS} \rangle, \bm{Q}_{c}^{'} ({u}_{sr} \oplus {u}_i), \bm{g} ({u}_{sr}), \bm{Q}_s ({u}_{sr})]$ for autoregressive generation, where $\oplus$ means the concatenation.

\subsubsection{Noro}

\begin{figure}[htbp]
\centering
\includegraphics[width=0.8\textwidth]{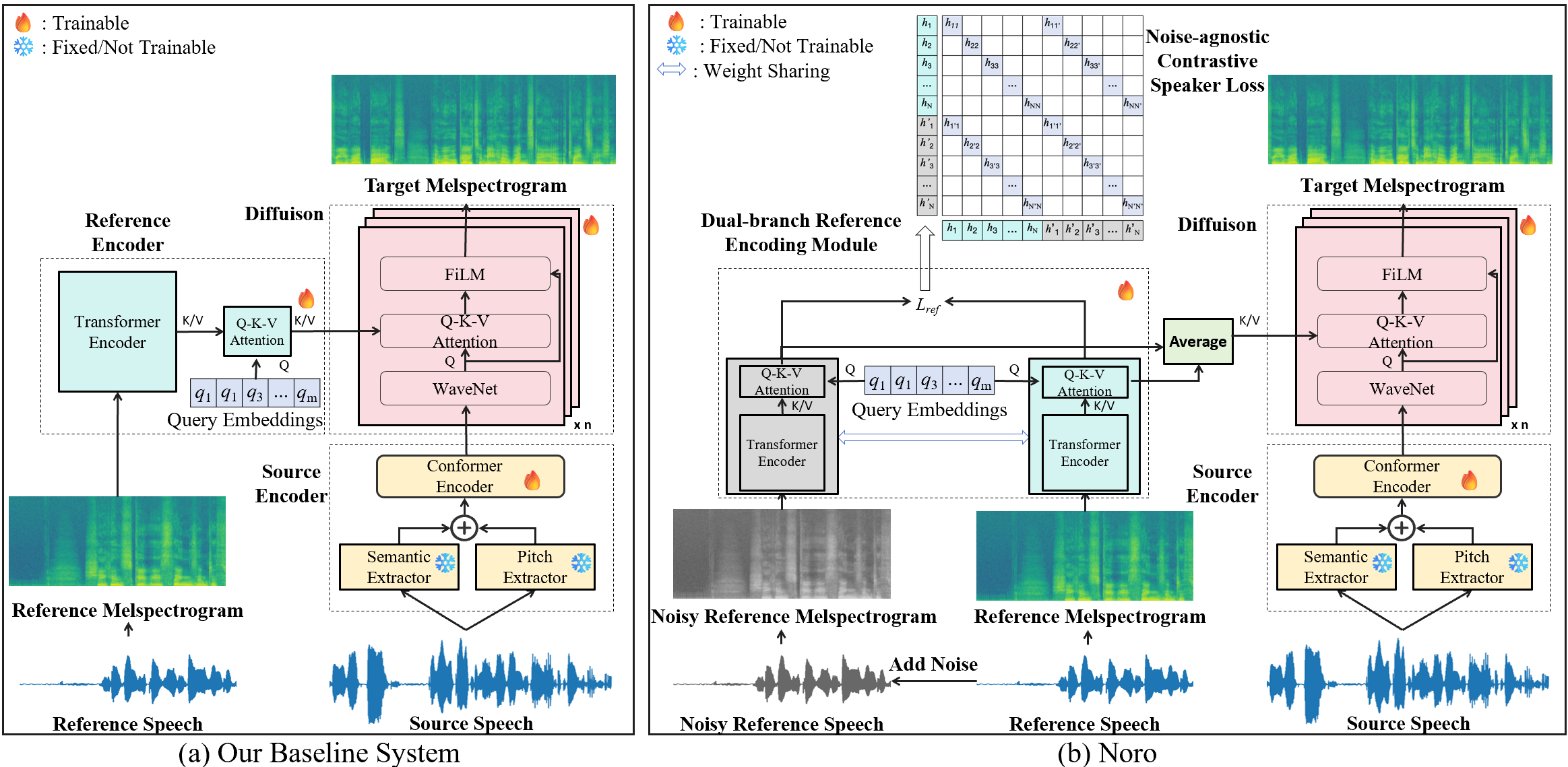}
\caption{Overview of the model architecture of our baseline system and Noro.}
\label{fig:noro} 
\end{figure}

Noro is a diffusion-based, noise-robust, one-shot voice conversion system featuring a dual-branch reference encoding module and a noise-agnostic contrastive speaker loss to enhance robustness in real-world scenarios, availble via this link~\footnote{\url{https://github.com/open-mmlab/Amphion/blob/main/egs/vc/Noro/}}. This section details the components and mechanisms of Noro. We begin by introducing our baseline one-shot voice conversion system. As illustrated in Fig.~\ref{fig:noro} (a), this system consists of a source encoder that simultaneously encodes semantic and pitch representations, a reference encoder that extracts speaker timbre representation from the reference speech, and a diffusion model that utilizes these representations as conditions to predict the mel-spectrogram of the target speech. To achieve noise robustness against noisy reference speeches, we propose Noro. An overview of Noro is provided in Fig.~\ref{fig:noro} (b). Noro builds upon the baseline system but replaces its original reference encoders with a dual-branch reference encoding module and incorporates a noise-agnostic contrastive speaker loss.

\paragraph{The baseline system of Noro}
(a) \textit{Source Encoder} To derive semantic representations from the source speech, our baseline system uses a pre-trained and frozen HuBERT model~\cite{HuBERT} as the semantic extractor. This model encodes the source speech into continuous embeddings. We then apply K-Means quantization to these embeddings. Instead of using K-Means cluster IDs (discrete semantic tokens), we use the quantized continuous embeddings, which preserve more information, for semantic representation. The resulting semantic representation for the source speech is denoted as $s_{src}$. For pitch representations, we use the open-source software PyWorld\footnote{\url{https://github.com/JeremyCCHsu/Python-Wrapper-for-World-Vocoder}} to extract frame-level F0 values from the source speech. We normalize these values and use them for pitch representation, denoted as $p_{src}$. After extracting the semantic and pitch representations, we employ a Conformer~\cite{conformer} encoder to simultaneously encode $s_{src}$ and $p_{src}$, obtaining the final source representation $h_{src}$.
(b) \textit{Reference Encoder}
To encode the speaker timbre in the reference speech, we leverage the speech prompting mechanism in NaturalSpeech 2~\cite{naturalspeech2}. Specifically, we first employ a Transformer based reference encoder to convert the melspectrogram of the reference speech into a hidden sequence. Instead of allowing the diffusion model to directly attend to this hidden sequence, we employ two attention blocks. In the first attention block, we use $m$ randomly initialized query embeddings to attend to the hidden sequence, resulting in $h_{ref}$, a hidden sequence of length $m$ that represents the reference speaker's timbre at the utterance level. In the second attention block, the output hidden sequence of the WaveNet Layer serves as the query, while $h_{ref}$ functions as both the key and value. The attention results of the second attention block are then used as conditional information for a FiLM layer in the final acoustic modeling process.
(c) \textit{Diffusion Model}
% \label{method-diffusion}
To enhance in-context learning for improved one-shot generation, each training sample is randomly segmented during the training phase. A segment comprising 25-45\% of the total sample is designated as the reference speech, while the remaining portion functions as both the source and target speech. We utilize a WaveNet-based diffusion model~\cite{naturalspeech2}, denoted as $s_\theta$, to predict the melspectrogram $z_{tgt}$ of the target speech. We describe both the diffusion (forward) process and the denoising (reverse) process using the stochastic differential equation. 
In forward, Gaussian noise is added to the melspectrogram $z_{tgt}$:
\begin{equation}
z_{tgt}^{t} = e^{-\frac{1}{2} \int_{0}^{t} \beta_s ds} z_{tgt} + (I - e^{- \int_{0}^{t} \beta_s ds}) \epsilon,
\end{equation}
where $\beta_t$ represents the noise schedule function, $\epsilon$ denotes randomly sampled Gaussian noise, and $t \in [0, 1]$.
The reverse process can be formulated as:
\begin{equation}
d z_{tgt}^t = -\left(\frac{1}{2} z_{tgt}^t + \nabla \log p_t(z_{tgt}^t)\right) \beta_t \, dt + \sqrt{\beta_t} \epsilon.
\end{equation}
$s_\theta$ takes the noised melspectrogram $z_{tgt}^t$, the time step $t$, and the conditions of the source representation $h_{src}$, containing both semantic and pitch information, and the reference representation $h_{ref}$, containing speaker timbre information, to estimate $\nabla \log p_t(z_{tgt}^t)$. The loss function of the diffusion model can be written as:
\begin{equation}
L_{diff} = \left\|s_\theta(z_{tgt}^t, t, h_{src}, h_{ref}) - \nabla \log p_t(z_{tgt}^t)\right\|_{1}.
\end{equation}
During the inference stage, we gradually denoise $z_{tgt}^t$ utilizing the estimated score $s_\theta(z_{tgt}^t, t, s_{src}, p_{src}, h_{ref})$. This process initiates with Gaussian noise, initially sampled as $z_{tgt}^1$.

\paragraph{Dual-branch Reference Encoding Module} 
To improve the noise robustness of the baseline system, we replaced its original reference encoder with a dual-branch reference encoding module. This module incorporates two reference encoders sharing identical model weights to form a dual-branch structure. For each clean training sample, we randomly mix clean reference speech with eight types of noise from the DEMAND database~\cite{demand} at SNRs following a normal distribution (0,20) dB to create noisy reference speeches. Subsequently, the clean and noisy reference speeches are processed by separate reference encoders using the same query embeddings. This process generates clean and noisy reference representations $h_{\text{ref}}$ and $h'_{\text{ref}}$, respectively. Considering the weight-sharing scheme used in this dual-branch reference encoding module, the method described here qualifies as a "training strategy" for enhancing our baseline system. Initially, we load the Noro system with weights from a pre-trained baseline system. During the training phase, the average of $h_{\text{ref}}$ and $h'_{\text{ref}}$ is fed into the diffusion model for acoustic modeling and acts as the keys and values in the second attention block. During the inference phase, only one reference encoder is utilized, and the Noro system's structure is the same as the baseline system.

\paragraph{Noise-agnostic Contrastive Speaker Loss} 
After obtaining the representations of the clean speeches $h_{\text{ref}}$ and their noisy counterparts $h'_{\text{ref}}$, we introduce a noise-agnostic contrastive speaker loss. This loss function aims to maximize the similarity between samples (clean or noisy) from the same speaker while minimizing it for samples from different speakers. First, as $h_{\text{ref}}$ and $h'_{\text{ref}}$ are hidden sequences of length $m$, we perform average pooling over the length dimension to form a comprehensive reference representation that captures the reference speaker's timbre at the utterance level. We then concatenate $h_{\text{ref}}$ and $h'_{\text{ref}}$ along the batch size dimension:
$h_{\text{all}} = \left[ h_{\text{ref}}; h'_{\text{ref}} \right].$
Given that the noisy speeches share the same speakers as the clean ones, we can express $y_{\text{all}}$ as:
$y_{\text{all}} = \left[ y_{\text{spk}}; y_{\text{spk}} \right].$

Finally, our proposed noise-agnostic contrastive speaker loss is formulated as:
\begin{equation}
\mathcal{L}_{\text{ref}} = \frac{1}{2N} \sum_{i=1}^{2N} \text{CrossEntropy} \left( \frac{h_{i} \cdot h_{j}^T}{\tau}, M_{i,j} \right),
\end{equation}
where $h_{i}$ and $h_{j}$ are the $i$-th and $j$-th reference representations in $h_{\text{all}}$, $\tau$ is the temperature parameter adjusting the scaling of the logits, $N$ is the batch size, and the mask matrix $M_{i,j}$ is defined as follows:
\begin{equation}
M_{i,j} =
\begin{cases} 
1 & \text{if } y_{i} = y_{j}, \\ 
0 & \text{otherwise},
\end{cases}
\end{equation}
where $y_{i}$ and $y_{j}$ are the speaker labels for $h_{i}$ and $h_{j}$ respectively.
This loss ensures that the reference encoders represent the timbre of different speakers regardless of noise interference, enhancing the robustness of the system.
The total loss function for the Noro system is defined as follows:
\begin{equation}
L_{total} = \alpha L_{diff} + \beta L_{ref},
\end{equation}
where \(\alpha\) and \(\beta\) are the weights assigned to each loss component. 

\subsection{Text to Audio}

\begin{figure*}[htbp]
  \centering
  \centerline{\includegraphics[width=0.55\linewidth]{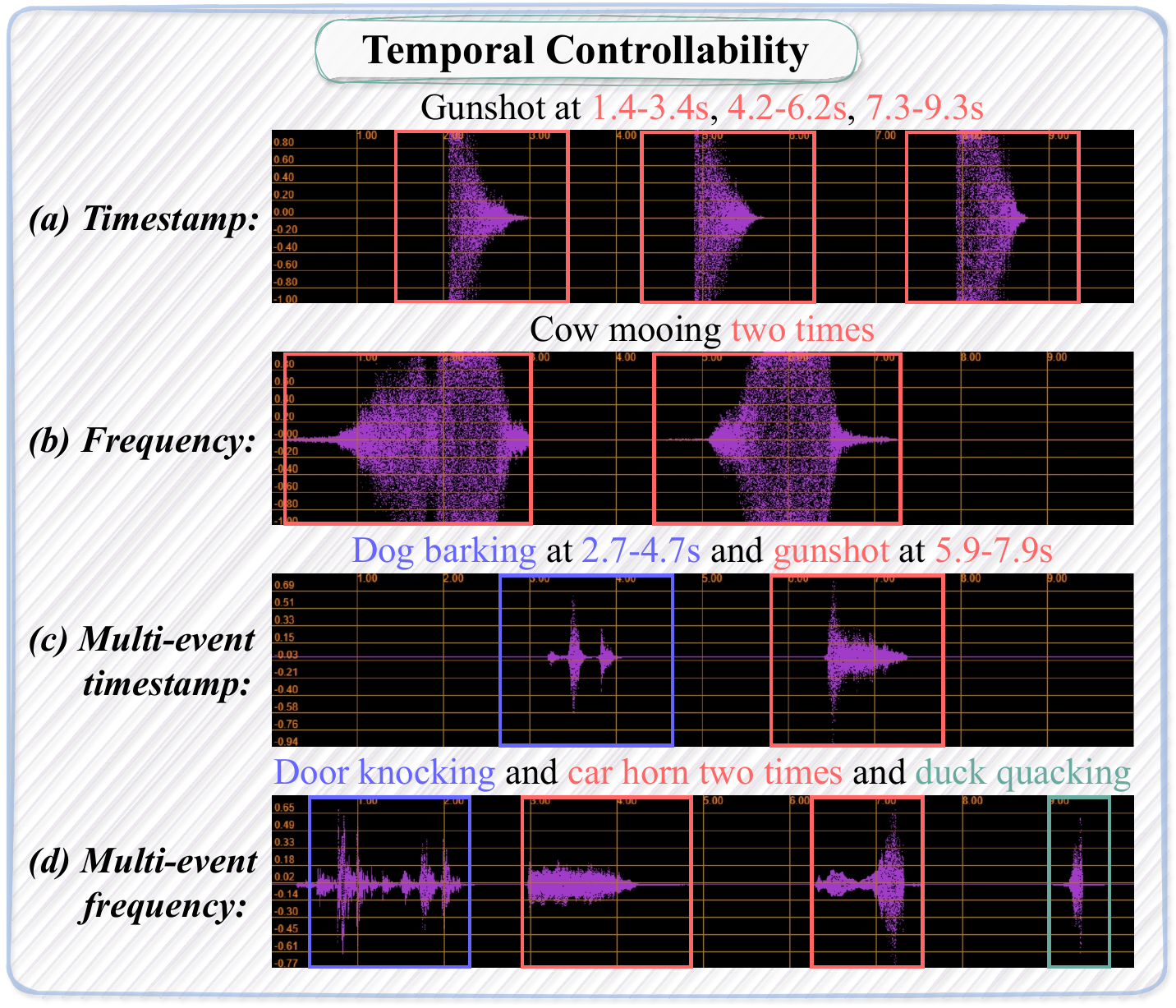}}
  \caption{Illustration of controlling timestamp / occurrence frequency of audio events.
  }
  \label{fig:tta_pico_example}
\end{figure*}
\begin{figure*}[htbp]
  \centering
  \centerline{\includegraphics[width=0.98\linewidth]{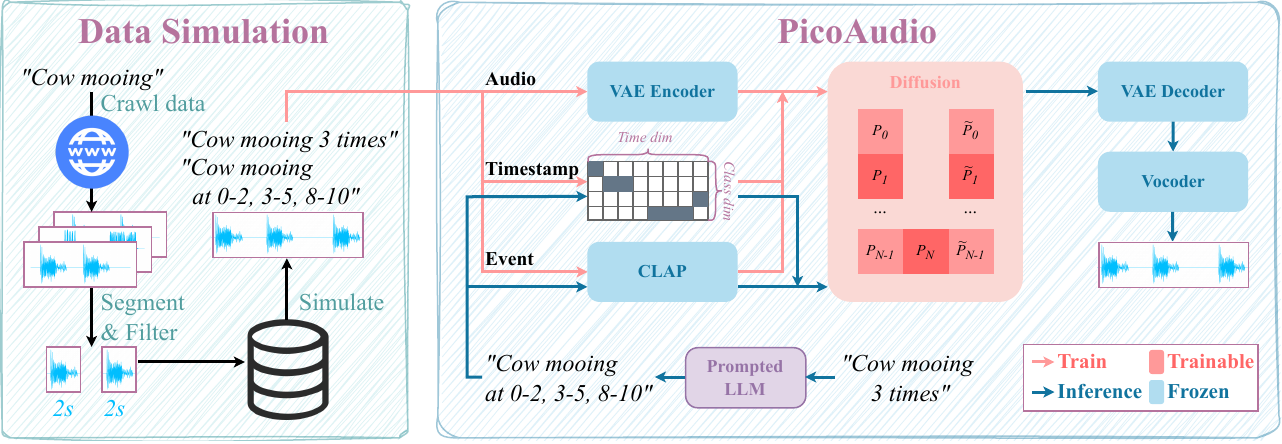}}
    \caption{
    An overview of PicoAudio. 
    (\textbf{Left}) illustrates the simulation pipeline, wherein data is crawled from the Internet, segmented and filtered, resulting in one-occurrence segments stored in a database.
    Pairs of audio, timestamp captions, and frequency captions are simulated from the database. 
    (\textbf{Right}) showcases the model framework. {\textcolor[RGB]{234,107,102}{Red}} arrows indicate the model training process. {\textcolor[RGB]{30,120,255}{Blue}} arrows indicate inference based on timestamp or frequency captions, where an LLM is prompted with the simulated training data.
    }
    \label{fig:tta_pico_model}
\end{figure*}

%TODO
Text-to-audio (TTA) aims to generate sound clips based on given text.
Recently, with significant progress in audio generation, it has become possible to synthesize high-fidelity audio clips~\cite{wang2023audit, kreuk2022audiogen,yang2023diffsound, liu2023audioldm2,ghosal2023text}.
However, temporal relations—a critical feature for audio content—remain underrepresented in mainstream models, resulting in imprecise temporal controllability.
As illustrated in Figure~\ref{fig:tta_pico_example}, a common demand is the ability to precisely control the timestamps or occurrence frequencies of audio events during generation.
To address this, we propose \textbf{PicoAudio}\footnote{Online inference is available on Hugging Face at \href{https://huggingface.co/spaces/amphion/PicoAudio}{\textcolor{cyan}{\textit{Space-Amphion-PicoAudio}}}.
Code and simulated datasets for training and evaluation are available at \href{https://github.com/open-mmlab/Amphion/tree/main/models/tta/picoaudio}{\textcolor{cyan}{\textit{Amphion-PicoAudio}}}.}, which enables \textbf{P}recise t\textbf{I}mestamp and frequency \textbf{CO}ntrollability of audio events.
An overview of PicoAudio is shown in Figure~\ref{fig:tta_pico_model}. It:
(1) leverages data crawling, segmentation, and filtering to simulate fine-grained temporally-aligned audio-text data; and
(2) integrates temporal information to guide audio generation through tailored model design and preprocessing with LLM.

% PicoAudio utilizes a data simulation pipeline~\cite{xu2024detailed} to simulate temporally-aligned audio-text annotations.
% The pipeline entails data crawling, segmenting and filtering audio clips to gather high-quality audio segments for simulation. 
% PicoAudio introduces tailored modules for temporal control.
% (a) Timestamp control is achieved by incorporating customized input, namely timestamp caption.
% With the assistance of LLMs~\cite{achiam2023gpt}, (b) frequency control, (c) ordering via multi-event timestamp control and (d) multi-event frequency control can be implemented, as shown in~\ref{fig:intro_example}.
% Beyond (a) - (d), \textbf{PicoAudio can achieve arbitrary precise temporal control} as long as LLMs are capable of converting the prompt into timestamp captions, which is straightforward for LLMs when prompted with simulated data.
% Our contributions encompass the following:  
% \begin{enumerate}
% \item A data simulation pipeline tailored specifically for temporal controllable audio generation frameworks;
% \item A timestamp and frequency controllable generation framework, enabling precise control over sound events;
% \item Achieving any temporal control by integrating LLMs.
% \end{enumerate}

% \paragraph{Temporal Controllable Model}
% To enable temporal control in audio generation, we first design a simulation pipeline that automatically acquires data and a tailored text processor to enhance the temporal awareness of audio generation models, as shown in Figure \ref{fig:simulation_model}. 

\paragraph{Temporally-aligned Data Simulation}

(a) \textit{Data crawling, segmentation \& filtering} Audios are crawled from the Internet using event tags as search keywords.
A text-to-audio grounding (TAG) model~\cite{xu2024towards} and a CLAP model~\cite{laionclap2023} are utilized for further filtering to obtain a database with a substantial number of high-quality one-occurrence segments.
% These weakly annotated clips possess only sound event tags and may contain noise.
% A text-to-audio grounding (TAG) model~\cite{xu2024towards}, which aims at temporally locating events described by text, is employed to segment crawled data. 
% Each localized segment encompasses \textbf{one occurrence} of a sound event, such as a ``\textit{2-seconds cow mooing}'' segment.
% We define a burst of continuous short sound segments as \textbf{one occurrence}, such as a burst of ``keyboard typing'' or ``door knocking''.
% To ensure data quality, a contrastive language-audio pre-training (CLAP) model~\cite{laionclap2023} is utilized for further filtering. 
% Thus, we obtain a database with a substantial number of high-quality one-occurrence segments. 
(b) \textit{Simulation} We randomly select events from the database and synthesize audio by randomly assigning occurrence on-set.
A simulated pair comprises an audio clip and a timestamp caption formatted as ``event-1 at timing-1 ... and event-N at timing-N'', as well as a occurrence frequency caption formatted as ``event-1 j times ... and event-N k times''.

\paragraph{Temporal Controllable Model}
(a) \textit{Text Processor} 
LLM is adopted to convert various input formats into standard format.
For example, the input ``a dog barking occurred between two and three seconds'' is transformed into the timestamp caption format ``dog barking at 2-3''.
The standard format makes rule-based transformations very straightforward.
The one-hot timestamp matrix $\mathcal{O} \in \mathbb{R}^{C \times T}$ is derived from the timestamp caption, where $C$ and $T$ denote the number of sound events and the frame number, respectively.
Futhermore, PicoAudio employs CLAP~\cite{laionclap2023} to extract audio event information beyond timestamp, denoted as event embedding $\mathcal{I}$.

\begin{equation}
 \mathcal{O}_{c, t}=\left\{
    \begin{aligned}
    &1, \text{ if event $c$ occurs at $t$-th frame }\\
    &0, \text{ otherwise}\\
    \end{aligned}
\right
.
\end{equation}

(b) \textit{Audio Representation}
PicoAudio utilizes a variational autoencoder (VAE) for audio representation to reduce computational complexity.
The VAE compresses the audio spectrogram 
$\mathcal{A} \in \mathbb{R}^{T\times M}$ into the latent representation $\mathcal{P}$.
A vocoder following the VAE decoder converts the spectrogram back into a waveform.
(C) \textit{Diffusion Model}
PicoAudio utilizes a diffusion model to predict 
$\mathcal{\tilde{P}}$ based on the timestamp condition 
$\mathcal{O}$ and event condition $\mathcal{I}$.
The model is trained to estimate the noise based on the noisy representation $\mathcal{P}_n$, the condition $\mathcal{O}$, $\mathcal{I}$, and a weight $\lambda_n$ related to the signal-to-noise ratio:
\begin{equation}
  \label{eqn:diffusion_r1}
        \mathcal{L}=\sum_{n=1}^{N}{\lambda_n \mathbb{E}_{\epsilon_n, \mathcal{P}_0}||\epsilon_n - \epsilon_{\theta}([\mathcal{P}_n,\mathcal{O}], \mathcal{I}) ||}
\end{equation}
where $[\cdot,\cdot]$ denotes concatenation.

\newpage
\section{Tutorial Examples}
\subsection{Data Preparation with Emilia-Pipe}

% \wcr{brief introduction of Emilia-Pipe -> how to prepare data -> how to set up Emilia-Pipe env -> how to run code -> what's the output -> common questions and issues}

The Emilia-Pipe is an open-source preprocessing pipeline designed to transform raw speech data into high-quality, annotated datasets for speech generation tasks. This subsection provides step-by-step guidance on how to prepare your data using Emilia-Pipe, ensuring reproducibility and optimal results. For further details, refer to the full documentation at \href{https://arxiv.org/abs/2407.05361}{our paper}.

\subsubsection{Setup and Installation}

\paragraph{1. Install Python and CUDA.}  
Ensure that you have Python 3.9 and CUDA installed on your system. Create and activate a Conda environment with the following commands:

\begin{lstlisting}
git clone https://github.com/open-mmlab/Amphion
cd preprocessors/Emilia

conda create -y -n AudioPipeline python=3.9 
conda activate AudioPipeline

bash env.sh
\end{lstlisting}

\paragraph{2. Download Pre-trained Models.}  
Download the necessary pre-trained models manually from the following locations:
\begin{itemize}
    \item \textbf{UVR-MDX-NET-Inst\_HQ\_3:} \href{https://github.com/TRvlvr/model_repo/releases/download/all_public_uvr_models/UVR-MDX-NET-Inst_HQ_3.onnx}{Download here}.
    \item \textbf{DNSMOS P.835:} \href{https://github.com/microsoft/DNS-Challenge/blob/master/DNSMOS/DNSMOS/sig_bak_ovr.onnx}{Download here}.
\end{itemize}

Additionally, create a HuggingFace access token following the \href{https://huggingface.co/pyannote/speaker-diarization-3.1#requirements}{official guide} for the \texttt{pyannote/speaker-diarization-3.1} model. Ensure a stable internet connection to automatically download other dependencies such as Silero-VAD and WhisperX-medium during the first pipeline execution.

\paragraph{3. Filling in the Configuration File.}  
The core configuration for Emilia-Pipe is contained in the \texttt{config.json} file. Update the following fields to customize the pipeline for your dataset:

\begin{lstlisting}
{
    "language": {
        "multilingual": true,
        "supported": ["zh", "en", "fr", "ja", "ko", "de"]
    },
    "entrypoint": {
        "input_folder_path": "examples", // Path to input audio files
        "SAMPLE_RATE": 24000
    },
    "separate": {
        "step1": {
            "model_path": "/path/to/model/separate_model/UVR-MDX-NET-Inst_HQ_3.onnx",
            "denoise": true
        }
    },
    "mos_model": {
        "primary_model_path": "/path/to/model/mos_model/DNSMOS/sig_bak_ovr.onnx"
    },
    "huggingface_token": "<HUGGINGFACE_ACCESS_TOKEN>"
}
\end{lstlisting}

Key configuration fields to update:
\begin{enumerate}
    \item \texttt{input\_folder\_path}: Set the folder path where the raw audio files are stored.
    \item \texttt{model\_path}: Provide the paths to the downloaded UVR-MDX-NET and DNSMOS models.
    \item \texttt{huggingface\_token}: Input your HuggingFace access token.
\end{enumerate}

\subsubsection{Running the Pipeline}

\paragraph{Activating the Environment.}  
Activate the Conda environment and specify the GPU for execution:

\begin{lstlisting}
conda activate AudioPipeline
export CUDA_VISIBLE_DEVICES=0
\end{lstlisting}

\paragraph{Executing the Pipeline.}  
Run the pipeline using the following command:

\begin{lstlisting}
python main.py
\end{lstlisting}

Processed audio data will be saved in a new folder named \texttt{<input\_folder\_path>\_processed}. This folder contains both segmented audio files and corresponding metadata.

\paragraph{Output Format}

The output for each processed audio file includes:
\begin{itemize}
    \item \textbf{MP3 Files:} Segmented audio files named \texttt{<original\_name>\_<idx>.mp3}, where \texttt{idx} is the segment index.
    \item \textbf{JSON Metadata:} A \texttt{.json} file containing transcriptions, timestamps, language, and quality scores.
\end{itemize}

An example of the JSON structure is shown below:
\begin{lstlisting}
[
    {
        "text": "This is an example transcription.",
        "start": 67.18,
        "end": 74.41,
        "language": "en",
        "dnsmos": 3.44
    }
]
\end{lstlisting}

\subsection{TTS and Voice Conversion with Vevo}
We present our reproduction of Vevo~\cite{vevo}, a versatile zero-shot voice imitation framework with controllable timbre and style. We invite readers to explore the audio samples\footnote{\href{https://versavoice.github.io/}{https://versavoice.github.io/}} to experience Vevo's capabilities firsthand.

We have included the following pre-trained Vevo models at Amphion:
\begin{itemize}
    \item \textbf{Vevo-Timbre}: It can conduct style-preserved voice conversion.
    \item \textbf{Vevo-Style}: It can conduct style conversion, such as accent conversion and emotion conversion.
    \item \textbf{Vevo-Voice}: It can conduct style-converted voice conversion.
    \item \textbf{Vevo-TTS}: It can conduct style and timbre controllable TTS.
\end{itemize}

To run this model, you need to follow the steps below:
\begin{enumerate}
    \item Clone the repository and install the environment.
    \item Run the inference script.
\end{enumerate}

\subsubsection{Setup and Installation}

\begin{lstlisting}
# Clone the repository
git clone https://github.com/open-mmlab/Amphion.git
cd Amphion

# Install the environment
# For Debian-like distribution (e.g. Ubuntu, Mint, etc.)
sudo apt-get install espeak-ng
# For RedHat-like distribution (e.g. CentOS, Fedora, etc.) 
sudo yum install espeak-ng

# For Windows
# Please visit https://github.com/espeak-ng/espeak-ng/releases to download .msi installer

conda create -n vevo python=3.10
conda activate vevo
pip install -r models/vc/vevo/requirements.txt

\end{lstlisting}

\subsubsection{Running Inference}

\begin{lstlisting}
# Vevo-TTS
python -m models.vc.vevo.infer_vevotts

# Vevo-Timbre
python -m models.vc.vevo.infer_vevotimbre

# Vevo-Style
python -m models.vc.vevo.infer_vevostyle

# Vevo-Voice
python -m models.vc.vevo.infer_vevovoice
\end{lstlisting}

Running this will automatically download the pretrained model from HuggingFace\footnote{\href{https://huggingface.co/amphion/Vevo}{https://huggingface.co/amphion/Vevo}} and start the inference process. The result audio is by default saved in \textit{models/vc/vevo/wav/output*.wav}, you can change this in the scripts \textit{models/vc/vevo/infer\_vevo*.py}.

\subsection{Inference MaskGCT TTS Model}

The MaskGCT model significantly advances text-to-speech (TTS) systems by adopting a fully non-autoregressive architecture. This design eliminates the reliance on explicit alignment information between textual and acoustic features and phoneme-level duration prediction, thereby simplifying the modeling process. The framework operates in two distinct stages. In the first stage, semantic tokens are predicted from textual input using representations derived from speech self-supervised learning (SSL) models. Subsequently, in the second stage, acoustic tokens are generated conditionally on these semantic tokens. The training paradigm follows a mask-and-predict strategy, enabling the model to reconstruct masked semantic or acoustic tokens under specified conditions. During inference, this architecture facilitates efficient, parallelized token generation. We also briefly introduce MaskGCT in Section~\ref{sec:maskgct}.

\subsubsection{Setup and Installation}

To facilitate the deployment of the MaskGCT model, the following setup procedure should be adhered to:

\paragraph{1. Repository Cloning and Environment Setup}

Begin by cloning the official Amphion repository and configuring the runtime environment. A \texttt{conda} virtual environment is recommended for dependency isolation.

\begin{lstlisting}
git clone https://github.com/open-mmlab/Amphion.git
cd Amphion
conda create -n maskgct python=3.10
conda activate maskgct
pip install -r models/tts/maskgct/requirements.txt
\end{lstlisting}

\paragraph{2. Installation of eSpeak-NG}

Since MaskGCT employs the \texttt{phonemizer} library for text-to-phoneme conversion, it is imperative to install the eSpeak-NG dependency. For systems based on Debian, the installation command is as follows:

\begin{lstlisting}
sudo apt-get install espeak-ng
\end{lstlisting}

\subsubsection{Running Inference}
\paragraph{Pre-trained Model Acquisition}

MaskGCT provides several pre-trained models, including components for semantic codec, acoustic codec, text-to-semantic (T2S) prediction, and semantic-to-acoustic (S2A) generation. These resources can be downloaded from HuggingFace using the following script:

\begin{lstlisting}
from huggingface_hub import hf_hub_download

# Semantic codec
semantic_code_ckpt = hf_hub_download("amphion/MaskGCT",     
    filename="semantic_codec/model.safetensors")

# Acoustic codec
codec_encoder_ckpt = hf_hub_download("amphion/MaskGCT", 
    filename="acoustic_codec/model.safetensors")
codec_decoder_ckpt = hf_hub_download("amphion/MaskGCT", 
    filename="acoustic_codec/model_1.safetensors")

# T2S model
t2s_model_ckpt = hf_hub_download("amphion/MaskGCT", 
    filename="t2s_model/model.safetensors")

# S2A models
s2a_1layer_ckpt = hf_hub_download("amphion/MaskGCT", 
    filename="s2a_model/s2a_model_1layer/model.safetensors")
s2a_full_ckpt = hf_hub_download("amphion/MaskGCT", 
    filename="s2a_model/s2a_model_full/model.safetensors")
\end{lstlisting}

Upon successful acquisition and configuration of the pre-trained models, the inference process can be initiated using the provided script. The following command illustrates the execution of the text-to-speech generation:

\begin{lstlisting}
python -m models.tts.maskgct.maskgct_inference
\end{lstlisting}

By default, the generated audio is saved as \texttt{generated\_audio.wav}. Users may modify the output path within the script to suit specific requirements.

% \subsubsection{Conclusion}

% The MaskGCT model exemplifies a robust, non-autoregressive approach to TTS, achieving efficient token generation and enhanced scalability. By following the outlined setup and inference steps, researchers and practitioners can effectively leverage this model for high-quality speech synthesis. For further details and advanced configurations, reference the official documentation provided in the Amphion repository.

\subsection{Using PicoAudio for Temporally Controllable Audio Generation}

\subsubsection{Code and Data Preparation}

\begin{lstlisting}
# Clone the repository
git clone https://github.com/open-mmlab/Amphion.git
cd Amphion/models/tta/picoaudio/picoaudio

# Install the environment
pip install -r requirements.txt
\end{lstlisting}

Download the training data from HuggingFace-Datasets\footnote{\href{https://huggingface.co/datasets/amphion/PicoAudio/tree/main}{https://huggingface.co/datasets/amphion/PicoAudio/tree/main}} and extract it to the directory ~\textit{"/data"}.

\subsubsection{Training}

\begin{lstlisting}
# root_dir : Amphion/models/tta/picoaudio/picoaudio
# To start traning:
accelerate launch runner/controllable_train.py
\end{lstlisting}

\subsubsection{Inference}

\textbf{Online inference} is available on HuggingFace Space\footnote{\href{https://huggingface.co/spaces/amphion/PicoAudio}{https://huggingface.co/spaces/amphion/PicoAudio}}.

\textbf{Offline inference} files are available on HuggingFace\footnote{\href{https://huggingface.co/spaces/amphion/PicoAudio/tree/main}{https://huggingface.co/spaces/amphion/PicoAudio/tree/main}}.
HuggingFace inference uses Gemini as a preprocessor, and we also provide a GPT preprocessing script in \textit{"/llm\_preprocess.py"}
\begin{lstlisting}
# Utilize LLM for text preprocessing
python inference.py \
    --text "spraying two times then gunshot three times." \
    
# Specify timestamp captions
# Timestamp caption, formatted as "event_1 at onset1-offset1_onset2-offset2 and event_2 at onset1-offset1 ...".
python inference.py \
    --timestamp_caption "spraying at 0.38-1.176_3.06-3.856 and gunshot at 1.729-3.729_4.367-6.367_7.031-9.031." \
\end{lstlisting}

\section{Experiments}
\subsection{Datasets}
In this section, we present evaluation results to Amphion's 100K-hour Emilia dataset.
To evaluate quality, we compared Emilia with existing datasets using DNSMOS P.835 OVRL~\cite{dnsmos835} scores. This non-intrusive speech quality metric reflects the overall quality of the speech data and is highly correlated with human ratings~\cite{dnsmos835}. 
To quantify this diversity, we conducted analyses on both the acoustic and semantic feature space, comparing it with the MLS~\cite{mls} dataset.
Specifically, we randomly select 5,000 samples each from the English subset of MLS and Emilia, and leverage a pre-trained WavLM model\footnote{\url{https://huggingface.co/microsoft/wavlm-base-plus}} to extract acoustic representations~\cite{chen2022wavlm}. We then apply the PCA algorithm to reduce the dimensionality to two.
For the semantic diversity analysis, we employ a pre-trained Sentence-BERT model\footnote{\url{https://github.com/UKPLab/sentence-transformers}} to generate text representations for the transcripts of each speech data. 

Table~\ref{tab:dnsmos_results} presents the speech quality comparison between Emilia and several existing datasets.
Emilia achieves a DNSMOS P.835 OVRL score of 3.26, ranking third among all datasets. The results indicate that, despite being sourced from raw speech data in the wild, after preprocessing, the speech quality of the Emilia dataset is comparable to existing datasets sourced from studio recordings or audiobooks and outperforms all existing datasets sourced from in-the-wild speech data.

\begin{table}[htbp]
\centering
\caption{Quality comparison between Emilia and nine existing datasets. The scores for LJSpeech, AutoPrepWild, Aishell-3, and LibriTTS are derived from~\cite{AutoPrep}. The score for Libri-Light is computed from its official "small" subset, and the score for WenetSpeech4TTS is computed from its official "basic" subset. The scores for MLS and Emilia are computed from a randomly sampled 600-hour subset.}
\label{tab:dnsmos_results}
\resizebox{0.5\textwidth}{!}{
\begin{tabular}{cc}
\toprule
\textbf{Dataset} & \textbf{DNSMOS P.835 OVRL} \\
\midrule
LJSpeech~\cite{ljspeech} & 3.30 ± 0.17 \\
AutoPrepWild~\cite{AutoPrep} & 3.24 ± 0.21 \\
VCTK \cite{vctk} & 3.20 ± 0.18\\
Aishell-3~\cite{aishell3} & 3.15 ± 0.17 \\
LibriTTS~\cite{libritts} & 3.25 ± 0.19 \\
GigaSpeech~\cite{gigaspeech} & 2.52 ± 0.54 \\
WenetSpeech4TTS~\cite{wenetspeech4tts} & 3.18 ± 0.22 \\
MLS \cite{mls}~& \textbf{3.33 ± 0.19} \\
Libri-Light~\cite{librilight} & 3.25 ± 0.26 \\
\midrule
Emilia & 3.26 ± 0.14 \\
\bottomrule
\end{tabular}
}
\end{table}

\begin{figure}[htbp]
    \centering
    \subfloat[Acoustic diversity]{\includegraphics[width=.44\columnwidth]{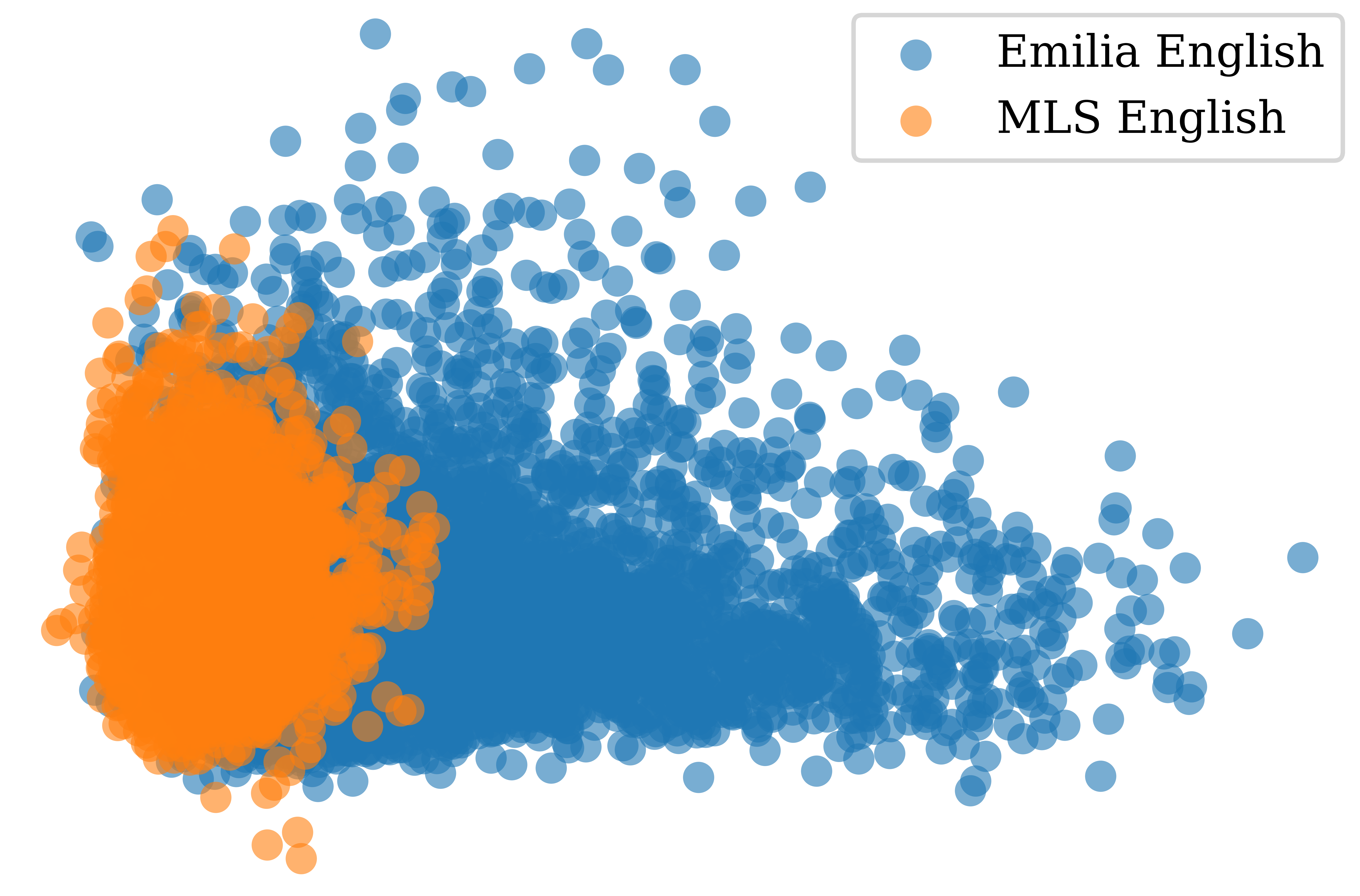}} \hspace{2pt}
    \subfloat[Semantic diversity]{\includegraphics[width=.44\columnwidth]{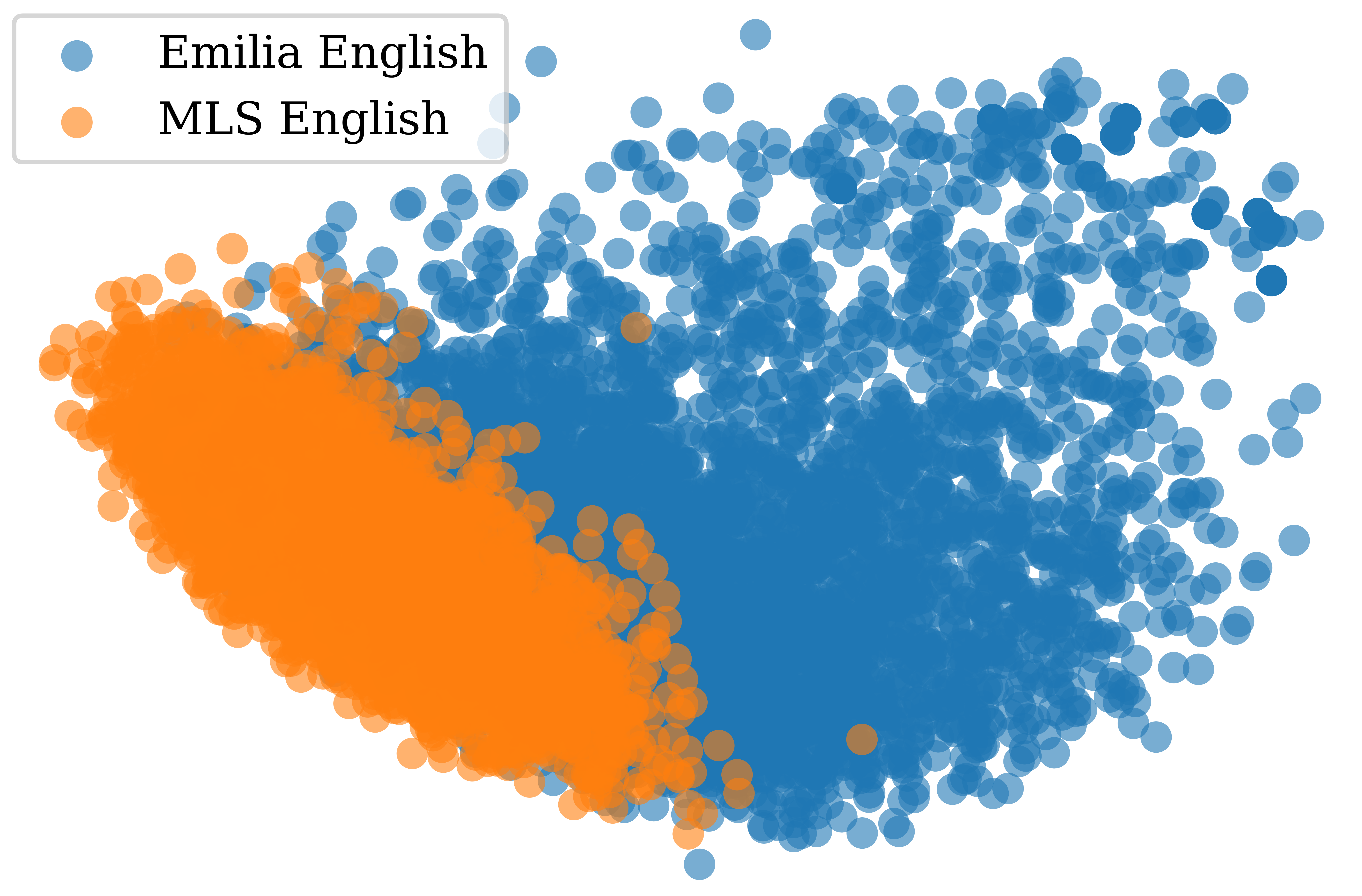}}
    \caption{A comparison of acoustic and semantic diversities between Emilia and MLS datasets.}
    \label{fig:combined}
\end{figure}

As shown in Fig.~\ref{fig:combined} (a), the Emilia dataset exhibits a broader dispersion, contrasting with MLS, which shows a more compact clustering. The more scattered pattern highlights the Emilia dataset as encompassing a richer acoustic characteristic coverage compared to the MLS dataset derived from audiobooks. 
As shown in Fig.~\ref{fig:combined} (b), the scatter of textual features indicates that the Emilia dataset covers a wide range of textual content, validating the significant diversity in Emilia's semantic coverage.

\subsection{Text to Speech}

\subsubsection{Evaluation Metrics} 

For the objective metrics, we evaluate the intelligibility (WER), speaker similarity (S-SIM), accent similarity (A-SIM), emotion similarity (E-SIM), and F0 correlation (FPC)~\cite{svcc-2023,multiple-contents-svc}. Specially, we calculate WER based on Whisper-large-v3~\cite{whisper,seedtts,maskgct}. For the three similarity metrics -- S-SIM, A-SIM, and E-SIM -- we calculate the cosine similarity between the embeddings (of speaker, accent, or emotion) of the generated sample and the reference. Specifically, we extract these embeddings using WavLM TDNN\footnote{\href{https://github.com/microsoft/UniSpeech/tree/main/downstreams/speaker\_verification}{https://github.com/microsoft/UniSpeech/tree/main/downstreams/speaker\_verification}}~\cite{chen2022wavlm,seedtts,maskgct} for speaker, CommonAccent\footnote{\href{https://huggingface.co/Jzuluaga/accent-id-commonaccent\_ecapa}{https://huggingface.co/Jzuluaga/accent-id-commonaccent\_ecapa}}~\cite{common-accent,convertandspeak} for accent, and emotion2vec\footnote{\href{https://github.com/ddlBoJack/emotion2vec}{https://github.com/ddlBoJack/emotion2vec}}~\cite{emotion2vec} for emotion, respectively. We also used CommonAccent and emotion2vec as the classifiers to measure the classification accuracy of accent and emotion (A-ACC and E-ACC). For subjective metrics, we use the Mean Opinion Score (MOS, rated from 1 to 5) to assess naturalness (N-MOS) and similarity in speaker, accent, emotion, and prosody (SS-MOS, AS-MOS, ES-MOS, and PS-MOS). SS MOS, AS-MOS, and ES-MOS evaluate the similarity between the generated sample and the \textit{reference}, while PS-MOS assesses the similarity between the generated sample and the \textit{source}. Additionally, we employ Comparative MOS (CMOS, rated from -3 to 3) to evaluate naturalness (N-CMOS), accentedness (A-CMOS), and emotiveness (E-CMOS).

\subsubsection{MaskGCT and Vevo-TTS}\label{sec:results-maskgct-vevo}

\begin{table}[t]
\caption{Amphion v0.2 results of MaskGCT and Vevo-TTS on zero-shot TTS task.}
\label{tab:maskgct-vevo-tts-results}
\centering
\begin{subtable}{\textwidth}
    % \caption{\note{LibriTTS test-clean and SeedEval-TTS}}
    \label{tab:results-voice-imitation-synthesis-libritts}
    \resizebox{\textwidth}{!}{%
    \begin{tabular}{l|c|c|cc|cc}
    \toprule
    \multicolumn{1}{c|}{\textbf{Model}} & \makecell[c]{\textbf{AR?}} & \makecell[c]{\textbf{Training} \textbf{Data}} & \textbf{WER ($\downarrow$)} & \makecell[c]{\textbf{Speaker}\\ \textbf{SIM ($\uparrow$)}} & \makecell[c]{\textbf{Naturalness}\\ \textbf{CMOS ($\uparrow$)}} & \makecell[c]{\textbf{Speaker}\\ \textbf{S-MOS ($\uparrow$)}} \\
    \midrule \midrule
    \multicolumn{7}{c}{\textbf{LibriTTS \textit{test-clean}}} \\
    \midrule \midrule
    Ground Truth & - & - & 2.845 & 0.763 & 0.00 \textcolor{white}{$_{\scriptscriptstyle \pm \text{0.00}}$} & - \\ \midrule
    VALL-E (Amphion v0.1) & \ding{51} & 45K hours, MLS English~\cite{mls}  & 8.204 & 0.551 & -0.95 $_{\scriptscriptstyle \pm \text{0.39}}$ & 3.27 $_{\scriptscriptstyle \pm \text{0.25}}$ \\
    Voicebox~\cite{voicebox} & \ding{55}  & 60K hours, Audiobook  & \textbf{3.175} & {0.631} &  -0.55 $_{\scriptscriptstyle \pm \text{0.15}}$ & {3.49} $_{\scriptscriptstyle \pm \text{0.14}}$  \\
    VoiceCraft~\cite{voicecraft} & \ding{51} & 10K hours, Gigaspeech~\cite{gigaspeech} & 4.737  & {0.570} & {-0.41} $_{\scriptscriptstyle \pm \text{0.18}}$ & 3.41 $_{\scriptscriptstyle \pm \text{0.13}}$  \\
    CosyVoice~\cite{cosyvoice} & \ding{51} & 171k hours, In-the-wild  & \underline{3.647} & \underline{0.727} & -0.44 $_{\scriptscriptstyle \pm \text{0.16}}$ & \underline{4.17} $_{\scriptscriptstyle \pm \text{0.15}}$ \\
    Vevo-TTS~\cite{vevo} & \ding{51} & 60K hours, Audiobook & {3.672} & {0.593} & \underline{-0.31} $_{\scriptscriptstyle \pm \text{0.14}}$ & {3.58} $_{\scriptscriptstyle \pm \text{0.15}}$ \\ \midrule
    \textbf{Vevo-TTS} (Amphion v0.2) & \ding{51} & Emilia~\cite{emilia} & 5.791 & 0.699 & - & - \\
    \textbf{MaskGCT} (Amphion v0.2) & \ding{55} & Emilia~\cite{emilia}  & 3.841  & \textbf{0.781} & \textbf{-0.21} $_{\scriptscriptstyle \pm \text{0.08}}$ & \textbf{4.30} $_{\scriptscriptstyle \pm \text{0.22}}$ \\
    \midrule \midrule
    \multicolumn{7}{c}{\textbf{SeedTTS \textit{test-en}}} \\
    \midrule \midrule
    Ground Truth & - & - & 1.426 & 0.723 & 0.00 \textcolor{white}{$_{\scriptscriptstyle \pm \text{0.00}}$} & - \\ \midrule
    VALL-E (Amphion v0.1) & \ding{51} & 45K hours, MLS English~\cite{mls}  & 6.129 & 0.433 & -1.02 $_{\scriptscriptstyle \pm \text{0.36}}$ & 2.68 $_{\scriptscriptstyle \pm \text{0.52}}$  \\
    Voicebox~\cite{voicebox} & \ding{55}  & 60K hours, Audiobook  & \textbf{2.129} & 0.500 & -0.12 $_{\scriptscriptstyle \pm \text{0.19}}$ & 2.91 $_{\scriptscriptstyle \pm \text{0.08}}$   \\
    VoiceCraft~\cite{voicecraft} & \ding{51} & 10K hours, Gigaspeech~\cite{gigaspeech} & 6.353 & {0.446} & {-0.10} $_{\scriptscriptstyle \pm \text{0.25}}$ & {3.02} $_{\scriptscriptstyle \pm \text{0.21}}$ \\
    CosyVoice~\cite{cosyvoice} & \ding{51} & 171k hours, In-the-wild  & 3.500 & \underline{0.627} & \textbf{0.11} $_{\scriptscriptstyle \pm \text{0.19}}$ & \underline{3.88} $_{\scriptscriptstyle \pm \text{0.12}}$  \\
    Vevo-TTS~\cite{vevo} & \ding{51} & 60K hours, Audiobook & {2.687} & {0.513} & {-0.11} $_{\scriptscriptstyle \pm \text{0.19}}$ & {3.83} $_{\scriptscriptstyle \pm \text{0.18}}$ \\ \midrule
    \textbf{Vevo-TTS} (Amphion v0.2) & \ding{51} & Emilia~\cite{emilia} & 7.328 & 0.598 & - & - \\
    \textbf{MaskGCT} (Amphion v0.2) & \ding{55} & Emilia~\cite{emilia}  & \underline{2.573}  & \textbf{0.688} & \underline{0.08} $_{\scriptscriptstyle \pm \text{0.25}}$ & \textbf{4.33} $_{\scriptscriptstyle \pm \text{0.14}}$ \\
    % \textbf{DualCodec-VALL-E 25Hz} (Amphion v0.2) & \ding{51} & Emilia~\cite{emilia} & \underline{2.49} & 0.57 & - & - \\
    % \textbf{DualCodec-VALL-E 12.5Hz} (Amphion v0.2) & \ding{51} & Emilia~\cite{emilia} & 4.90 & 0.54 & - & - \\
    % \textbf{DualCodec-MaskGCT 12.5Hz} (Amphion v0.2) & \ding{51} & Emilia~\cite{emilia} & 5.07 & 0.59 & - & - \\
    \end{tabular}%
    }    
\end{subtable}
\hfill
\begin{subtable}{\textwidth}
    % \caption{\note{Accented and Emotional Corpus}}
    \label{tab:results-voice-imitation-synthesis-accent}
    \resizebox{\textwidth}{!}{%
    \begin{threeparttable}
            \begin{tabular}{l|c|ccc|ccc}
    \toprule \midrule
    \multicolumn{8}{c}{\textbf{L2Arctic \textit{(Accented corpus)}}} \\
    \midrule \midrule
    \multicolumn{1}{c|}{\textbf{Model}} & \makecell[c]{\textbf{AR?}} & \textbf{WER ($\downarrow$)} & \makecell[c]{\textbf{Speaker}\\ \textbf{SIM ($\uparrow$)}} & \makecell[c]{\textbf{Accent}\\ \textbf{SIM ($\uparrow$)}} & \makecell[c]{\textbf{Naturalness}\\ \textbf{CMOS ($\uparrow$)}} & \makecell[c]{\textbf{Speaker}\\ \textbf{S-MOS ($\uparrow$)}} & \makecell[c]{\textbf{Accent}\\ \textbf{S-MOS ($\uparrow$)}} \\
    \midrule
    Ground Truth & - & 10.903 & 0.747 & 0.633 & 0.00 \textcolor{white}{$_{\scriptscriptstyle \pm \text{0.00}}$} & - & - \\ \midrule
    VALL-E (Amphion v0.1) & \ding{51}  & 10.721 & 0.403 & 0.485 & -1.04 $_{\scriptscriptstyle \pm \text{0.50}}$ & 3.12 $_{\scriptscriptstyle \pm \text{0.41}}$ & 2.77 $_{\scriptscriptstyle \pm \text{0.45}}$ \\
    Voicebox~\cite{voicebox} & \ding{55}  & \textbf{6.181} & 0.475 & {0.575} & {-0.55} $_{\scriptscriptstyle \pm \text{0.22}}$ & {3.93} $_{\scriptscriptstyle \pm \text{0.25}}$ & {3.49} $_{\scriptscriptstyle \pm \text{0.29}}$ \\
    VoiceCraft~\cite{voicecraft} & \ding{51} & 10.072 & 0.438 & 0.517 & -0.39 $_{\scriptscriptstyle \pm \text{0.22}}$ & 3.51 $_{\scriptscriptstyle \pm \text{0.33}}$ & 3.29 $_{\scriptscriptstyle \pm \text{0.28}}$ \\
    CosyVoice~\cite{cosyvoice} & \ding{51} & 6.660 & {0.653} & \underline{0.640} & 0.10 $_{\scriptscriptstyle \pm \text{0.19}}$ & \underline{4.23} $_{\scriptscriptstyle \pm \text{0.18}}$ & 3.99 $_{\scriptscriptstyle \pm \text{0.23}}$ \\
    Vevo-TTS~\cite{vevo} & \ding{51} & {9.673} & {0.544} & {0.579} & \underline{0.12} $_{\scriptscriptstyle \pm \text{0.20}}$ & {4.11} $_{\scriptscriptstyle \pm \text{0.20}}$ & \underline{4.12} $_{\scriptscriptstyle \pm \text{0.21}}$ \\
    \midrule
    \textbf{Vevo-TTS} (Amphion v0.2) & \ding{51} & 9.884 & \underline{0.660} & 0.639 & - & - & - \\
    \textbf{MaskGCT} (Amphion v0.2) & \ding{55} & \underline{6.382} & \textbf{0.717} & \textbf{0.645} & \textbf{0.23} $_{\scriptscriptstyle \pm \text{0.20}}$ & \textbf{4.24} $_{\scriptscriptstyle \pm \text{0.16}}$ & \textbf{4.38} $_{\scriptscriptstyle \pm \text{0.14}}$ \\
    \midrule \midrule
    \multicolumn{8}{c}{\textbf{ESD \textit{(Emotional corpus)}}} \\
    \midrule \midrule
    \multicolumn{1}{c|}{\textbf{Model}} & \makecell[c]{\textbf{AR?}} & \textbf{WER ($\downarrow$)} & \makecell[c]{\textbf{Speaker}\\ \textbf{SIM ($\uparrow$)}} & \makecell[c]{\textbf{Emotion}\\ \textbf{SIM ($\uparrow$)}} & \makecell[c]{\textbf{Naturalness}\\ \textbf{CMOS ($\uparrow$)}} & \makecell[c]{\textbf{Speaker}\\ \textbf{S-MOS ($\uparrow$)}} & \makecell[c]{\textbf{Emotion}\\ \textbf{S-MOS ($\uparrow$)}} \\
    \midrule
    Ground Truth & - & 11.792 & 0.673 & 0.936 & 0.00 \textcolor{white}{$_{\scriptscriptstyle \pm \text{0.00}}$} & - & - \\
    \midrule
    VALL-E (Amphion v0.1) & \ding{51}  & 15.731 & 0.396 & 0.735 & -1.43 $_{\scriptscriptstyle \pm \text{0.33}}$ & 2.52 $_{\scriptscriptstyle \pm \text{0.38}}$ & 2.63 $_{\scriptscriptstyle \pm \text{0.36}}$ \\
    Voicebox~\cite{voicebox} & \ding{55}  & {12.647} & {0.451} & {0.811} & -0.65 $_{\scriptscriptstyle \pm \text{0.20}}$ & {3.81} $_{\scriptscriptstyle \pm \text{0.16}}$ & {3.61} $_{\scriptscriptstyle \pm \text{0.19}}$ \\
    VoiceCraft~\cite{voicecraft} & \ding{51} & 16.042 & 0.345 & 0.788 & {-0.60} $_{\scriptscriptstyle \pm \text{0.24}}$ & 3.42 $_{\scriptscriptstyle \pm \text{0.31}}$ & 3.52 $_{\scriptscriptstyle \pm \text{0.25}}$ \\
    CosyVoice~\cite{cosyvoice} & \ding{51} & \textbf{10.139} & \underline{0.575} & \underline{0.839} & -0.45 $_{\scriptscriptstyle \pm \text{0.18}}$ & 3.98 $_{\scriptscriptstyle \pm \text{0.19}}$ & 3.66 $_{\scriptscriptstyle \pm \text{0.19}}$ \\
    Vevo-TTS~\cite{vevo} & \ding{51} & {14.458} & {0.466} & \textbf{0.840} & \underline{-0.39} $_{\scriptscriptstyle \pm \text{0.15}}$ & \underline{3.99} $_{\scriptscriptstyle \pm \text{0.22}}$ & \textbf{4.03} $_{\scriptscriptstyle \pm \text{0.19}}$  \\ \midrule
    \textbf{Vevo-TTS} (Amphion v0.2) & \ding{51} & 17.350 & 0.494 & 0.823 & - & - & - \\
    \textbf{MaskGCT} (Amphion v0.2) & \ding{55} & \underline{12.502} & \textbf{0.600} & 0.822 & \textbf{-0.31} $_{\scriptscriptstyle \pm \text{0.17}}$ & \textbf{4.07} $_{\scriptscriptstyle \pm \text{0.16}}$ & \underline{3.76} $_{\scriptscriptstyle \pm \text{0.25}}$ \\
    \bottomrule
    \end{tabular}%
    \begin{tablenotes}
        \footnotesize{
        \item[*] The best and the second best results are shown in \textbf{bold} and by \underline{underlined}.
        }
    \end{tablenotes}
    \end{threeparttable}
    }    
\end{subtable}
\end{table}

We consider various evaluation settings to construct the evaluation set: (1) For clean data, such as recordings made in studio environments, we select audiobook speech data. Specifically, we adopt LibriTTS test-clean~\cite{libritts} as the evaluation samples. (2) For noisy data, which may include in-the-wild recordings and diverse recording devices, we use the Common Voice English dataset (CV)~\cite{common-voice}. Specifically, we adopt the samples of SeedTTS test-en~\cite{seedtts} to conduct the evaluation. (3) Additionally, to introduce more stylized and expressive data, we use L2Arctic~\cite{l2arctic} as the accented corpus and use ESD~\cite{esd} as the emotional corpus. The experimental results of MaskGCT and Vevo-TTS in Amphion v0.1 can be seen in Table~\ref{tab:maskgct-vevo-tts-results}.

\subsection{Speech Enhancement}

\subsubsection{Evaluation Metrics}

Traditional signal-based metrics like PESQ, STOI, and SDR are inadequate for evaluating generative-model-based enhancement due to the lack of waveform alignment~\cite{li2024masksr,wang2024selm,tang2024tselm}. Instead, we use DNSMOS~\cite{dnsmos835}, a reference-free perceptual quality estimator for 16 kHz audio, which outputs SIG, BAK, and OVRL scores (1–5); NISQA~\cite{nisqa2021}, a reference-free quality estimator for 48 kHz audio that provides a single overall score (1–5); SpeechBERTScore~\cite{saeki2024speechbertscore}, which evaluates semantic similarity using HuBERT-base\footnote{\url{https://huggingface.co/facebook/hubert-base-ls960}} features; and speaker cosine similarity calculated with embeddings extracted by WavLM\footnote{\url{https://huggingface.co/microsoft/wavlm-base-plus-sv}}.

\begin{table*}[!ht]
    \centering
    \caption{Evaluation on various enhancement tasks, *indicates the evaluation dataset is at 16kHz.}
    \label{tab:result-main}
    \vspace{-5pt}
    \resizebox{\textwidth}{!}{
    \begin{tabular}{lllcccccc}
        \toprule
        \textbf{Group} & \textbf{Dataset} & \textbf{Model} & \textbf{SIG} & \textbf{BAK} & \textbf{OVRL} & \textbf{NISQA} & \textbf{SpeechBERTScore} & \textbf{Similarity} \\
        \midrule
        \multirow{20}{*}{GSR} &
        \multirow{6}{*}{Voicefixer GSR}
        & TFGridNet & 3.253 & 3.906 & 2.945 & 3.643 & 0.782 & 0.613 \\
        & & NSNet2 & 3.011 & 3.969 & 2.758 & 3.433 & 0.728 & 0.615 \\
        & & Voicefixer (checkpoint) & 3.299 & 3.971 & 3.003 & 4.16 & 0.797 & 0.882 \\
        & & Voicefixer (retrained) & 3.3 & 3.984 & 2.996 & 4.054 & 0.818 & 0.884 \\
        & & MaskSR & \textbf{3.408} & 4.041 & 3.122 & \textbf{4.335} & \textbf{0.832} & 0.916 \\
        & & AnyEnhance (Amphion) & 3.406 & \textbf{4.073} & \textbf{3.136} & 4.308 & 0.829 & \textbf{0.924} \\
        \cmidrule{2-9}
        & \multirow{6}{*}{Librivox GSR}
        & TF-GridNet & 3.274 & 3.872 & 2.951 & 3.138 & 0.77 & 0.931 \\
        & & NSNet2 & 2.895 & 3.866 & 2.589 & 2.735 & 0.7 & 0.892 \\
        & & Voicefixer (checkpoint) & 3.365 & 3.971 & 3.085 & 3.77 & 0.706 & 0.864 \\
        & & Voicefixer (retrained) & 3.35 & 4.024 & 3.069 & 3.63 & 0.758 & 0.897 \\
        & & MaskSR & 3.499 & 4.133 & 3.258 & 4.155 & 0.779 & 0.94 \\
        & & AnyEnhance (Amphion)& \textbf{3.546} & \textbf{4.142} & \textbf{3.308} & \textbf{4.346} & \textbf{0.822} & \textbf{0.955} \\
        \cmidrule{2-9}
        & \multirow{6}{*}{CCMusic GSR}
        & TF-GridNet & 2.764 & 3.37 & 2.362 & 2.396 & 0.57 & 0.81 \\
        & & NSNet2 & 2.608 & 3.645 & 2.226 & 2.439 & 0.574 & 0.798 \\
        & & Voicefixer (checkpoint) & 2.75 & 3.094 & 2.354 & 2.917 & 0.636 & 0.823 \\
        & & Voicefixer (retrained) & 2.948 & 3.349 & 2.551 & 3.224 & 0.738 & 0.881 \\
        & & MaskSR & 3.153 & 3.483 & 2.715 & 3.157 & 0.77 & 0.889 \\
        & & AnyEnhance (Amphion)& \textbf{3.243} & \textbf{3.547} & \textbf{2.797} & \textbf{3.345} & \textbf{0.811} & \textbf{0.915} \\
        \midrule
        \multirow{23}{*}{SE} &
        \multirow{11}{*}{DNS No Reverb$^{*}$}
        & DEMUCS & 3.533 & 4.157 & 3.31 & 3.742 & 0.877 & 0.984 \\
        & & FRCRN & 3.574 & 4.154 & 3.332 & 4.495 & \textbf{0.914} & \textbf{0.993} \\
        & & SGMSE & 3.501 & 3.710 & 3.137 & - & - & - \\
        & & StoRM & 3.514 & 3.941 & 3.205 & - & - & - \\
        & & SELM & 3.508 & 4.096 & 3.258 & - & - & - \\
        & & TFGridNet & 3.539 & 4.047 & 3.268 & 4.347 & 0.902 & 0.675 \\
        & & NSNet2 & 3.391 & 4.071 & 3.132 & 3.943 & 0.847 & 0.675 \\
        & & Voicefixer (checkpoint) & 3.504 & 4.109 & 3.253 & 4.274 & 0.819 & 0.956 \\
        & & Voicefixer (retrained) & 3.452 & 4.035 & 3.157 & 4.105 & 0.827 & 0.959 \\
        & & MaskSR & 3.616 & \textbf{4.183} & 3.393 & 4.754 & 0.875 & 0.983 \\
        & & AnyEnhance (Amphion)& \textbf{3.64} & 4.179 & \textbf{3.418} & \textbf{4.821} & 0.907 & 0.988 \\
        \cmidrule{2-9}
        & \multirow{11}{*}{DNS With Reverb$^{*}$}
        & DEMUCS & 2.937 & 3.844 & 2.615 & 2.188 & 0.725 & 0.93 \\
        & & FRCRN & 2.933 & 2.923 & 2.279 & 2.27 & 0.783 & \textbf{0.966} \\
        & & SGMSE & 2.730 & 2.741 & 2.430 & - & - & - \\
        & & StoRM & 2.947 & 3.141 & 2.516 & - & - & - \\
        & & SELM & 3.160 & 3.577 & 2.695 & - & - & - \\
        & & TFGridNet & 3.11 & 3.225 & 2.51 & 2.614 & \textbf{0.84} & 0.686 \\
        & & NSNet2 & 2.756 & 3.719 & 2.421 & 2.043 & 0.763 & 0.691 \\
        & & Voicefixer (checkpoint) & 3.43 & 4.016 & 3.132 & 3.822 & 0.711 & 0.91 \\
        & & Voicefixer (retrained) & 3.074 & 3.721 & 2.667 & 2.906 & 0.724 & 0.918 \\
        & & MaskSR & 3.396 & \textbf{4.043} & 3.085 & 3.353 & 0.701 & 0.946 \\
        & & AnyEnhance (Amphion)& \textbf{3.5} & 4.04 & \textbf{3.204} & \textbf{3.722} & 0.738 & 0.951 \\
        \midrule
        \multirow{11}{*}{SR} &
        \multirow{5}{*}{Voicefixer SR}
        & Voicefixer (checkpoint) & 3.405 & 4.029 & 3.11 & 4.131 & 0.873 & 0.882 \\
        & & Voicefixer (retrained) & 3.041 & 3.903 & 2.745 & 3.556 & 0.837 & 0.854 \\
        & & AudioSR & \textbf{3.492} & 4.002 & 3.18 & 4.255 & 0.913 & 0.911 \\
        & & MaskSR & 3.464 & 4.028 & 3.154 & \textbf{4.352} & 0.925 & 0.939 \\
        & & AnyEnhance (Amphion)& 3.449 & \textbf{4.063} & \textbf{3.156} & 4.201 & \textbf{0.941} & \textbf{0.943} \\
        \cmidrule{2-9}
        & \multirow{5}{*}{CCMusic SR}
        & Voicefixer (checkpoint) & 3.179 & 3.534 & 2.743 & \textbf{3.356} & 0.463 & 0.65 \\
        & & Voicefixer (retrained) & 3.108 & 3.504 & 2.692 & 3.218 & 0.768 & 0.864 \\
        & & AudioSR & 3.192 & 3.531 & 2.75 & 2.836 & 0.468 & 0.63 \\
        & & MaskSR & 3.308 & 3.588 & 2.857 & 3.173 & 0.813 & 0.892 \\
        & & AnyEnhance (Amphion)& \textbf{3.339} & \textbf{3.628} & \textbf{2.899} & 3.225 & \textbf{0.854} & \textbf{0.919} \\
        \midrule
        \multirow{7}{*}{TSE} &
        \multirow{3}{*}{Librimix TSE$^{*}$}
        & WeSep & 3.563 & 3.931 & 3.228 & 4.041 & \textbf{0.922} & \textbf{0.991} \\
        & & TSELM & 3.55 & \textbf{4.084} & 3.288 & 4.029 & 0.808 & 0.908 \\
        & & AnyEnhance (Amphion)& \textbf{3.638} & 4.066 & \textbf{3.353} & \textbf{4.277} & 0.735 & 0.914 \\
        \cmidrule{2-9}
        & \multirow{3}{*}{VCTK Noisy TSE}
        & WeSep & 2.483 & 2.191 & 1.933 & 1.959 & 0.568 & 0.856 \\
        & & TSELM & 3.345 & 3.875 & 3.004 & 3.388 & 0.58 & 0.81 \\
        & & AnyEnhance (Amphion)& \textbf{3.545} & \textbf{4.102} & \textbf{3.275} & \textbf{4.57} & \textbf{0.727} & \textbf{0.925} \\
        \bottomrule
    \end{tabular}
    }
\end{table*}
\subsubsection{Evaluation Datasets \& Baseline Models}

\textbf{Evaluation Datasets} We evaluate the performance of AnyEnhance across various speech and singing enhancement tasks using the following datasets: For the General Speech Restoration (GSR) task, we use the official Voicefixer GSR dataset~\cite{liu2022voicefixer} and a simulated GSR testset on Librivox~\cite{kearns2014librivox} and CCMusic acapella~\cite{ccmusic}, with inputs featuring simultaneous distortions such as noise, reverberation, clipping, and bandwidth limitation. For the Speech Enhancement (SE) task, we use the 2020 DNS Challenge dataset~\cite{reddy2020interspeech}, which includes 16 kHz synthetic noisy mixtures with and without reverberation. For the Super Resolution (SR) task, we use the official Voicefixer SR subset and a simulated SR testset on the CCMusic acapella dataset, with input bandwidths ranging from 2 kHz to 8 kHz. Finally, for the target speaker extraction (TSE) task, we evaluate on the official Librimix~\cite{cosentino2020librimix} testset, which contains no noise, and a simulated VCTK~\cite{vctk} testset, which includes noise and reverberation.

\textbf{Baseline Models} We compare the performance of AnyEnhance with several state-of-the-art models on the GSR, SE, SR, and TSE tasks. For GSR, we use TF-GridNet~\cite{wang2023tf}, NSNet2~\cite{braun2020data}, Voicefixer~\cite{liu2022voicefixer}, and MaskSR~\cite{li2024masksr}, with official checkpoints or retrained models as needed. For SE, we evaluate the same GSR baselines along with DEMUCS~\cite{defossez2019demucs}, FRCRN~\cite{zhao2022frcrn}, SGMSE~\cite{richter2023sgmse}, StoRM~\cite{lemercier2023storm}, and SELM~\cite{wang2024selm}. For SR, we compare AnyEnhance with Voicefixer, MaskSR, and AudioSR~\cite{liu2024audiosr}. For TSE, we use WeSep~\cite{wang2024wesep} and TSELM~\cite{tang2024tselm}.

\subsubsection{Evaluation Results}

As shown in Table~\ref{tab:result-main}, AnyEnhance demonstrates superior performance across various enhancement tasks compared to baseline models. In the general speech restoration (GSR) task, AnyEnhance consistently outperforms other models on all datasets, achieving the highest SIG, OVRL, and NISQA scores, as well as the best SpeechBERTScore and Similarity metrics. For example, on the Librivox GSR dataset, AnyEnhance achieves a SIG score of 3.546, an OVRL score of 3.308, and a NISQA score of 4.346, all of which surpass the competing models. Similarly, in the DNS No Reverb SE task, AnyEnhance attains the highest SIG (3.64), OVRL (3.418), and NISQA (4.821) scores, while maintaining competitive SpeechBERTScore and Similarity. In the super-resolution (SR) task, AnyEnhance achieves strong results, particularly on the CCMusic SR dataset, where it outperforms other methods in all metrics, including a SIG score of 3.339 and a Similarity score of 0.919. Additionally, in the target speaker extraction (TSE) task, AnyEnhance delivers state-of-the-art results on both Librimix TSE and VCTK Noisy TSE datasets, achieving the highest scores across SIG, OVRL, NISQA, and similarity metrics, particularly excelling in challenging noisy scenarios. Overall, AnyEnhance exhibits robust generalization and effectiveness across diverse enhancement tasks, validating its design and methodological innovations. More evaluation details and experimental results can be found in ~\cite{zhang2025anyenhance}.

\subsection{Voice Conversion}

We use the same evaluation metrics to Section~\ref{sec:results-maskgct-vevo}. For evaluation samples, in addition to SeedTTS test-en, L2Arctic, and ESD that is introduced in Section~\ref{sec:results-maskgct-vevo}, we also use 200 samples of VCTK~\cite{vctk} as the clean corpus. The results of FACodec, Vevo-Timbre, and Vevo-Voice in Amphion v0.2 are shown in Table~\ref{tab:results-vc}. Comparing Vevo-Timbre and Vevo-Voice, we observe that Vevo-Timbre's strength lies in preserving the style of the source (FPC, PS-MOS), whereas Vevo-Voice additionally excels in style imitation, resulting in higher speaker similarity (S-SIM, SS-MOS). However, due to the autoregressive design in content-style modeling stage, Vevo-Voice scores lower in intelligibility (WER) compared to Vevo-Timbre.

\begin{table}[t]
\caption{Amphion v0.2 results of FACodec, Vevo-Timbre, and Vevo-Voice. (ContRep/Model: Hours of training data for the used content representations and the model)
% \textbf{S/A/E-SIM}: Speaker/Accent/Emotion SIM. \textbf{N-MOS}: Naturalness MOS. \textbf{SS/PS/AS/ES-MOS}: Speaker/Prosody/Accent/Emotion Similarity MOS.)
}
% \vspace{-3mm}
\label{tab:results-vc}
\centering
\begin{subtable}{\textwidth}
    \resizebox{\textwidth}{!}{%
    \begin{tabular}{l|c|c|ccc|cccc}
    \toprule
    \midrule
    \multicolumn{9}{c}{\textbf{VCTK and SeedTTS test-en \textit{(General Corpus)}}}  \\
    \midrule\midrule
    \multicolumn{1}{c|}{\textbf{Model}} & \makecell[c]{\textbf{AR?}} & \makecell[c]{\textbf{Training Data} \\ \small{(ContRep / Model)} } & \makecell[c]{\textbf{WER}\\ ($\downarrow$)}  & \makecell[c]{\textbf{S-SIM} \\ ($\uparrow$)} & \makecell[c]{\textbf{FPC} \\  ($\uparrow$)} & \makecell[c]{\textbf{N-MOS}\\ \textbf{($\uparrow$)}} & \makecell[c]{\textbf{SS-MOS} \\ ($\uparrow$)} 
    & \makecell[c]{\textbf{PS-MOS} \\ ($\uparrow$)} 
    \\
    \midrule
    HierSpeech++~\cite{hierspeech++} & \ding{55} & 500K / 2.8K & 4.233 & 0.385 & \textbf{0.634} & 3.05 $_{\scriptscriptstyle \pm \text{0.23}}$ & 3.24 $_{\scriptscriptstyle \pm \text{0.25}}$  & 3.08 $_{\scriptscriptstyle \pm \text{0.26}}$  \\
    LM-VC~\cite{lmvc} & \ding{51} & 1K / 60K & 8.623 & 0.310 & 0.524 & 2.90 $_{\scriptscriptstyle \pm \text{0.11}}$ & 2.98 $_{\scriptscriptstyle \pm \text{0.18}}$ & 2.16 $_{\scriptscriptstyle \pm \text{0.26}}$   \\
    UniAudio~\cite{uniaudio} & \ding{51} & 1K / 100K & 7.241 & 0.264 & 0.575 & 3.04 $_{\scriptscriptstyle \pm \text{0.15}}$ & 2.47 $_{\scriptscriptstyle \pm \text{0.20}}$ & 2.51 $_{\scriptscriptstyle \pm \text{0.25}}$ \\
    CosyVoice-Timbre~\cite{cosyvoice} & \ding{55} & 400K / 171K  & 3.828 & 0.469 & 0.608 & - & - & - \\
    \text{Vevo-Timbre}~\cite{vevo} & \ding{55} & 60K / 60K & \textbf{2.968} & {0.420} & {0.536} & \textbf{3.35} $_{\scriptscriptstyle \pm \text{0.09}}$ & \underline{3.36} $_{\scriptscriptstyle \pm \text{0.16}}$ & \textbf{3.45} $_{\scriptscriptstyle \pm \text{0.17}}$  \\
    Vevo-Voice~\cite{vevo} & \ding{51} & 60K / 60K & 7.694 & {0.458} & 0.484 & \underline{3.09} $_{\scriptscriptstyle \pm \text{0.13}}$ & \textbf{3.51} $_{\scriptscriptstyle \pm \text{0.24}}$ & 2.60 $_{\scriptscriptstyle \pm \text{0.23}}$  \\
    \midrule
    \textbf{FACodec} (Amphion v0.2) & \ding{55} & 60K / 60K & {3.682} & 0.327 & \underline{0.611} & 2.50 $_{\scriptscriptstyle \pm \text{0.20}}$ & 3.10 $_{\scriptscriptstyle \pm \text{0.24}}$ &  \underline{3.10} $_{\scriptscriptstyle \pm \text{0.23}}$  \\
    \textbf{Vevo-Timbre} (Amphion v0.2) & \ding{55} & 101K / 101K & \underline{3.531} & \underline{0.481} & 0.585 & - & - & -  \\
    \textbf{Vevo-Voice} (Amphion v0.2) & \ding{51} & 101K / 101K & 7.855 & \textbf{0.543} & 0.546 & - & - & - \\
    % \midrule
    \end{tabular}%
    }
\end{subtable}
\hfill
\begin{subtable}{\textwidth}
    \resizebox{\textwidth}{!}{%
    \begin{threeparttable}
        \begin{tabular}{l|cccc|cccccccc}
        \midrule \midrule
        \multicolumn{9}{c}{\textbf{L2Arctic and ESD \textit{(Expressive corpus)}}} \\
        \midrule\midrule
        \multicolumn{1}{c|}{\textbf{Model}} & \makecell[c]{\textbf{WER}\\ ($\downarrow$)}  & \makecell[c]{\textbf{S-SIM} \\ ($\uparrow$)} & \makecell[c]{\textbf{A-SIM} \\  ($\uparrow$)} & \makecell[c]{\textbf{E-SIM} \\  ($\uparrow$)} & \makecell[c]{\textbf{N-MOS}\\ \textbf{($\uparrow$)}} & \makecell[c]{\textbf{SS-MOS} \\ ($\uparrow$)} & \makecell[c]{\textbf{AS-MOS} \\ ($\uparrow$)}
        & \makecell[c]{\textbf{ES-MOS} \\ ($\uparrow$)} 
        \\
        \midrule
        Ground Truth & 10.917 & 0.762 & 0.763 & 0.965 & - & - & - & - \\
        \midrule
        HierSpeech++~\cite{hierspeech++} & 12.921 & 0.466 & 0.526 & 0.658 & 3.04 $_{\scriptscriptstyle \pm \text{0.14}}$ & 3.15 $_{\scriptscriptstyle \pm \text{0.23}}$ & 3.13 $_{\scriptscriptstyle \pm \text{0.22}}$ & 2.55 $_{\scriptscriptstyle \pm \text{0.19}}$   \\
        LM-VC~\cite{lmvc} &  20.353 & 0.312 & 0.426 & 0.649 & 2.40 $_{\scriptscriptstyle \pm \text{0.10}}$ & 2.56 $_{\scriptscriptstyle \pm \text{0.15}}$ & 3.02 $_{\scriptscriptstyle \pm \text{0.19}}$ & 2.46 $_{\scriptscriptstyle \pm \text{0.17}}$ \\
        UniAudio~\cite{uniaudio} & 15.751 & 0.311 & 0.486 & 0.611 & 2.95 $_{\scriptscriptstyle \pm \text{0.11}}$ & 2.39 $_{\scriptscriptstyle \pm \text{0.17}}$ & 2.42 $_{\scriptscriptstyle \pm \text{0.15}}$ & 2.41 $_{\scriptscriptstyle \pm \text{0.26}}$ \\
        CosyVoice-Timbre~\cite{cosyvoice} & 12.449 & 0.601 & 0.602 & 0.711 & - & - & - & - \\
        Vevo-Timbre~\cite{vevo} & \underline{12.351} & {0.486} & {0.567} & {0.612} & \textbf{3.43} $_{\scriptscriptstyle \pm \text{0.09}}$ & \underline{3.46} $_{\scriptscriptstyle \pm \text{0.15}}$ & \underline{3.55} $_{\scriptscriptstyle \pm \text{0.25}}$ & \underline{2.66} $_{\scriptscriptstyle \pm \text{0.26}}$  \\
        \text{Vevo-Voice}~\cite{vevo} & 15.214 & {0.517} & \underline{0.614} & \textbf{0.753} & \underline{3.24} $_{\scriptscriptstyle \pm \text{0.11}}$ & \textbf{3.70} $_{\scriptscriptstyle \pm \text{0.24}}$ & \textbf{3.90} $_{\scriptscriptstyle \pm \text{0.19}}$ & \textbf{3.20} $_{\scriptscriptstyle \pm \text{0.16}}$  \\ \midrule
        \textbf{FACodec} (Amphion v0.2) & \underline{12.731} & 0.434 & 0.514 & 0.688 & 2.36 $_{\scriptscriptstyle \pm \text{0.18}}$ & 3.19 $_{\scriptscriptstyle \pm \text{0.22}}$ & 3.01 $_{\scriptscriptstyle \pm \text{0.16}}$ & 2.30 $_{\scriptscriptstyle \pm \text{0.22}}$ \\
        \textbf{Vevo-Timbre} (Amphion v0.2) & \textbf{12.192} & \underline{0.609} & \textbf{0.630} & 0.582 & - & - & - & - \\
        \textbf{Vevo-Voice} (Amphion v0.2) & 17.106 & \textbf{0.674} & \underline{0.614} & \underline{0.718} & - & - & - & - \\
        \bottomrule
        \end{tabular}%
        \begin{tablenotes}
        \footnotesize{
            \item[1] We use CosyVoice-Timbre to notate only the flow-matching model of the whoe CosyVoice pipeline, which is a non-AR model~\cite{cosyvoice}.
            \item[2]  {PS-MOS}, {E-SIM}, and {ES-MOS} are evaluated only on ESD. {A-SIM} and {AS-MOS} are evaluated only on ACCENT. The best and the second best result is shown in \textbf{bold} and by \underline{underlined}.
        }
        \end{tablenotes}
    \end{threeparttable}
    }
\end{subtable}
\end{table}

\subsection{Speech LLMs Evaluation using SD-Eval}
In this section, we introduce details about how to evaluate Speech LLMs using SD-Eval.
\subsubsection{Evaluation Metrics}
\paragraph{Objective Evaluation}

We introduce a reference-free metric based on LLMs for evaluating responses.
In particular, we design tailored prompts for each evaluation subset.
These prompts direct the LLM judge to consider (a) the response’s naturalness, coherence, engagement, and groundedness, and (b) whether the response is appropriate and adequately accounts for factors such as the input speech’s emotion, accent, age, or background sound.
The LLM judge then assigns a direct score—for example, 5 on a 1–10 scale—to each response.

For comparison, we also report results from n-gram-based metrics (ROUGE-L \cite{lin-2004-rouge}, BLEU-4 \cite{papineni-etal-2002-bleu}, and METEOR \cite{banerjee-lavie-2005-meteor}), as well as embedding-based metrics (BERTScore \cite{bert-score}\footnote{We use \href{https://huggingface.co/spaces/evaluate-metric/bertscore}{Hugging Face Evaluate} for scoring, with the \textit{roberta-large} model.}).

\paragraph{Subjective Evaluation}
In addition, we conduct a human evaluation on 200 randomly selected utterances drawn from four different subsets, with each subset contributing 50 utterances. Each sample is rated by at least three human evaluators, who are instructed to assess the generated responses.
For each utterance, three distinct responses are collected, corresponding to outputs from Cascade LLM, VS-LLM, and the LLM (Upper Bound), respectively.
We ensure that every valid sample is reviewed by at least three human annotators.
As a result, each subset contains no fewer than 120 valid samples.

\begin{table}
    \centering
    \caption{Main results of six models on four subsets of SD-Eval. \label{table:main_result}}
    \resizebox{\textwidth}{!}{
    \begin{tabular}{lccccccccc}
    \toprule
     \multirow{2}{*}{\textbf{Model}} & \multirow{2}{*}{\textbf{BLEU-4}} & \multirow{2}{*}{\textbf{ROUGE-L}} & \multirow{2}{*}{\textbf{METEOR}} & \multirow{2}{*}{\textbf{BERTScore}} & \multicolumn{4}{c}{\textbf{LLM Judges}}  & \textbf{Human} \\
     \cmidrule(l){6-9}
     & & & & & \textbf{Yi-1.5} & \textbf{Qwen2} & \textbf{Gemma} & \textbf{GPT-4o}  & \textbf{Evaluation} $\dagger$ \\
    \midrule
    \midrule
    \rowcolor{light-gray}
    \multicolumn{10}{c}{\textbf{\textit{test-emo / Emotion}}} \\
    \midrule
    SALMONN \cite{tang2024salmonn} &2.48 &16.57 &18.97 &86.20 & 4.98 & 3.35 & 2.32 &2.61& - \\
    Qwen-Audio \cite{chu2023qwen} & 3.93 & 19.02 & 16.82 & 86.59 & 4.19 & 2.35 & 2.02 & 2.24 &  - \\
    Qwen2-Audio-AA \cite{chu2024qwen2} &3.01& 16.82& 17.51& 86.17& 4.75 & 2.52 & 2.21 & 2.33& - \\
    Qwen2-Audio-VC \cite{chu2024qwen2}&2.21& 14.57& 22.08& 85.41& 5.88 & 3.83 & 2.93 & 3.25& - \\
      Cascade LLM & 4.66 & 21.98 & 21.70 & 87.93 & 5.67 & 3.86 & 2.35 & 4.47 & 5.05 \\
      VS-LLM & 8.29 & 25.52 & 27.23 & 89.48 & 6.40 & 4.56 & 4.03 & 5.30 & 6.31 \\
      LLM (Upper Bound) & \textbf{12.35} & \textbf{26.08} & \textbf{28.27} & \textbf{89.77} & \textbf{7.03} & \textbf{5.82} & \textbf{6.46} & \textbf{6.74} & \textbf{7.29} \\
      \midrule
      \midrule
      \rowcolor{light-gray}
    \multicolumn{10}{c}{\textbf{\textit{test-acc / Accent}}} \\
    \midrule
    SALMONN\cite{tang2024salmonn} &7.50&22.22&21.23&87.53&5.27 & 6.16 & 3.16&2.93& - \\
    Qwen-Audio \cite{chu2023qwen}  & 4.52 & 17.15 & 17.78 & 85.59 &3.48 &3.45 & 1.86 & 1.72 & - \\
    Qwen2-Audio-AA \cite{chu2024qwen2}&7.26&21.80&19.68&87.68&5.04 & 6.13 & 3.01&2.54& - \\
    Qwen2-Audio-VC \cite{chu2024qwen2}&3.47&17.46&23.77&86.26&5.96 & 6.20 & 3.94&4.37& - \\
      Cascade LLM & 14.51 & 30.53 & 34.13 & 89.66 & 7.23 & 7.32 & 5.65 & 6.62 & 6.71 \\
      VS-LLM & 17.98 & 33.06& 37.65 & 90.08 & 7.82 & 7.65& 6.59 & 7.85 & 7.95 \\
      LLM (Upper Bound) & \textbf{18.35} & \textbf{33.48}  & \textbf{38.27} & \textbf{90.23} & \textbf{7.85} & \textbf{7.75} & \textbf{6.73} & \textbf{8.02} & \textbf{8.30}\\
      \midrule
      \midrule
      \rowcolor{light-gray}
    \multicolumn{10}{c}{\textbf{\textit{test-age / Age}}} \\
    \midrule
    SALMONN\cite{tang2024salmonn} &10.03&24.95&23.55&88.10&5.41 & 4.66 & 3.14&3.35& - \\
    Qwen-Audio \cite{chu2023qwen}  & 7.28 & 23.09 & 21.80 & 86.72 &4.43 &3.98 & 2.25& 2.50 & - \\
    Qwen2-Audio-AA \cite{chu2024qwen2}&6.81&22.72&20.51&87.47&5.19 & 4.58 & 3.01&3.14& - \\
    Qwen2-Audio-VC \cite{chu2024qwen2} &5.64&18.90&28.23&86.70&7.03 & 5.92 & 4.48&5.06& - \\
      Cascade LLM & 15.36 & 31.96 & 31.99 & 90.08 & 7.22 & 7.16 & 6.46 & 4.47 & 6.51\\
      VS-LLM & 17.22 & 34.17 & 33.78 & 90.63 & 7.74 & 7.39 & 7.25 & 7.95 & 7.11 \\
      LLM (Upper Bound) & \textbf{18.78} & \textbf{35.62} & \textbf{36.01} & \textbf{91.00} & \textbf{7.82} & \textbf{7.54} & \textbf{7.40} & \textbf{8.25} & \textbf{7.44} \\
      \midrule
      \midrule
      \rowcolor{light-gray}
    \multicolumn{10}{c}{\textbf{\textit{test-env / Environment}}} \\
    \midrule
    SALMONN\cite{tang2024salmonn} &2.87&16.53&21.37&86.71&4.70 & 5.00 & 3.40&3.56& - \\
    Qwen-Audio \cite{chu2023qwen}  & 2.37 & 16.83 & 17.50 & 85.81 & 3.77 & 1.86 & 2.16 & 2.14 & - \\
    Qwen2-Audio-AA \cite{chu2024qwen2} &2.97&16.32&19.84&86.50&4.52 & 5.02 & 3.49&3.50& - \\
    Qwen2-Audio-VC \cite{chu2024qwen2} &2.06&12.35&23.40&85.17&6.30 & 6.21 & 4.85&5.30& - \\
      Cascade LLM &5.44 & 21.75& 26.41 & 88.22 & 6.03 & 5.84 & 5.31 &  5.66 & 6.62 \\
      VS-LLM & 9.42 & 25.85 & 28.27 & 89.23 & 6.14 & 5.88 & 5.10 & 5.82 & 7.11 \\
      LLM (Upper Bound) & \textbf{11.72} & \textbf{27.95}  & \textbf{31.50} & \textbf{89.73} & \textbf{7.14} & \textbf{7.14} & \textbf{6.25} & \textbf{7.40} & \textbf{8.13}\\
    \bottomrule
    \end{tabular}
    }
    \label{tab:emotion}
\end{table}

\subsubsection{Main Results}
Table \ref{table:main_result} presents the main results of all models on SD-Eval.
First, across all four test sets, VS-LLM consistently outperforms Cascade LLM on every metric. This finding indicates that using speech as a direct input enables VS-LLM to implicitly learn paralinguistic and environmental information.
Second, VS-LLM performs worse than LLM (Upper Bound). One possible explanation is that VS-LLM implicitly acquires both content and paralinguistic/environmental information directly from speech, whereas the LLM (Upper Bound) uses ground-truth transcripts and labels. This suggests that the method of processing input data is critical for model performance.

Regarding the open-source models, while SALMONN \cite{tang2024salmonn} achieves better or comparable results relative to Qwen2-Audio-AA \cite{chu2024qwen2} and Qwen-Audio \cite{chu2023qwen}, Qwen2-Audio-VC \cite{chu2024qwen2} significantly outperforms the other open-source models because its voice chat mode is more suitable for conversational tasks. Nevertheless, the overall performance of open-source models on SD-Eval is not particularly impressive. This highlights the current lack of well-defined tasks and datasets in this domain.

\subsection{Text to Audio}

\subsubsection{Evaluation Metrics}
\paragraph*{Subjective Evaluation} Mean Opinion Score (MOS) are conducted from two aspects: audio quality and temporal controllability. 
Audio quality considers the naturalness, distortion, and event accuracy of the generated audio. 
Temporal controllability evaluates the accuracy of timestamp / frequency control.
For each task, $5$ audio clips from each model are rated by $10$ evaluators, and the mean score is calculated.

\paragraph*{Objective Evaluation}
The commonly used FAD in audio generation tasks is utilized to assess the quality of generated audio~\cite{kilgour2019frechet}.
The temporal condition in the timestamp / frequency caption is used as the ground truth for evaluation. 
TAG~\cite{xu2024towards} is employed to detect the on- and off-sets of segments in the generated audio.
(a) For the timestamp control task, the accuracy of the detected segments is assessed by the segment F$_\text{1}$ score~\cite{mesaros2016metrics}, a commonly used metric in sound event detection. 
(b) For the frequency control task, accuracy is measured by the absolute difference between the specified frequency in the caption and the detected occurrence frequency in the audio. 
The difference is averaged on test samples $N$ and number of class $C$, denoted as $L_1^{\text{freq}}$:
\begin{equation}
L_1^{\text{freq}} = \frac{1}{N*C}\sum_{n=1}^{N}{\sum_{c=1}^{C}{|\#specified - \#detected|}}
\end{equation}

Simulated audios in the test set are utilized as the ground truth to obtain an objective upper bound, since there are inevitable errors in TAG detecting results.
% TAG cannot detect and localize audio events with $100\%$ accuracy. 

\begin{table*}[t]
\renewcommand{\arraystretch}{1.1}
    \centering
    \caption{ 
    Main results of temporally controllable audio generation.     
    F1$_{\text{segment}}$ / $\bm{L_1^{\text{freq}}}$ respectively measures the \textbf{timestamp alignment} / \textbf{occurrence frequencies error} between generated audio and input conditions.
    \textbf{FAD}  measures the audio quality.
    \textbf{MOS} denotes subjective metrics.
    Ablation study: \textbf{``w/o T''} indicates that the model does not utilize timestamp matrix $\mathcal{O}$, which shares a similar framework with the baseline models.
    }
    \resizebox{\textwidth}{!}{
    \begin{tabular}{c|cc|cc|cc|cc}
    \toprule
    \multicolumn{1}{c|}{Condition}&\multicolumn{4}{c|}{Timestamp}&\multicolumn{4}{c}{Occurrence Frequency}  \\
    \midrule
    \multicolumn{1}{c|}{Metrics} &F1$_{\text{segment}}$  &MOS$_{\text{control}}$&FAD$\downarrow$&MOS$_{\text{quality}}$&$L_1^{\text{freq}}\downarrow$&MOS$_{\text{control}}$&FAD$\downarrow$&MOS$_{\text{quality}}$ \\
    \midrule
    \midrule
    \multicolumn{9}{c}{Single Event}\\
    \midrule
    \midrule
    %\multirow{5}{4em}{Single Event} &
    Ground Truth   &0.797 & 4.78 &0 & 4.44 &0.302& 4.9 &0& 4.38 \\
    AudioLDM2     &0.675& 2.14 &10.853& 3.34    &2.408& 2.3  &20.677& 3.68 \\
    Audit       &0.566& 1.98 &11.774&2.82  &2.060& 2.22 &11.999 & 3.54   \\
    PicoAudio w/o T & 0.694 & 2.78 &  5.926 & \textbf{4.2} & 1.25 & 2.92 & 5.923 & \textbf{4.2} \\
    PicoAudio (Ours) &\textbf{0.783} &\textbf{4.58} &\textbf{3.175} &4.16 &\textbf{0.537} &\textbf{4.92} &\textbf{2.295} &4.1\\
    \midrule
    \midrule
    \multicolumn{9}{c}{Multiple Events}\\
    \midrule
    \midrule
    %\multirow{5}{4em}{Multiple Events} &
    Ground Truth   &0.787& 4.6 &0& 4.38 &0.447& 4.68 &0& 4.56\\
    AudioLDM2      &0.593&1.82&10.112&2.36&2.046&2.14&18.334  &2.3  \\
    Audit        &0.520&2.2&10.979 &2.72   &1.851&2.48&11.769 &3.24    \\
    PicoAudio w/o T & 0.614 & 2.12 & 5.218 & 3.42 & 1.216 &2.1 &  5.215 &3.3 \\
    PicoAudio (Ours) &\textbf{0.772}&\textbf{4.84}&\textbf{2.863}&\textbf{4.12}&\textbf{0.713}&\textbf{4.6}   &\textbf{2.1823}&\textbf{4.38}\\
    \bottomrule
    \end{tabular}
    }
    \label{tab:pico_result}
\end{table*}

\subsubsection{Main Results}
The control of timestamp and occurrence frequency are evaluated separately on both single-event and multiple-event test sets.
The results are presented in table~\ref{tab:pico_result}.
Two mainstream audio generation models, AudioLDM2~\cite{liu2023audioldm2} and Audit~\cite{wang2023audit}, are employed as baselines.
Both subjective and objective metrics demonstrate that PicoAudio surpasses baseline models.

%\paragraph*{Timestamp \& Occurrence Frequency Control}
The timestamp controlled audio clips generated by PicoAudio are very close to the ground truth (upper bound), demonstrating the precision of control, whether in single-event or multi-event tasks.
PicoAudio introduces tailored modules to convert the textual timestamp information into a timestamp matrix, achieving exact control of timestamp in the generated audio at a time resolution of $40$ ms.
Equipped with GPT-4, PicoAudio demonstrates outstanding performance in the frequency error metric $L_1^{\text{freq}}$.
Even in the presence of TAG errors, it achieves an average error rate of $0.537$ / $0.713$ occurrences per sound event on the single-event / multi-event tasks, respectively.
Achieving $L_1^{\text{freq}}$ less than $1$, which is close to the ground truth, PicoAudio has demonstrated practicality in frequency controlling.

Baseline text-to-audio generation models, however, fall short in performance.
They obtain lower F1$_{\text{segment}}$ scores and produce a frequency error around $2$ times per event, as they tend to excessively repeat events when the input text contains temporal conditions.
Furthermore, the ablation study employs a model trained on simulated data without using timestamp matrix $\mathcal{O}$, which shares a similar framework with baseline model. 
The ablation results lie between baseline models and PicoAudio, indicating that achieving precise control requires not only temporally-aligned audio-text data but also specific model design.

\section{Conclusion}
In this report, we presented an overview of Amphion with it latest developments in v0.2.
It is an open-source toolkit designed to advance the fields of audio, music, and speech generation. Building upon its initial release, Amphion v0.2 introduces significant enhancements, including a 100K-hour multilingual speech dataset, tailored datasets for specific downstream tasks, and expanded support for state-of-the-art generative models across various applications.
Future developments aim to further enhance Amphion’s capabilities by incorporating additional generative tasks, improving code readability, and growing Amphion's open-source community. With Amphion v0.2, we hope to lower the barriers for entry and inspire new advancements in audio, music, and speech generation technologies.

\bibliographystyle{plain}
\bibliography{references}

\end{document}